\documentclass[prd,aps,twocolumn,a4paper,showkeys,nofootinbib]{revtex4-1}
\usepackage{amsmath}
\usepackage{amsfonts}
\usepackage{amssymb}	
\usepackage{graphicx}
\usepackage{amsbsy}
\usepackage{color}
\usepackage{commath}
\allowdisplaybreaks

\usepackage{hyperref}
\usepackage{makecell}
\usepackage{multirow}

\def\GMc2{G M_{\odot} c^{-2}}

\def\O{\mathcal{O}}

\def\F{{\cal F}}
\def\lm{{\ell m}}

\def\lm{{\ell m}}

\def\lm{{\ell m}}

\def\l{{\ell }}

\def\F{{\cal F}}
\def\Fr{{\hat{\F}_r}}
\def\Fphi{{\hat{\F}_{\varphi}}}

\def\O{{\cal O}}

\def\ha{{\hat{a}}}

\def\TEOBResumS{\texttt{TEOBResumS}}

\def\SEOBNR{{\texttt{SEOBNRv4HM}}}
\def\SEOBNRe{{\texttt{SEOBNRv4EHM}}}
\def\Teukode{{\texttt{Teukode}}}

\def\Jteuk{{\dot{J}_{\rm teuk}}}
\def\Eteuk{{\dot{E}_{\rm teuk}}}

\def\HKerr{{\hat{H}^{\rm eq}_{\rm Kerr}}}

\def\Fphiqc{{\hat{\F}_\varphi^{\rm QC}}}
\def\Frqc{{\hat{\F}_r^{\rm QC}}}
\def\Fphiecc{{\hat{\F}_\varphi^{\rm ecc}}}
\def\Frecc{{\hat{\F}_r^{\rm ecc}}}

\def\FphiK{{\Fphi^{\rm \QC2PN{}}}}
\def\FrK{{\Fr^{\rm \QC2PN{}}}}

\newcommand\fnp[1]{{\hat{f}_{\varphi #1}^{\rm N _{nc}}}}
\def\NP22{{\fnp{,22}}}

\newcommand{\QC}{QC}
\newcommand{\QCtPN}{QC2PN}
\newcommand{\NCN}[1]{NCN{#1}}

\newcommand\be{\begin{equation}}
\newcommand\ee{\end{equation}}

\usepackage{color}
\definecolor{cyan}{rgb}{0,0.9,0.9}
\definecolor{orange}{rgb}{0.9,0.5,0}
\definecolor{magenta}{rgb}{1,0,1}
\definecolor{purple}{rgb}{0.8,0.4,0.8}
\definecolor{gray}{rgb}{0.8242,0.8242,0.8242}
\definecolor{dodgerblue}{rgb}{0.12, 0.56, 1.0}
\definecolor{darkgrey}{rgb}{0.5,0.5,0.5}
\definecolor{darkgreen}{rgb}{0,0.65,0}

\definecolor{colortab1}{rgb}{0.1, 0.1, 1.0}
\definecolor{colortab2}{rgb}{0.9,0,0.1}

\begin{document}

\title{
  Assessment of effective-one-body radiation reactions for generic planar orbits
  }

\author{Simone \surname{Albanesi}${}^{1,2}$}
\author{Alessandro \surname{Nagar}${}^{2,3}$}
\author{Sebastiano \surname{Bernuzzi}${}^{4}$}
\author{Andrea \surname{Placidi}${}^{5,6}$}
\author{Marta \surname{Orselli}${}^{5,6}$}
\affiliation{${}^{1}$ Dipartimento di Fisica, Universit\`a di Torino, via P. Giuria 1, 
10125 Torino, Italy}
\affiliation{${}^2$INFN Sezione di Torino, Via P. Giuria 1, 10125 Torino, Italy} 
\affiliation{${}^3$Institut des Hautes Etudes Scientifiques, 91440 Bures-sur-Yvette, France}
\affiliation{${}^4$Theoretisch-Physikalisches Institut, Friedrich-Schiller-Universit{\"a}t 
Jena, 07743, Jena, Germany}  
\affiliation{${}^5$Dipartimento di Fisica e Geologia, Universit\`a di Perugia,
INFN Sezione di Perugia, Via A. Pascoli, 06123 Perugia, Italia}
\affiliation{${}^6$Niels Bohr Institute, Copenhagen University,  Blegdamsvej 17, DK-2100 Copenhagen \O{}, Denmark}

\begin{abstract}
In this paper we assess the performances of different analytical
prescriptions for the effective-one-body (EOB) radiation reaction
along generic planar orbits using exact numerical result in the test-mass limit.
We consider three prescriptions put forward in the recent literature: 
(i) the quasicircular prescription (\QC{}), 
(ii) the \QC{} with second post-Newtonian (2PN) order noncircular corrections (\QC2PN{}), and
(iii) the \QC{} corrected by the noncircular Newtonian prefactor (\NCN{}).
The analytical fluxes are then compared against the exact fluxes that are computed by solving 
the Teukolsky equation with a test-mass source in geodesic motion. 
We find that the \NCN{} prescription is the most accurate for both
eccentric and hyperbolic orbits and it is in robust agreement also for
large values of the eccentricity. 
This result carries over to the comparable masses, as we discuss for a
numerical-relativity (NR) case study. 
We also demonstrate that, while the EOB/NR waveform
unfaithfulness is a necessary check for the precision  
of EOB models, the direct comparison of EOB/NR fluxes is a more
stringent and informative test to select the best prescription.
Finally, we propose an improved radiation reaction, \NCN{2PN}, that
includes noncircular 2PN corrections, in resummed form, as a further
multiplicative contribution and that is valid for any mass ratio.
\end{abstract}
\date{\today}

\maketitle

%===========================================================================================
\section{Introduction}
%===========================================================================================
\label{sec:introduction}
Motivated by recent observational hints for eccentric and hyperbolic binary black hole mergers in some LIGO-Virgo 
events \cite{LIGOScientific:2020iuh,Gayathri:2020coq,Gamba:2021gap} and by future prospect of detecting binary sources on generic orbits with 
LISA~\cite{LISA:2017pwj,Babak:2017tow,Gair:2017ynp}, the analytical relativity community has intensified the efforts
in building up accurate gravitational-wave (GW) models for {\it noncircularized} coalescing compact binaries~\cite{Hinder:2017sxy,Hinderer:2017jcs,Chiaramello:2020ehz,Nagar:2020xsk,Islam:2021mha,Nagar:2021gss,Nagar:2021xnh,Albanesi:2021rby,Liu:2021pkr,Yun:2021jnh,Tucker:2021mvo,Setyawati:2021gom,Khalil:2021txt,Ramos-Buades:2021adz,Placidi:2021rkh}.
Among the possible approaches, the effective-one-body (EOB) framework~\cite{Buonanno:1998gg,Buonanno:2000ef,Damour:2000we,Damour:2001tu,Damour:2015isa} 
 offers a theoretically complete method for a unified description of different classes of astrophysical binaries, from the comparable masses to the intermediate and extreme-mass-ratio regimes, and for incorporating the fast motion and ringdown dynamics. Together with the Hamiltonian, the key building block of an EOB model is the radiation reaction force, which is currently prescribed by sophisticated analytically resummed expressions building on the circular, factorized EOB waveform of Ref.~\cite{Damour:2008gu}. The purpose of this work is to assess the performances of different prescriptions for the EOB radiation reaction along generic planar orbits using exact 
numerical result in the test-mass limit.

The generalization of the quasicircular EOB model \TEOBResumS{}~\cite{Nagar:2020pcj,Riemenschneider:2021ppj} 
to nonquasicircular configurations~\cite{Chiaramello:2020ehz,Nagar:2020xsk,Nagar:2021gss} 
has allowed the construction of the first, and currently only, 
waveform model for spin-aligned black hole binaries that can accurately deal 
with both hyperbolic captures~\cite{East:2012xq,Gold:2012tk,Nagar:2020xsk,Nagar:2021gss,Nagar:2021xnh}
and eccentric inspirals~\cite{Chiaramello:2020ehz,Nagar:2021gss,Nagar:2021xnh}.
The model of Refs.~\cite{Chiaramello:2020ehz,Nagar:2021gss} has been used to analyze the 
GW source GW190521~\cite{LIGOScientific:2020iuh,LIGOScientific:2020ufj} under the hypothesis 
of a hyperbolic capture~\cite{Gamba:2021gap}. 
In its latest avatar~\cite{Nagar:2021xnh}, the \TEOBResumS{} model for generic, but planar, orbital
motion displays: (i) an excellent agreement
with the 28 numerical relativity (NR) public simulations of  eccentric
inspirals of the Simulating eXtreme Spacetimes (SXS) catalog 
(EOB/NR unfaithfulness mostly around $0.1\%$ except a single outlier $\sim 1\%$)
and (ii) a high consistency (maximum disagreement $\sim 4\%$ for the most relativistic configurations) 
with the only available NR calculation of the scattering angle~\cite{Damour:2014afa}.
Parallel work has extended the quasicircular \SEOBNR{} model to to eccentric inspirals~\cite{Cao:2017ndf,Liu:2019jpg,Liu:2021pkr,Yun:2021jnh} 
and presented qualitative results for EOB waveforms for hyperbolic encounters~\cite{Ramos-Buades:2021adz}. 
In particular \SEOBNRe~\cite{Ramos-Buades:2021adz} showed a EOB/NR unfaithfulness mostly around $0.1\%$
for all the 28 numerical relativity datasets available from the SXS collaboration.
The EOB radiation reaction consists of an azimuthal component ${\cal F}_\varphi$ and a radial component ${\cal F}_r$. 
An accurate analytical representation of $({\cal F}_\varphi,{\cal F}_{r})$ is of paramount importance to correctly 
track the orbital phase during the inspiral. In the quasicircular case, many years of work eventually brought  
a highly sophisticated analytical expression of $\cal{F}_\varphi$, as used in \TEOBResumS{}, 
that incorporates various levels of resummation~\cite{Damour:2008gu,Pan:2010hz,Nagar:2016ayt,Messina:2018ghh,Nagar:2019wrt}.
Key benchmarks for the development of these resummed analytical
expressions are the exact fluxes in the test-mass limit that can be numerically computed by solving the
Regge-Wheeler or Teukolsky perturbation equations~\cite{Cutler:1993vq,Bernuzzi:2011aj,Harms:2014dqa}. 
Remarkably, the prescriptions developed in that limit proved reliable {\it also} when compared with NR-computed fluxes for
binaries of comparable masses~\cite{Albertini:2021tbt}.
Note that such systematic and progressive development in the test-mass 
regime is absent in the \SEOBNR{} model, whose radiation reaction force
has a less sophisticated analytical structure than 
the \TEOBResumS{} one. This might be one of the reasons behind the 
fact that the \TEOBResumS{} EOB/NR faithfulness for quasicircular, spin-aligned,
binaries is, on average, larger by approximately 1 order of magnitude than 
the \SEOBNR{} EOB/NR faithfulness. Indeed the larger faithfulness values are 
obtained for small masses, i.e. during the radiation reaction-dominated inspiral,
see Fig.~18 of Ref.~\cite{Albertini:2021tbt}.

The noncircular radiation reaction of \TEOBResumS{} has been developed
with the same rationale as the quasicircular one by systematically using the test-mass benchmark~\cite{Chiaramello:2020ehz}.
The studies of test-mass flux data of Refs.~\cite{Chiaramello:2020ehz,Albanesi:2021rby} 
found that the most reliable analytical representation is given by {\it dressing} the 
quasicircular (resummed) quadrupolar flux mode with
the leading-order, Newtonian, {\it generic} (i.e., noncircular) correction factor.
This noncircular, Newtonian prescription is indicated here as \NCN{}.
An analogous approach can be followed, multipole by multipole, for the waveform~\cite{Chiaramello:2020ehz},
which can be further improved by incorporating higher post-Newtonian (PN) corrections as suitably resummed multiplicative 
factors, as done in Ref.~\cite{Placidi:2021rkh} at 2PN accuracy.
% 
%============================
% Fig.1: geodesic motion and flux test
%============================
\begin{figure*}[t]
  \center
  \includegraphics[width=0.28\textwidth]{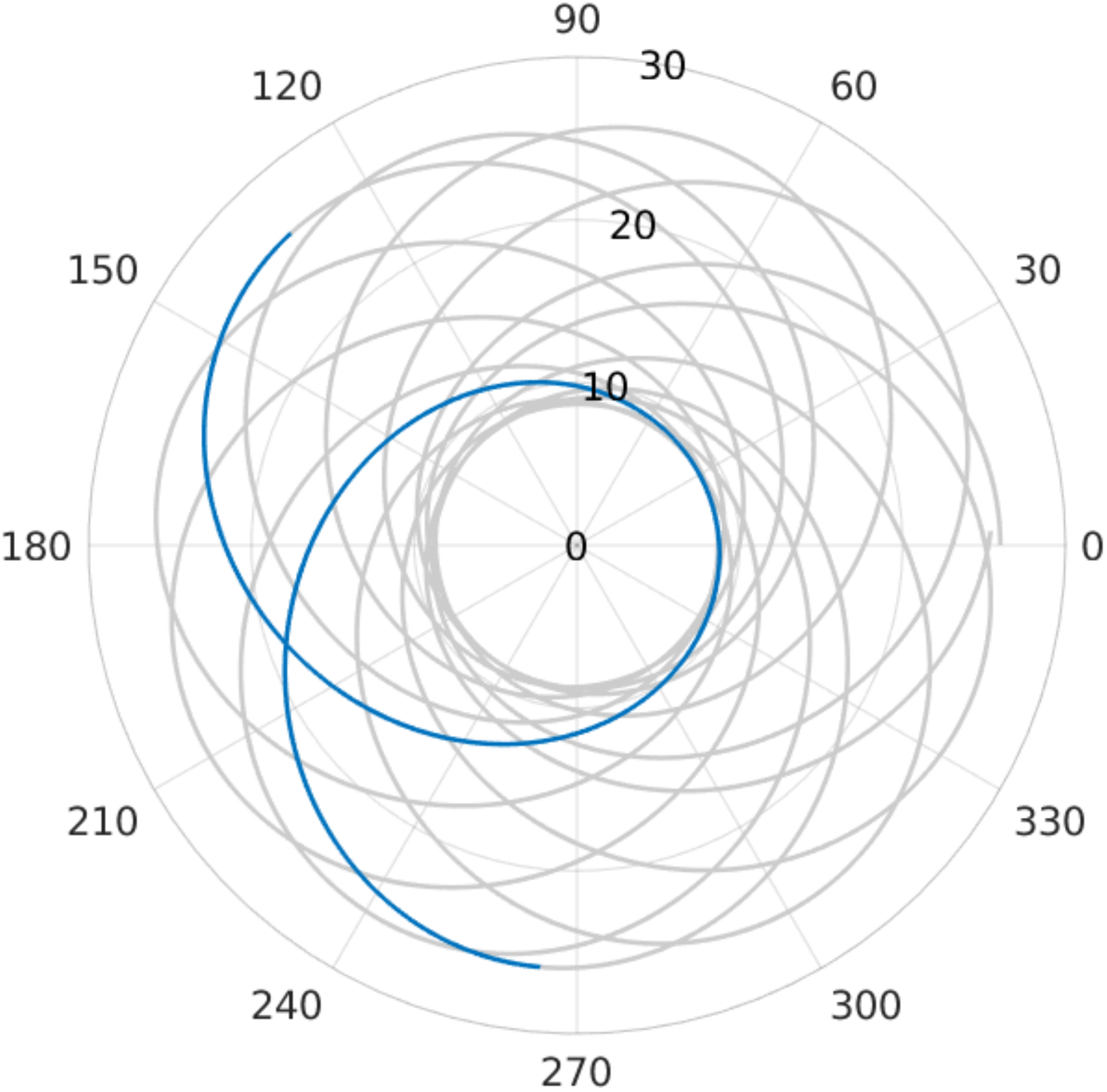}
  \includegraphics[width=0.32\textwidth]{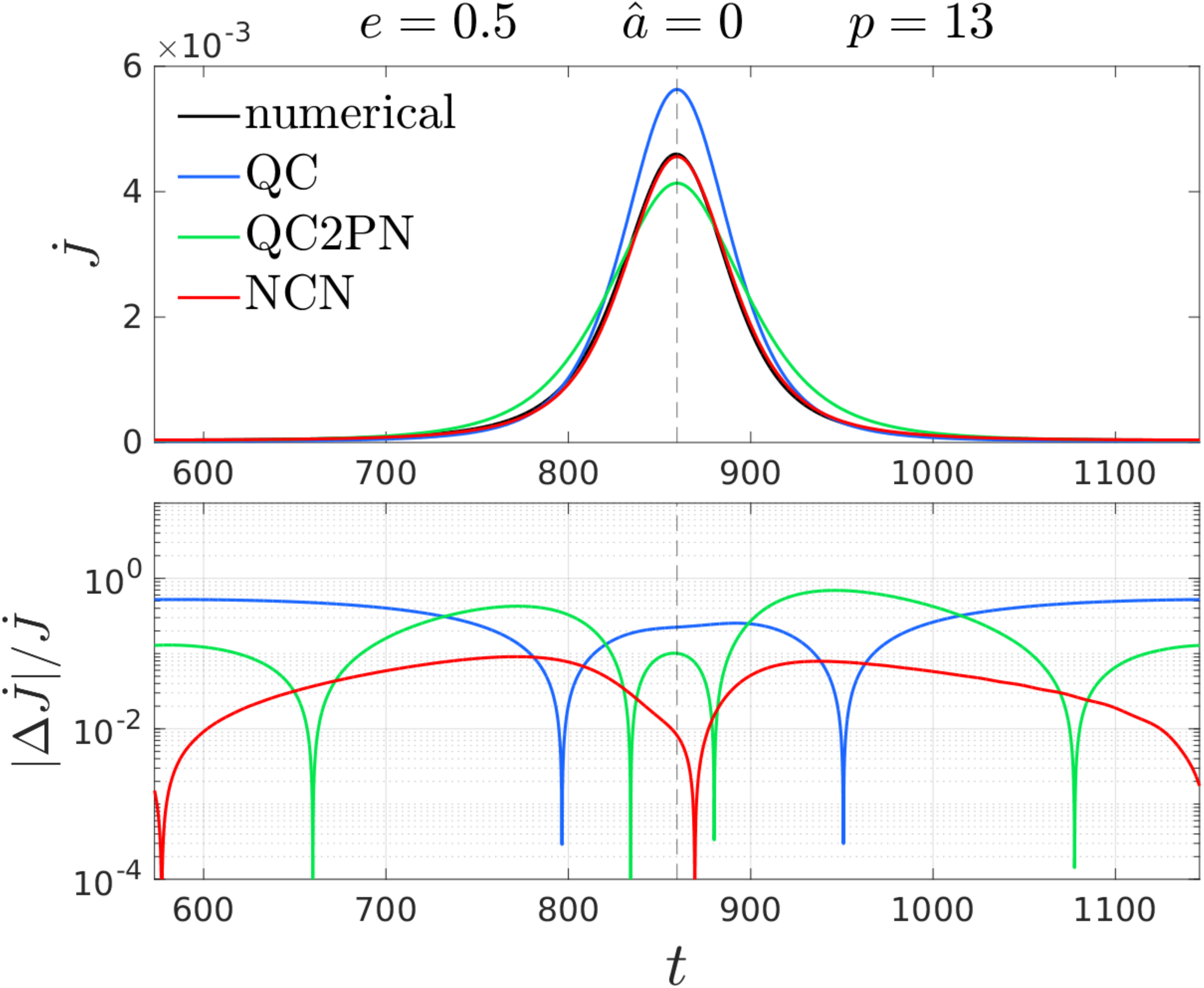}
  \includegraphics[width=0.32\textwidth]{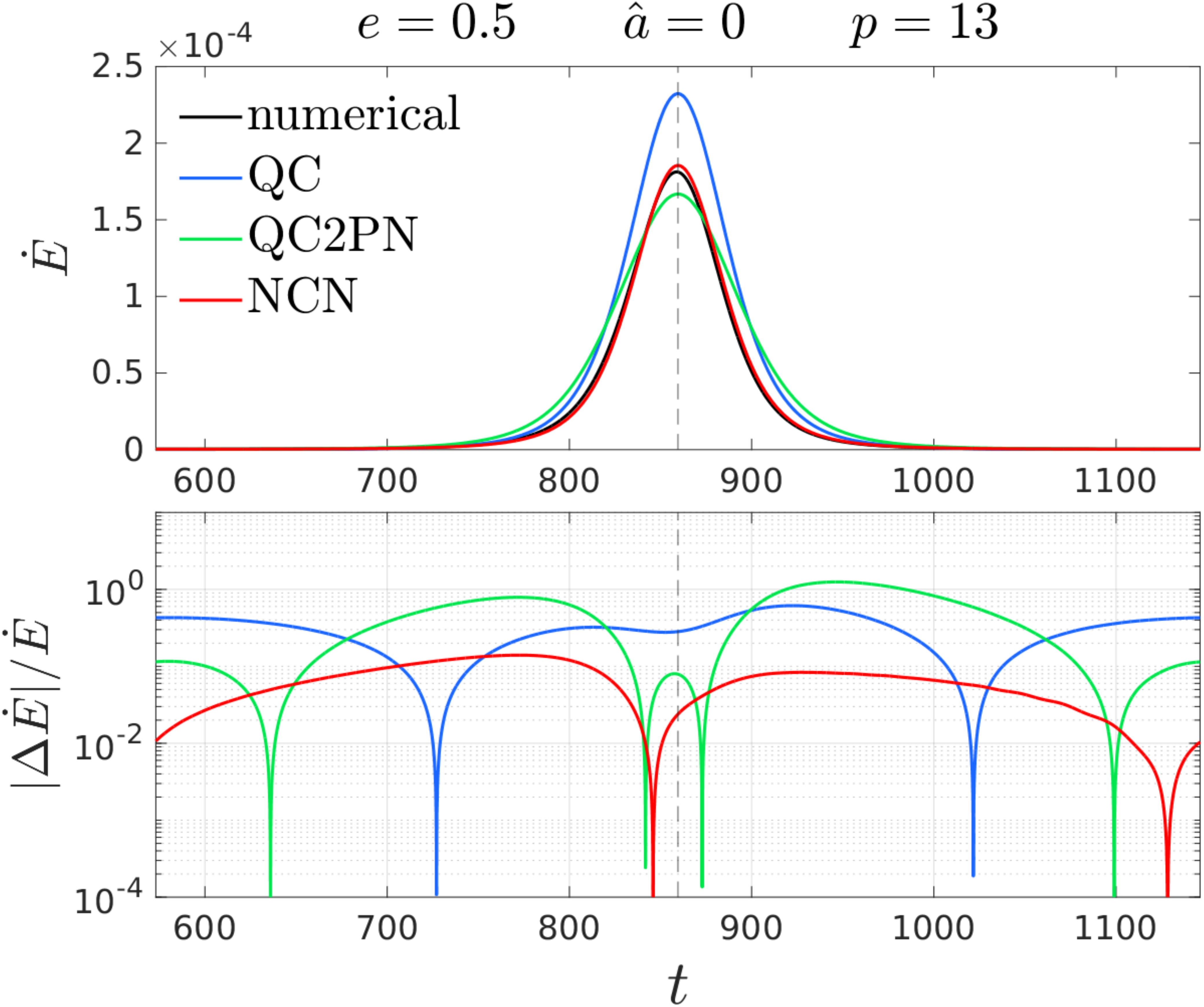}
  \caption{\label{fig:overview} 
  Test particle orbiting around a Schwarzschild black hole ($\hat{a}=0$) on a geodesic 
  with eccentricity $e=0.5$ and semilatus rectum $p=13$. One radial period is highlighted
  on the trajectory (left panel) and we show the corresponding fluxes. We contrast
  the numerical fluxes (black) and the three different analytical fluxes considered in this 
  work: (i) the \NCN{} flux (red), as used in {\TEOBResumS}, obtained
  from Eqs.~\eqref{eq:FphiTEOB} and~\eqref{eq:FrTEOB}; (ii) the \QC{} flux
  (blue) computed from Eq.~\eqref{eq:RRqc}, which 
  is a proxy of the fluxes of {\SEOBNRe}~\cite{Ramos-Buades:2021adz};
  and (iii) the \QC2PN{} flux (green) computed using  Eqs.~\eqref{eq:FKhalil} as 
  taken from Khalil {\it et al.}~\cite{Khalil:2021txt}. The analytical/numerical relative
  differences are shown in the bottom panels. The \NCN{} analytical fluxes deliver the
  closer agreement with the numerical ones either at apastron and periastron. 
  See Sec.~\ref{sec:RR} and~\ref{sec:test} for full discussion.}
\end{figure*}
%
%======================
% Fig.2: effect on the dynamics
%======================
\begin{figure}[t]
  \center
  \includegraphics[width=0.47\textwidth]{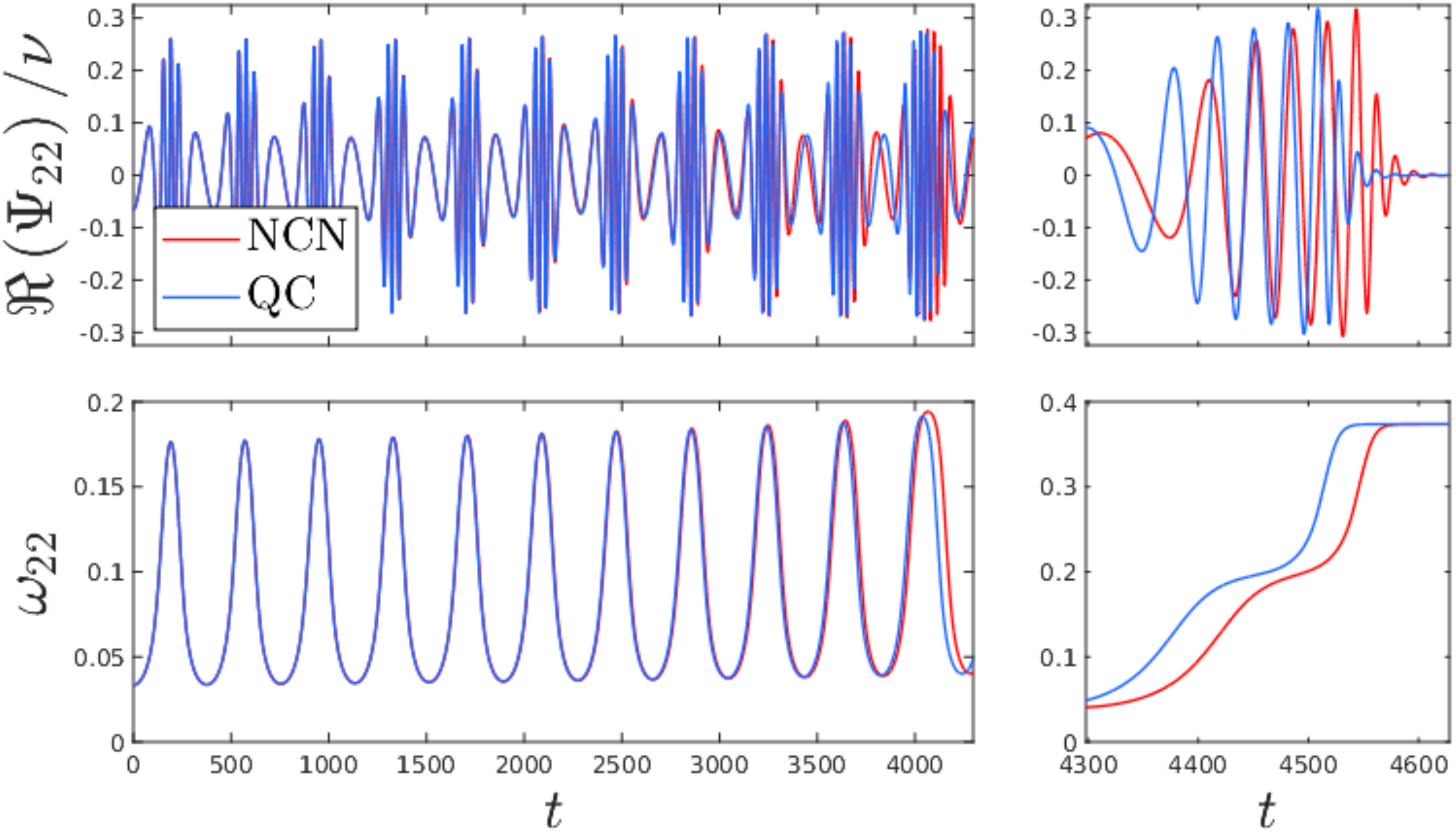}\\
  \caption{\label{fig:overview_wave} EOB $\ell=m=2$ Regge-Wheeler-Zerilli 
  normalized strain waveforms of 
  a test-particle inspiraling and plunging around a Schwarzschild black hole. Initial eccentricity 
  $e_0$ and semilatus rectum $p_0$ are $(e_0,p_0)=(0.5,7.6)$.
  The inspiral behind the red waveform is driven by the \NCN{} flux, while that 
  of the blue one by the \QC{} flux. 
  In the bottom panels we show the corresponding quadrupolar frequencies.
  The fact that the \QC{} flux overestimates the correct answer (see Fig.~\ref{fig:overview}) 
  at periastron yields an {\it unphysical} acceleration of the inspiral related to the (small 
  but non-negligible) 
  circularization of the orbit.}
\end{figure}
By contrast, a different approach was taken in the {\SEOBNRe} model~\cite{Ramos-Buades:2021adz},
that uses the standard quasicircular (\QC{}) azimuthal radiation reaction force $\cal{F}_\varphi$
inherited from the quasicircular \SEOBNR{} model~\cite{Bohe:2016gbl,Cotesta:2018fcv}, 
without taking care of explicit nonradial contributions to the loss of angular
momentum. Despite the small EOB/NR unfaithfulness 
found in Ref.~\cite{Ramos-Buades:2021adz} for (mild) eccentricities, 
no dedicated study of the \QC{} radiation reaction in eccentric inspirals 
was undertaken.
Reference~\cite{Khalil:2021txt} computed noncircular corrections up to 2PN
order in the waveform that, once incorporated into the \QC{} waveform,
constitute another possible prescription for the EOB radiation
reaction. We refer to this prescription as \QCtPN{}.
The 2PN corrections, though available, were not included 
Ref.~\cite{Ramos-Buades:2021adz} and thus the \QCtPN{} 
choice remains so far totally untested against numerical data.

This paper aims at assessing the three prescriptions for the EOB radiation
reaction: 
(i) the \QC{} in the noncircular case (as used in {\SEOBNRe}~\cite{Ramos-Buades:2021adz}),
(ii) the \QCtPN{} proposed in Ref.~\cite{Khalil:2021txt} (not
previously implemented in any complete waveform model),
and (iii) \NCN{} of Refs.~\cite{Chiaramello:2020ehz,Nagar:2021gss,Nagar:2021xnh} (used in
\TEOBResumS{}).
The analytical fluxes are compared to the exact numerical fluxes
emitted at null infinity by a test-particle orbiting along 
eccentric equatorial orbits in Kerr spacetime, as computed solving the
2+1 Teukolsky equation on hyperboloidal slices~\cite{Harms:2014dqa}.

\subsection{Overview of the results}

The main result of the paper is illustrated in Fig.~\ref{fig:overview}
for a representative case in Schwarzschild spacetime for eccentricity
$e=0.5$. The figure shows the angular momentum and energy fluxes
(middle and right panels) computed along a portion of the geodesic trajectory
(left panel). 
The \QC{} prescription (blue line) is the least accurate, 
as {\it a priori} expected. The \QCtPN{} of Ref.~\cite{Khalil:2021txt} is more 
reliable than the \QC{} prescription, though differences up to several
percent with respect to numerical data are found.
The \NCN{} prescription performs best in reproducing the numerical
results; this is a feature of almost all the cases analyzed in this work.
To be more quantitative, we considered the analytical/numerical relative differences
between orbital-averaged fluxes. For the \NCN{}
angular momentum fluxes we get $0.6\%$, for the \QCtPN{}  
fluxes we obtain $9.8\%$, and  $14.5\%$ for the \QC{}. For the energy fluxes the hierarchy is the 
same and we get $0.8\%$, $18.1\%$, and $33.4\%$, respectively.
At lower eccentricities, the discrepancies between analytical expressions are milder, while for higher 
eccentricities the differences are more evident. In particular, for high
eccentricity the  \QC{} and \QCtPN{} prescriptions exhibit 
differences up to a factor 2 with the numerical results at apastron.
The effects on the dynamics, and thus on the waveform, of using different radiation 
reactions can be seen for a representative case in Fig.~\ref{fig:overview_wave}. 
As usual, we adopt Regge-Wheeler-Zerilli normalized multipolar waveforms 
$\Psi_\lm=h_\lm/\sqrt{(\ell+2)(\ell+1)\ell(\ell-1)}$ where the strain multipoles are defined as
\be
h_+-ih_\times=\dfrac{1}{D_L}\sum_{\ell=2}^{\infty}\sum_{m=-\ell}^{\ell}h_\lm {}_{-2}Y_{\lm},
\ee
where $D_L$ is the luminosity distance and ${}_{-2}Y_{\lm}$ are the $s=-2$ spin-weighted spherical harmonics. 
In the top panel of Fig.~\ref{fig:overview_wave}, we contrast the real part of the $\ell=m=2$
waveform obtained with the dynamics driven by the \NCN{}
radiation reaction (red) with that obtained with the
\QC{} prescription (blue).  The bottom panel of the figure displays the instantaneous gravitational
wave frequency, defined as $\omega_{22}=\dot{\phi}_{22}$ from the amplitude and phase 
decomposition of the waveform $\Psi_{22}=A_{22}\exp(-i\phi_{22})$.
The effects of the different radiation reaction start to be evident after 6-7 orbits, 
with the \QC{} flux, which is large at the apastron, that progressively circularizes the 
orbits more and eventually drives a faster transition from inspiral to plunge and merger.
The stronger circularization effect of the \QC{} flux is visualized by the frequency evolution
in the bottom panel of Fig.~\ref{fig:overview_wave}, with the amplitude of the \QC{} oscillation
that is getting progressively smaller than the \NCN{} one.
We will see that in the comparable-mass case the action of the radiation reaction on the 
dynamics is stronger and thus the discrepancies can show up already on 
a shorter and less eccentric inspiral, as discussed in Sec.~\ref{sec:comp_mass}.

The reminder of the paper is structured as follows. In Sec.~\ref{sec:EOB} 
we recall the main elements of the EOB model and write explicitly the test-particle
limit.
Section~\ref{sec:RR} describes in detail  the different analytical prescriptions 
used in current EOB models that are finally tested in Sec.~\ref{sec:test} in
the test-mass limit.
Section~\ref{sec:comp_mass} studies the effects of using different radiation
reactions in the comparable-mass case. Finally, in Sec.~\ref{sec:2PN_resum} 
we present a new Newtonian-factorized radiation reaction with 2PN noncircular
corrections.
Note that a more detailed discussion of the topics presented in Sec.~\ref{sec:EOB} 
can be found in Ref.~\cite{Albanesi:2021rby}.

Throughout this paper, we use geometrized units $G=c=1$, even if sometimes
we will write $c$ explicitly to highlight the PN orders. Moreover, the time and the 
phase-space variables used in this work are related to the physical ones by $t=T/(GM)$, 
$r=R/(GM)$, $p_{r}=P_{R}/\mu$ and $p_\varphi=P_\varphi/(\mu GM)$,
where $M$ is the total mass of the system and $\mu$ its reduced mass.

%===========================================================================================
\section{EOB dynamics and test-mass limit}
%===========================================================================================
\label{sec:EOB}
The EOB model maps the relative motion of two bodies with masses $m_1$ and $m_2$ 
into the motion of an effective body into an effective metric that is a continuous $\nu$-deformation 
of the Kerr metric, with $\nu \equiv m_1 m_2/(m_1 + m_2)$ being the symmetric mass ratio. 
Setting $\nu=0$ we recover the Kerr metric and therefore we describe the motion of a 
test particle with mass $\mu=\nu M$ in Kerr spacetime, where $M$ is the mass of the Kerr
black hole. The EOB model is based on three building blocks:
(i) the EOB Hamiltonian $\hat{H}_{\rm EOB}$~\cite{Damour:2014sva,Rettegno:2019tzh} 
that describes the conservative part of the dynamics;
(ii) a radiation reaction that consists in an azimuthal and a radial component $(\F_\varphi, \F_r)$ 
and accounts for the loss of energy and angular momentum due to the emission of GWs;
(iii) a prescription for the multipolar waveform $h_\lm$ at infinity.
Since the description of comparable-mass binaries within the EOB framework is continuously 
connected to the motion of a test particle by the symmetric mass ratio $\nu$, any analytical
structure of the model must be robust and accurate in the test particle limit. Seen the other 
way around, the test particle limit is a useful {\it laboratory} to learn and develop new ideas, 
as originally advocated for in seminal EOB works~\cite{Nagar:2006xv,Damour:2007xr,Damour:2008gu}.

Here we consider eccentric (and hyperbolic) equatorial orbits of a test particle around
a Kerr black hole with dimensionless spin parameter $\ha$. Dynamics of this kind have been 
already discussed in Sec.~IIA of Ref.~\cite{Albanesi:2021rby} to
validate the \NCN{} 
radiation reaction, but we report here a brief summary for completeness.
The equatorial dynamics is obtained solving Hamilton's equations
\begin{align}
\dot{r} &=\left(\frac{A}{B}\right)^{1 / 2} \frac{\partial \HKerr }{\partial p_{r_{*}}}  , \\ 
\dot{\varphi} &=\frac{\partial \HKerr}{\partial p_{\varphi}} \equiv \Omega  , \label{eq:freq} \\ 
\dot{p}_{r_*} &=\left(\frac{A}{B}\right)^{1 / 2} \left( \hat{\F}_r -  \frac{\partial \HKerr}{\partial r} \right)  , \\
\dot{p}_\varphi &=\hat{\F}_{\varphi}  ,
\end{align}
where $\hat{\F}_{r,\varphi}= \F_{r,\varphi}/\nu$ and $\HKerr$ is the equatorial
$\mu$-normalized Kerr Hamiltonian
\begin{equation}
\HKerr =  \frac{2\ha p_\varphi}{r r_c^2} + \sqrt{A \del{r} \del{1 + \frac{p_{\varphi}^2}{r_{c}^2}}+p_{r_*}^2}
\end{equation}
written in terms of the gauge-invariant centrifugal radius $r_c$~\cite{Damour:2014sva}
\begin{equation}
r_c^2=r^2+\ha^2+\frac{2\ha^2}{r}\,,
\end{equation}
where $\hat a$ is the dimensionless Kerr spin parameter.

The metric functions $A(r)$ and $B(r)$ are
\begin{align}
A(r) & = \frac{1+2 u_c}{1+2 u}\left(1- 2 u_c\right) \ , \\
B(r) & = \frac{1}{1-2 u + \ha^2 u^2}\,,
\end{align}
where we used $u=1/r$ and $u_c = 1/r_c$.
In addition, $p_{r_*}$ is the momentum conjugate to the tortoise coordinate $r_*$, 
defined as $p_{r_*} = \sqrt{A/B}\;p_r$. Orbits are characterized using the eccentricity 
$e$ and the semilatus rectum $p$, whose definitions are given in terms of 
the apastron $r_+$ and the periastron $r_-$ as
\begin{equation}
\label{eq:ep_definition}
e  = \frac{r_+ - r_-}{r_+ + r_-}, \quad \quad p  = \frac{2 r_+ r_-}{r_+ + r_-}.
\end{equation}
For stable orbits,  $p$ must satisfy the condition 
$p>p_s(e,\ha)$ where $p_s(e,\ha)$ is the separatrix given by a
root of~\cite{OShaughnessy:2002tbu,Stein:2019buj}
\begin{align}
& p_s^2 (p_s - 6 - 2 e)^2 + \ha^4 (e-3)^2 (e+1)^2  \\ 
& - 2 \ha^2 (1+e) p_s \left[14 + 2 e^2 + p_s (3 - e)\right] = 0 . \nonumber
\end{align} 
Geodesics correspond to $\Fphi=\Fr=0$ and $e$, $p$ are constants of motion.
Note that $e$ and $p$ are defined only for bound motion; therefore, eccentricity
and semilatus rectum are not defined for hyperbolic configurations. In that
case, we use the energy and the angular momentum to characterize the orbits.

%===========================================================================================
\section{EOB radiation reaction}
%===========================================================================================
\label{sec:RR}
In this section we recall the prescriptions for the radiation reaction 
forces $(\Fphi, \Fr)$ that are currently used in the 
\TEOBResumS{} model (\NCN)~\cite{Nagar:2021gss,Nagar:2021xnh,Placidi:2021rkh}, 
in the \SEOBNRe{}~\cite{Ramos-Buades:2021adz} model (\QC) and proposed, but not  
implemented yet, in Ref.~\cite{Khalil:2021txt} (\QCtPN).

\subsection{Noncircular Newtonian prefactor (\TEOBResumS{})}
\label{subsec:RRteob}
The comprehensive work of Ref.~\cite{Albanesi:2021rby} clearly indicates that, among
various analytical possibilities, incorporating noncircular effects in the radiation reaction via
the Newtonian correction factor to the quadrupolar flux mode already yields an excellent
agreement ($\sim 1\%$) with the numerical fluxes up to mild values of the eccentricity.
Before discussing it explicitly, let us recall its structure for quasicircular orbits.
In this case the radial component can be
set to zero since $\Fr=0$, as this is equivalent to a gauge
choice~\cite{Buonanno:2000ef}.
Therefore, we have only the angular component, 
\begin{equation}
\label{eq:Fphiqc}
\Fphi = - \frac{32}{5} \nu r_\Omega^4 \Omega^5 \hat{f},
\end{equation}
where $r_\Omega = r \left( 1 + \ha r^{-3/2} \right)^{2/3}$ and $\hat{f}$ is the 
Newton-normalized flux that includes all the (resummed) PN corrections and reads
\begin{equation}
\label{eq:hatf}
\hat{f}  = \sum_{\l=2}^{8} \sum_{m=1}^{\l} (F_{22}^N)^{-1} F_{\lm}  = \sum_{\l=2}^{8} \sum_{m=1}^{\l} \hat{F}_\lm .
\end{equation}
The flux contributions $\hat{F}_\lm$ are written in terms of the multipolar
waveform $h_\lm$ and the variable $x =  v_\varphi^2 =r_\Omega^2 \Omega^2$:
\begin{align}
F_\lm &  = \frac{1}{8\pi} m^2 \Omega^2 |h_{\lm}(x)|^2,  \label{eq:Flm} \\
F_{22}^N & = \frac{32}{5} x^5.
\end{align} 
The $h_\lm$ multipoles are factorized and resummed building upon the original prescription of Ref.~\cite{Damour:2008gu}.
Here we use the latest development of resummed circular flux in the test particle limit as discussed in Ref.~\cite{Albanesi:2021rby}, 
which yields considerable improvements with respect to previous implementations. In particular, (i) it includes (relative) 
6PN  orbital information  in all multipoles up to $\ell=8$ included and (ii) the residual amplitude corrections are  resummed 
following the approach of Refs.~\cite{Nagar:2016ayt,Messina:2018ghh}, that is based on the separate treatment of the 
orbital and spin dependent terms. The accuracy of our circular analytical flux with respect to numerical data is recalled 
in Appendix~\ref{appendix:hatf}. It must be emphasized that this specific flux is slightly different with respect to the test particle
limit of the \TEOBResumS{} flux~\cite{Nagar:2020pcj}, e.g. in the treatment of the $\ell=m=2$ mode and in the modes with
$\ell>6$. These differences are, however, irrelevant for our current purposes.

The angular component of the generic radiation reaction used in 
{\TEOBResumS}~\cite{Nagar:2020xsk,Nagar:2021xnh,Placidi:2021rkh} for noncircular binaries 
is obtained dressing the quadrupolar contribution in $\Fphiqc$ with the corresponding 
Newtonian noncircular correction~\cite{Chiaramello:2020ehz}, explicitly
\begin{equation}
\hat{{\cal F}}_\varphi^{\rm \NCN{}} =  - \frac{32}{5} \nu r_\Omega^4 \Omega^5  \hat{f}_{{\rm nc}_{22}},\\
\label{eq:FphiTEOB}
\end{equation}
where $\hat{f}_{{\rm nc}_{22}}$ contains the noncircular corrections $\fnp{,22}$:
\begin{align}
%\hat{f}_{{\rm nc}_{22}}  & \equiv \hat{F}_{22} \NP22 + \sum_{\l} \sum_{m\neq 2}^{\l} \hat{F}_{\lm},
\label{eq:hatfnc22}
\hat{f}_{{\rm nc}_{22}}  & \equiv \hat{F}_{22} \NP22 + \hat{F}_{21} + \sum_{\l\geq 3}^8 \sum_{m=1}^{\l} \hat{F}_{\lm}, \\
\label{eq:NewtPref22_flux}
%----------------------------------------l2m2----------------------------------------
\fnp{,22} & = 1+\frac{3 \dot{r}^4}{4 r^4 \Omega ^4}+\frac{3 \dot{r}^3 {\dot{\Omega}}}{4 r^3 \Omega ^5}+\frac{3 \ddot{r}^2}{4 r^2 \Omega ^4}+\frac{3 \ddot{r} \dot{r} {\dot{\Omega}}}{8 r^2 \Omega ^5} \nonumber \\
&-\frac{{r^{(3)}} \dot{r}}{2 r^2 \Omega ^4}+\frac{\dot{r}^2 {\ddot{\Omega}}}{8 r^2 \Omega ^5}+\frac{4 \dot{r}^2}{r^2 \Omega ^2}+\frac{\ddot{r} {\ddot{\Omega}}}{8 r \Omega ^5} \nonumber \\
&-\frac{2 \ddot{r}}{r \Omega ^2}-\frac{{r^{(3)}} {\dot{\Omega}}}{8 r \Omega ^5}+\frac{3 \dot{r} {\dot{\Omega}}}{r \Omega ^3}+\frac{3 {\dot{\Omega}}^2}{4 \Omega ^4}-\frac{{\ddot{\Omega}}}{4 \Omega ^3} .
\end{align}
For the radial component of the radiation reaction $\hat{\F}_r$ we 
adopt the Pad\'e resummed expression of Refs.~\cite{Chiaramello:2020ehz,Nagar:2021gss}
that relies on the 2PN-accurate results of Ref.~\cite{Bini:2012ji},
\begin{align}
\label{eq:FrTEOB}
\hat{{\cal F}}_r^{\rm \NCN{}}& = \frac{32}{3} \nu \frac{p_{r_*}}{r^4} P^0_2 [ \hat{f}_{r}^{\rm N} + \hat{f}_{r}^{\rm 1PN} + \hat{f}_{r}^{\rm 2PN} ] ,
\end{align}
where
\begin{subequations}
\label{eq:FrPN}
\begin{align}
\hat{f}_{r}^{\rm N} =& -\frac{8}{15} + \frac{56}{5} \frac{p_{\varphi}^{2}}{r}, \\
\hat{f}_{r}^{\rm 1PN} =& -\frac{1228}{105} p_{r_*}^2-\frac{1984}{105} \frac{1}{r}-\frac{124}{105} \frac{p_{r_*}^2 p_{\varphi}^2}{r}+ \nonumber \\ 
&+\frac{1252}{105} \frac{p_{\varphi}^4}{r^3} -\frac{1696}{35} \frac{p_{\varphi}^2}{r^2},  \\
\hat{f}_{r}^{\rm 2PN} =& \frac{323}{315}p_{r_*}^4+\frac{59554}{2835}{r^2}-\frac{1774}{21} \frac{p_{r_*}^2 p_{\varphi}^2}{r^2} + \nonumber \\
& -\frac{628}{105} \frac{p_{r_*}^2 p_{\varphi}^4}{r^3}-\frac{29438}{315} \frac{p_{\varphi}^2}{r^3} -\frac{461}{315} \frac{p_{r_*}^4 p_{\varphi}^2}{r}+ \nonumber \\
&+\frac{20666}{315} \frac{p_{r_*}^2}{r}-\frac{3229}{315} \frac{p_{\varphi}^6}{r^5}-\frac{35209}{315} \frac{p_{\varphi}^4}{r^4}. 
\end{align}
\end{subequations}
Before applying a Pad\'e approximant is useful to factorize the 
leading-order term, so that Eq.~\eqref{eq:FrTEOB}
actually reads 
$\hat{{\cal F}}_r^{\rm \NCN{}} = \frac{32}{3} \nu \frac{p_{r_*}}{r^4} \hat{f}_{r}^{\rm N} 
P^0_2[1 + \hat{f}_{r}^{\rm 1PN}/\hat{f}_{r}^{\rm N} + \hat{f}_{r}^{\rm 2PN}/\hat{f}_{r}^{\rm N}]$ .

\subsection{Quasicircular with $\boldsymbol{\Fr\neq 0}$ (\SEOBNRe{})}
\label{subsec:RRqc}

Let us move now to describing the \QC{} choice, which mimics what it is implemented 
in \SEOBNRe{}~\cite{Ramos-Buades:2021adz}.
While in {\TEOBResumS} the radial component of the radiation reaction is set to zero 
for the quasicircular inspiral, it is not the case for {\SEOBNR} that has
\begin{subequations}
\label{eq:RRqc}
\begin{align}
\Fphiqc & = - \frac{32}{5} \nu r_\Omega^4 \Omega^5 \hat{f}, \\
\Frqc   & = \Fphiqc \frac{p_r}{p_\varphi}.
\end{align}
\end{subequations}
As mentioned in the Introduction, this quasicircular radiation is also used 
in {\SEOBNRe}~\cite{Ramos-Buades:2021adz}, the eccentric extension 
of {\SEOBNR}.
We note, however, two aspects. The first one is that Refs.~\cite{Khalil:2021txt,Ramos-Buades:2021adz} 
(and not even Refs.~\cite{Bohe:2016gbl,Cotesta:2018fcv}, which define the \SEOBNR{} model)
do not write explicitly the leading-order (Newtonian) contribution to $\Fphi$. 
In particular it is not stated explicitly whether Kepler's constraint is relaxed as originally proposed 
in Refs.~\cite{Nagar:2006xv} to have a more faithful representation of the angular momentum losses 
during the plunge. This may induce the reader to  think that the dominant term of the angular 
component is $\Fphi^{\rm qc, LO} = -32/5 \nu \Omega^{7/3}$. 
Nonetheless, inspecting the public code it is possible to see that also in the 
{\SEOBNR} models the Keplerian constraint is relaxed so that 
$\Fphi^{\rm qc, LO} = -32/5 \nu r_\Omega^4 \Omega^5 $, as in \TEOBResumS{}.
The second aspect to consider is that, consistently with what we wrote in Sec.~\ref{subsec:RRteob} 
concerning the circular flux, through this paper we will consider $\hat{f}$ as implemented 
in the test-mass version of {\TEOBResumS}. While the term is formally the same of  {\SEOBNR}, 
there are some differences in PN orders adopted and in the related resummations that 
enter $\hat{f}$ and follow Refs.~\cite{Nagar:2016ayt,Messina:2018ghh,Albanesi:2021rby}.
However, the reliability of the term $\hat{f}$ that we will use in this work is shown in 
Appendix~\ref{appendix:hatf}. With the aforementioned caveats, the radiation 
reaction of Eq.~\eqref{eq:RRqc} is a proxy of the one used in the actual {\SEOBNRe} model.

\subsection{Quasicircular with 2PN noncircular corrections} 
Let us finally turn to the \QC2PN{} radiation reaction, that contains 2PN noncircular
correction in factorized form as introduced by Khalil {\it et al.}~\cite{Khalil:2021txt}. 
The radiation reaction forces are factorized in quasicircular and eccentric parts:
\begin{subequations}
\label{eq:FKhalil}
\begin{align}
\FphiK &= \Fphiqc \Fphiecc, \\
\FrK &= \Frqc \Frecc,
\end{align} 
\end{subequations}
where $\Fphiqc$ and $\Fphiqc$ are the quasicircular terms of Eq.~\eqref{eq:RRqc}, 
while $\Fphiecc$ and $\Frecc$ are the noncircular corrections up to 2PN and 
reduce to unity in the circular case. Their explicit expressions in the test-mass limit,
written in terms of $(r, p_r,\dot{p}_r)$, are reported in 
Appendix~\ref{appendix:Fecc}.
%===========================================================================================
\section{Testing the reliability of the EOB radiation reaction}
\label{sec:test}
%===========================================================================================
In this section we test the reliability of the radiation reactions discussed in 
Sec.~\ref{sec:RR} comparing the corresponding fluxes with numerical results. 

\subsection{Method}
The method we use is discussed in Ref.~\cite{Albanesi:2021rby}, but we report here the main idea 
since it is crucial for this paper. 
First, we consider geodesic orbits, so that the radiation reaction does not influence the dynamics. 
Then, from a given dynamics one computes the energy and angular momentum analytical 
fluxes at infinity, $\dot{E}$ and $\dot{J}$, using the balance equations~\cite{Bini:2012ji}
\begin{subequations}
\label{eq:enbalance}
\begin{align}
\label{eq:Fr}
\dot{E} &= -\dot{r}\F_r - \Omega \F_\varphi - \dot{E}_{\rm Schott} ,\\
\label{eq:Fphi}
\dot{J} &= -\F_\varphi,
\end{align}
\end{subequations}
where $\dot{E}_{\rm Schott}$ is the time derivative of the Schott
energy. The latter is linked to the interaction of the source with the local 
field~\cite{Bini:2012ji}. Following Ref.~\cite{Chiaramello:2020ehz}, 
the Schott energy is factorized in a circular (c) and noncircular (nc) 
contribution and each term is additionally resummed, yielding
\begin{equation}
\label{eq:ESchott}
E_{\rm Schott}  = \frac{16}{5} \frac{p_{r_*}}{r^3} P^0_2[E_{\rm Schott}^{\rm c}] P^0_2[E_{\rm Schott}^{\rm nc}]  ,
\end{equation}
where the circular and noncircular parts are at 2PN accuracy and $P^0_2$ is 
the $(0,2)$ Pad\'e approximant.
The two contributions explicitly read
\begin{align}
E_{\rm Schott}^{\rm c} =& 1 - \frac{157}{56} \frac{1}{r} -\frac{3421}{756}\frac{1}{r^2},  \\
E_{\rm Schott}^{\rm nc} =& \frac{p_{\varphi}^{2}}{r} -\frac{3}{2} p_{r_*}^2+\frac{2}{21} \frac{1}{r}-\frac{1}{2} \frac{p_{r_*}^2 p_{\varphi}^2}{r}  \nonumber \\
& +\frac{55}{168} \frac{p_{\varphi}^4}{r^3} -\frac{575}{168}
   \frac{p_{\varphi}^2}{r^2} + \frac{5}{8}
   p_{r*}^4  \nonumber \\
   &-\frac{2143}{5292} \frac{1}{r^2}-\frac{61}{48} \frac{p_{r_*}^2 p_{\varphi}^2}{r^2}-\frac{13}{168} \frac{p_{r_*}^2 p_{\varphi}^4}{r^3} \nonumber \\ 
   &+\frac{370189}{84672} \frac{p_{\varphi}^2}{r^3} + \frac{3}{8}
   \frac{p_{r_*}^4 p_{\varphi}^2}{r}-\frac{181}{112} \frac{p_{r*}^2}{r}  \nonumber \\ 
   &-\frac{25}{504} \frac{p_{\varphi}^6}{r^5}-\frac{130223}{28224} \frac{p_{\varphi}^4}{r^4} . 
\end{align}
The analogous contribution to the angular momentum, $\dot{J}_{\rm Schott}$, 
can be eliminated with a suitable gauge choice~\cite{Bini:2012ji}.
Given an analytical prescription for the radiation reaction, we compute the
analytical fluxes at infinity  $(\dot{E},\dot{J})$ and compare them with the corresponding 
numerical results. 

\subsection{Numerical fluxes}
We use numerical multipolar
waveforms at {\it null-infinity} $h_\lm^{\rm teuk}$ 
obtained solving the 2+1 Teukolsky 
equation~\cite{Teukolsky:1972my, Teukolsky:1973ha} 
with the hyperboloidal time-domain code
{\Teukode}~\cite{Harms:2014dqa}.
Explicitly, the numerical fluxes are given by
\begin{subequations}
 \label{eq:fluxes_infty}
\begin{align}
\label{eq:fluxes_infty_E}
\Eteuk &= \;\;\,\frac{1}{16 \pi} \sum_{\l=2}^8\sum_{m=-\l}^{\l} | \dot{h}^{\rm teuk}_{\lm} |^2  ,\\
\Jteuk &= -\frac{1}{16 \pi} \sum_{\l=2}^8 \sum_{m=-\l}^{\l} m \Im \left( \dot{h}^{\rm teuk}_{\lm} \left(h_{\lm}^{\rm teuk}\right)^* \right) \  . \label{eq:fluxes_infty_J}
\end{align}
\end{subequations}
We use the same geodesic numerical simulations presented in Ref.~\cite{Albanesi:2021rby};
therefore, we consider spins in the range $\ha\in[-0.9,0.9]$ and eccentricities up
to $e=0.9$. For each pair $(e, \ha)$ we compute the semilatus rectum $p$ according to the 
separatrix $p_s(e,\ha)$.
More precisely, we consider \textit{intermediate} and \textit{distant} configurations
that have semilatus rectum $p=p_{\rm schw} p_s(e,\ha)/p_s(e,0)$, where $p_{\rm schw}$ is 9 
and 13, respectively. 
In Ref.~\cite{Albanesi:2021rby} we considered also \textit{near} simulations 
that had $p(e,\ha) = p_s(e,\ha)+0.01$; however, we will not consider
them in this work since they show a strong 
zoom-whirl behavior and thus are less significant for testing the noncircular terms.
More details on the numerical simulations can be found in Sec.~II of Ref.~\cite{Albanesi:2021rby}.

\subsection{Instantaneous fluxes for eccentric orbits}
\label{sec:ecc_inst}
\begin{figure*}[]
  \center
  \includegraphics[width=0.26\textwidth]{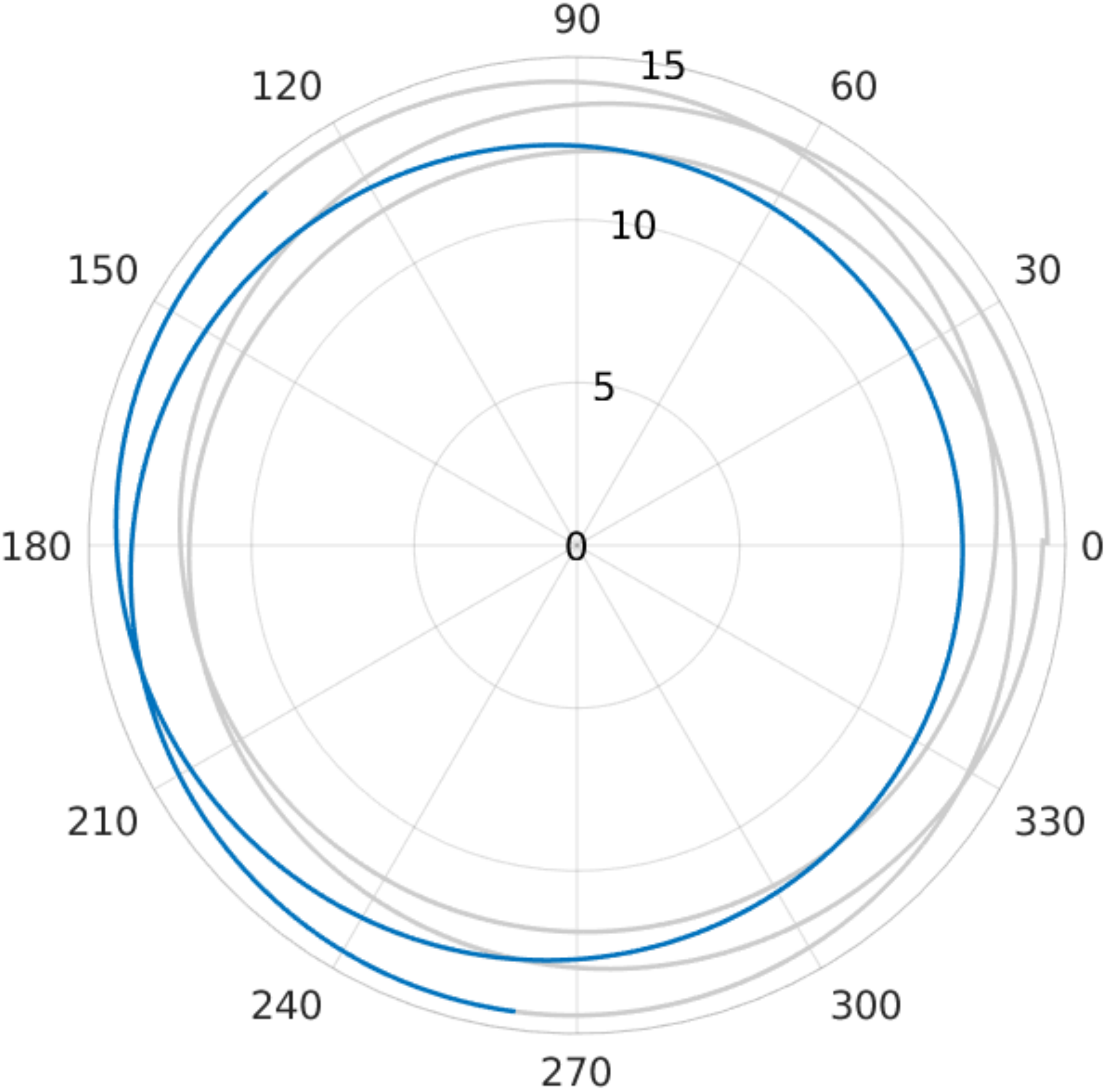}
  \hspace{0.2cm}
  \includegraphics[width=0.32\textwidth]{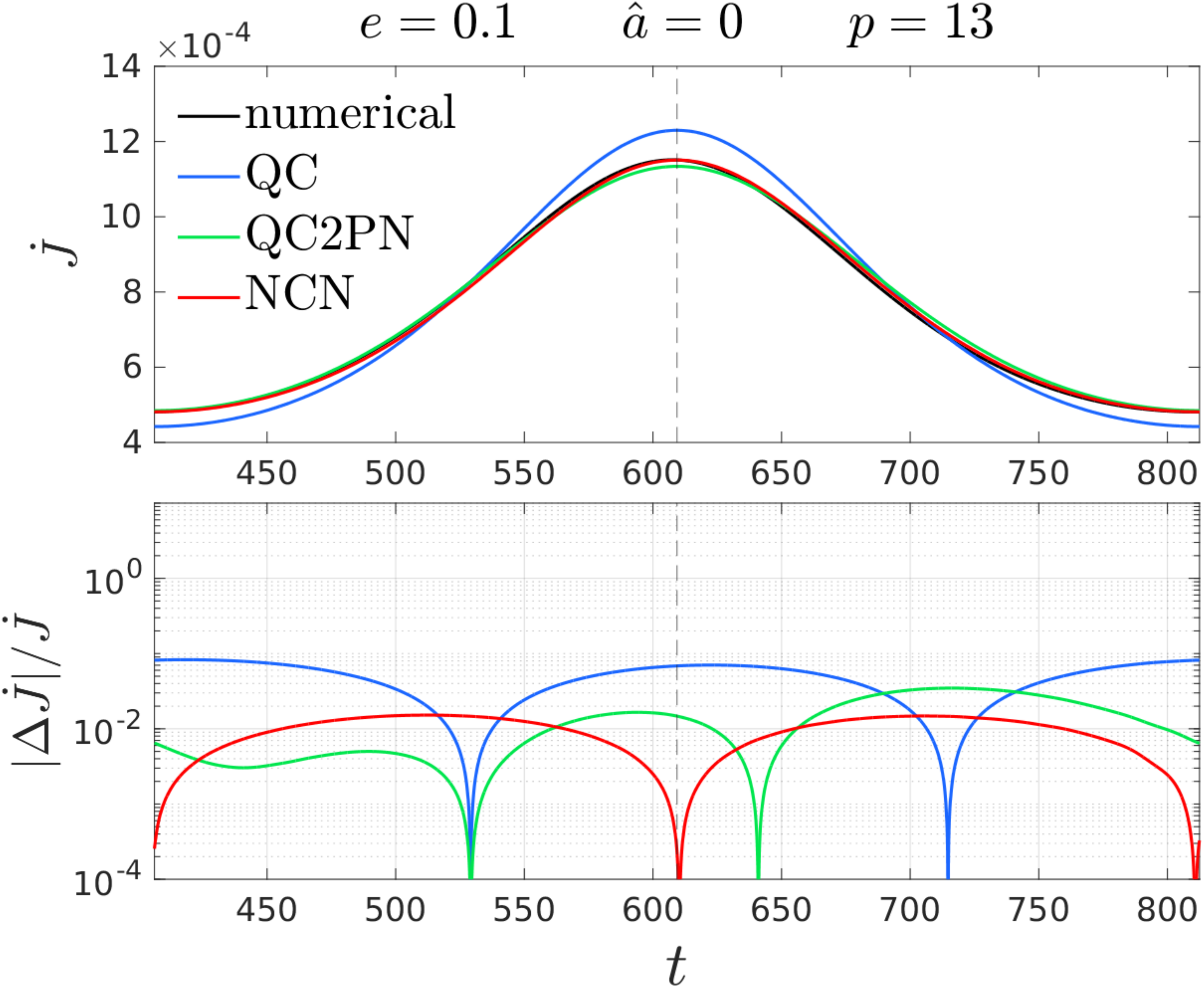}
  \hspace{0.2cm}
  \includegraphics[width=0.32\textwidth]{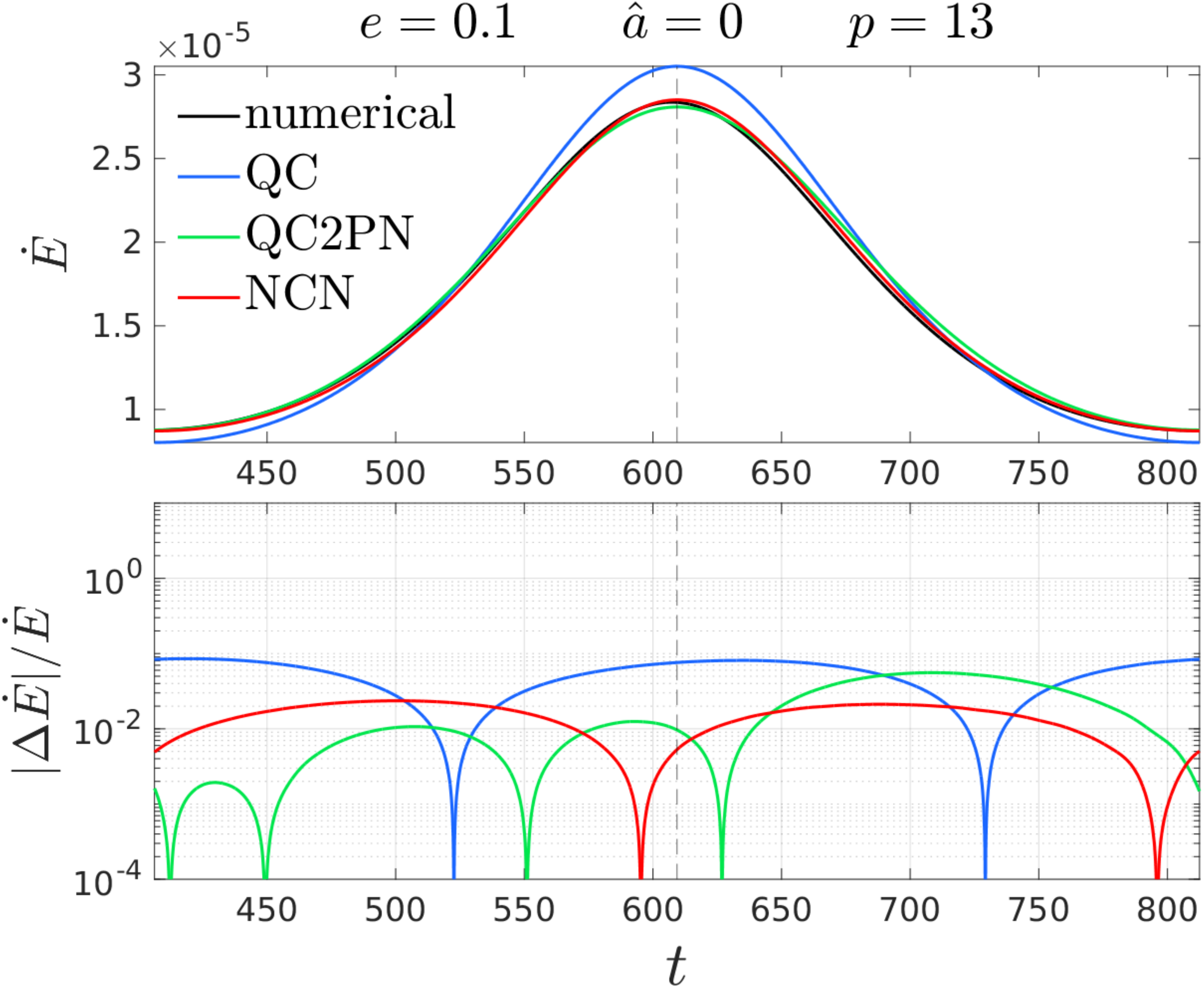}\\
  \vspace{0.5cm}
  \includegraphics[width=0.26\textwidth]{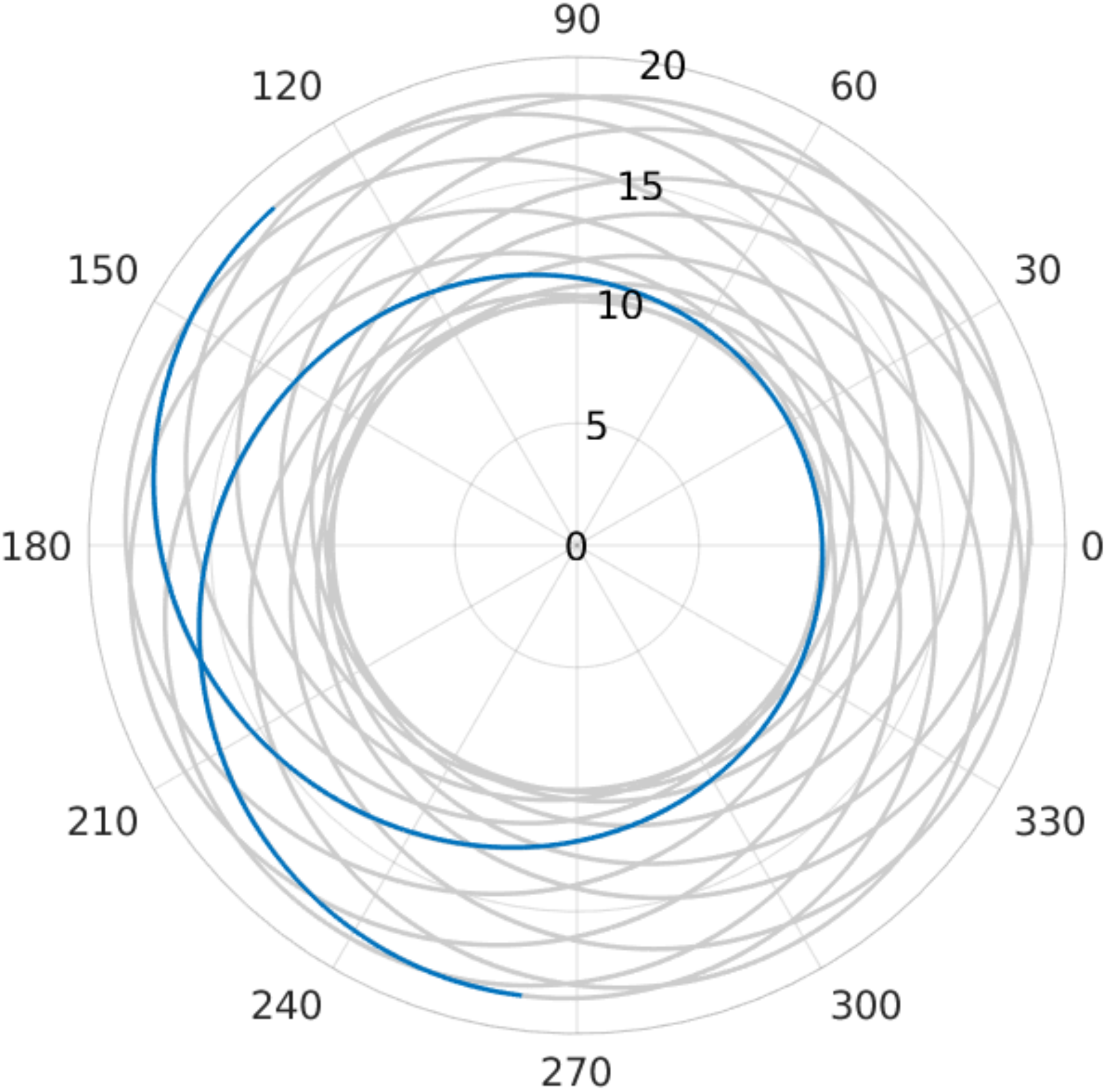}
  \hspace{0.2cm}
  \includegraphics[width=0.32\textwidth]{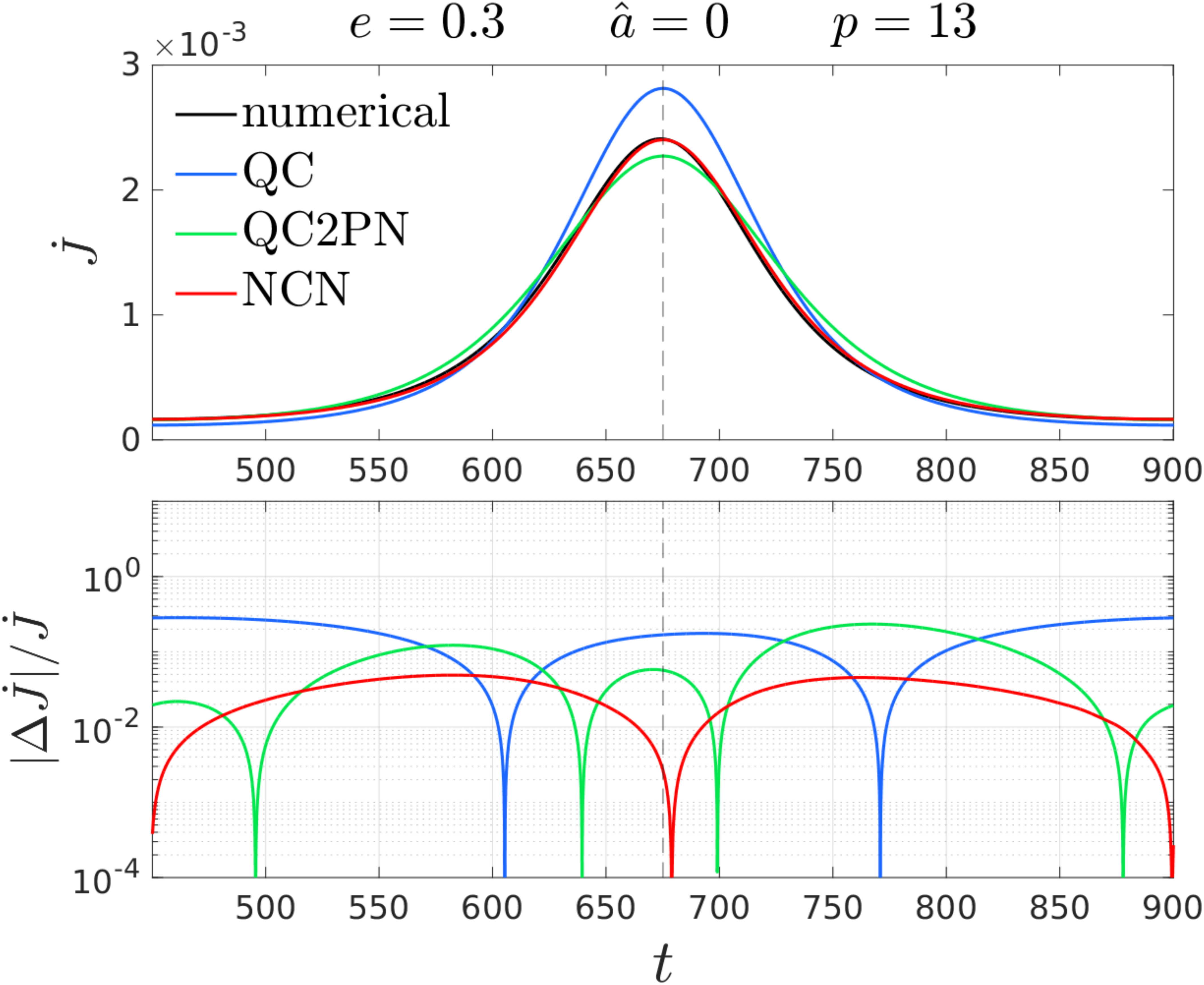}
  \hspace{0.2cm}
  \includegraphics[width=0.32\textwidth]{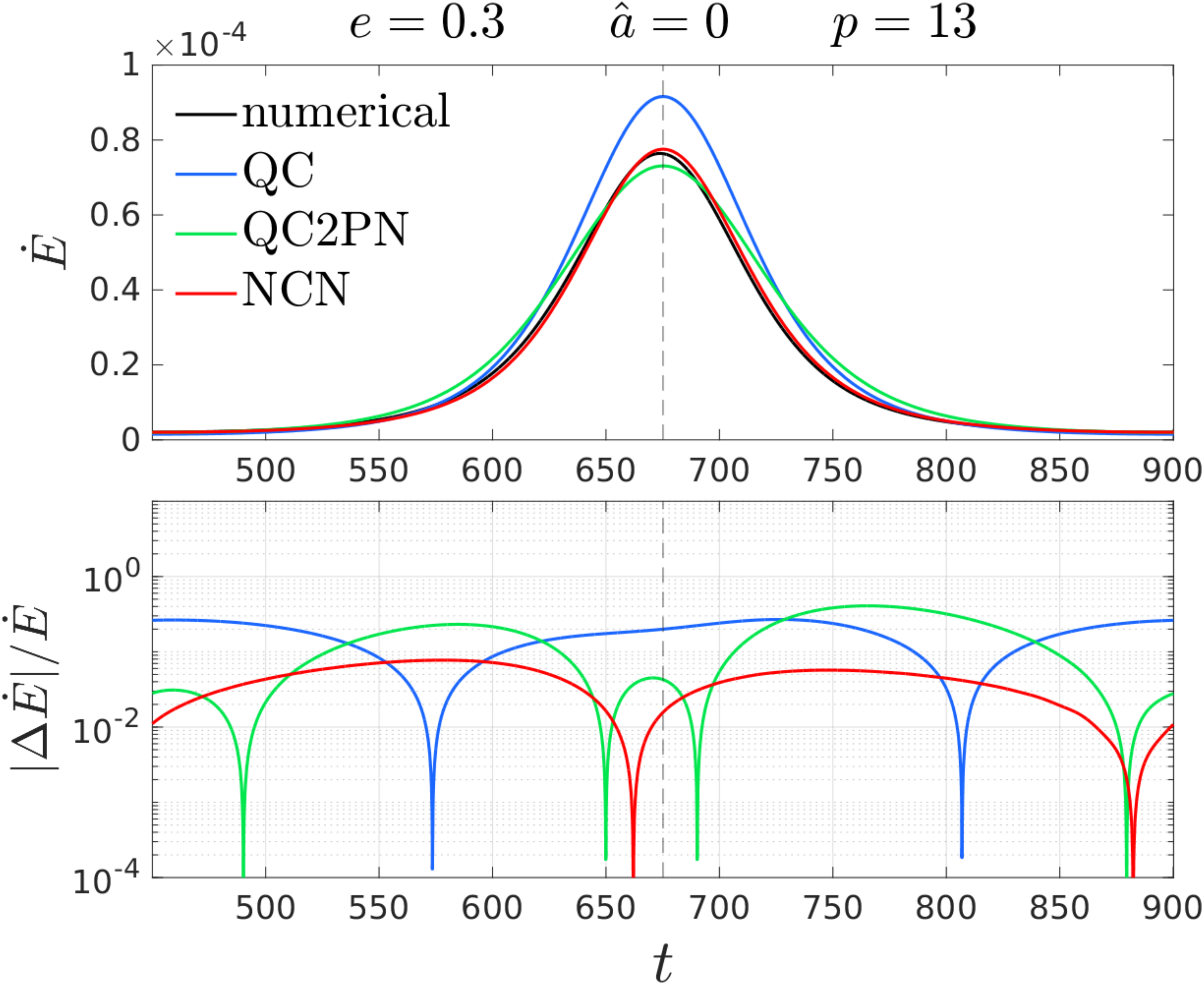}\\
  \vspace{0.5cm}
  \includegraphics[width=0.26\textwidth]{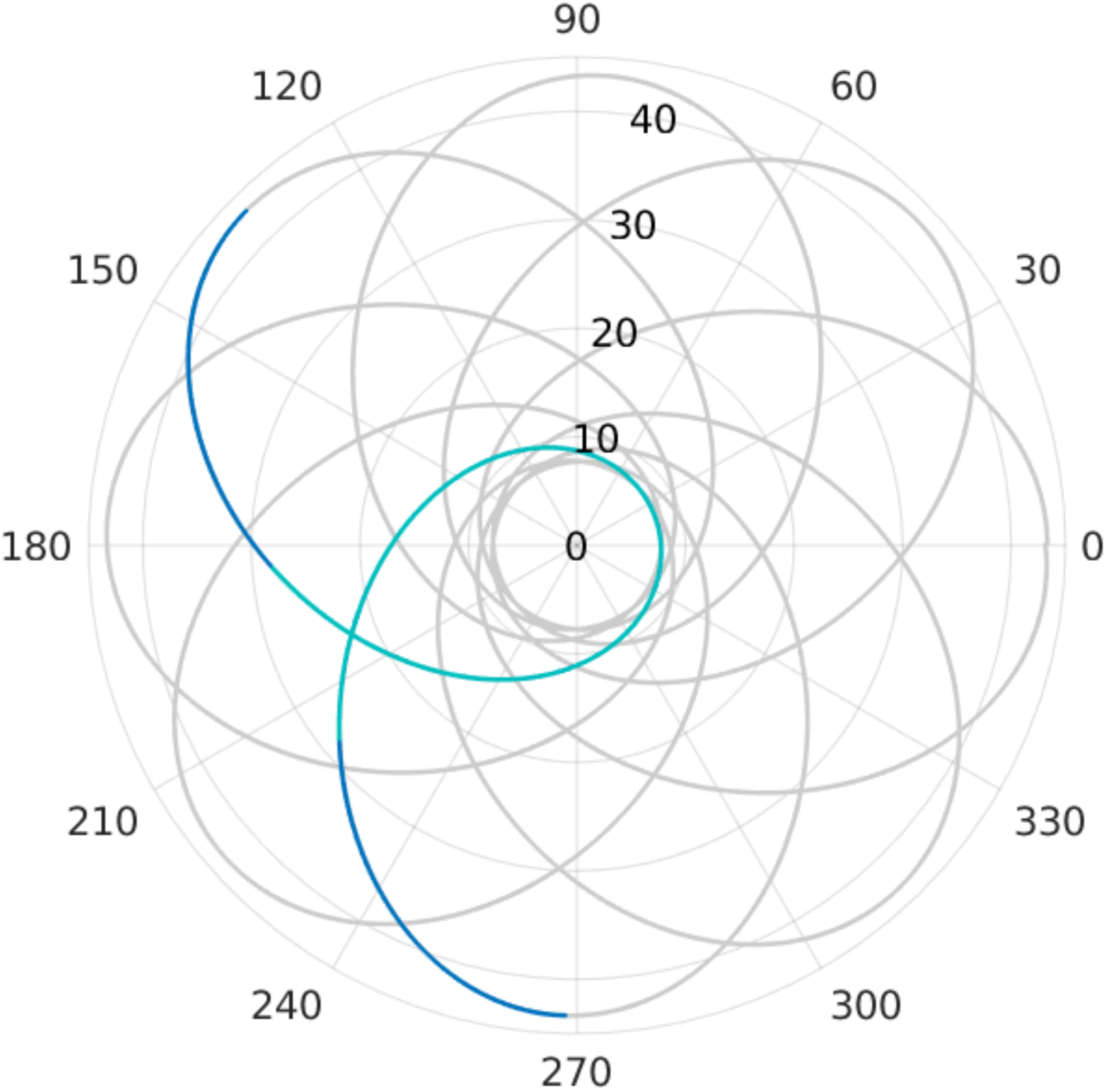}
  \hspace{0.2cm}
  \includegraphics[width=0.32\textwidth]{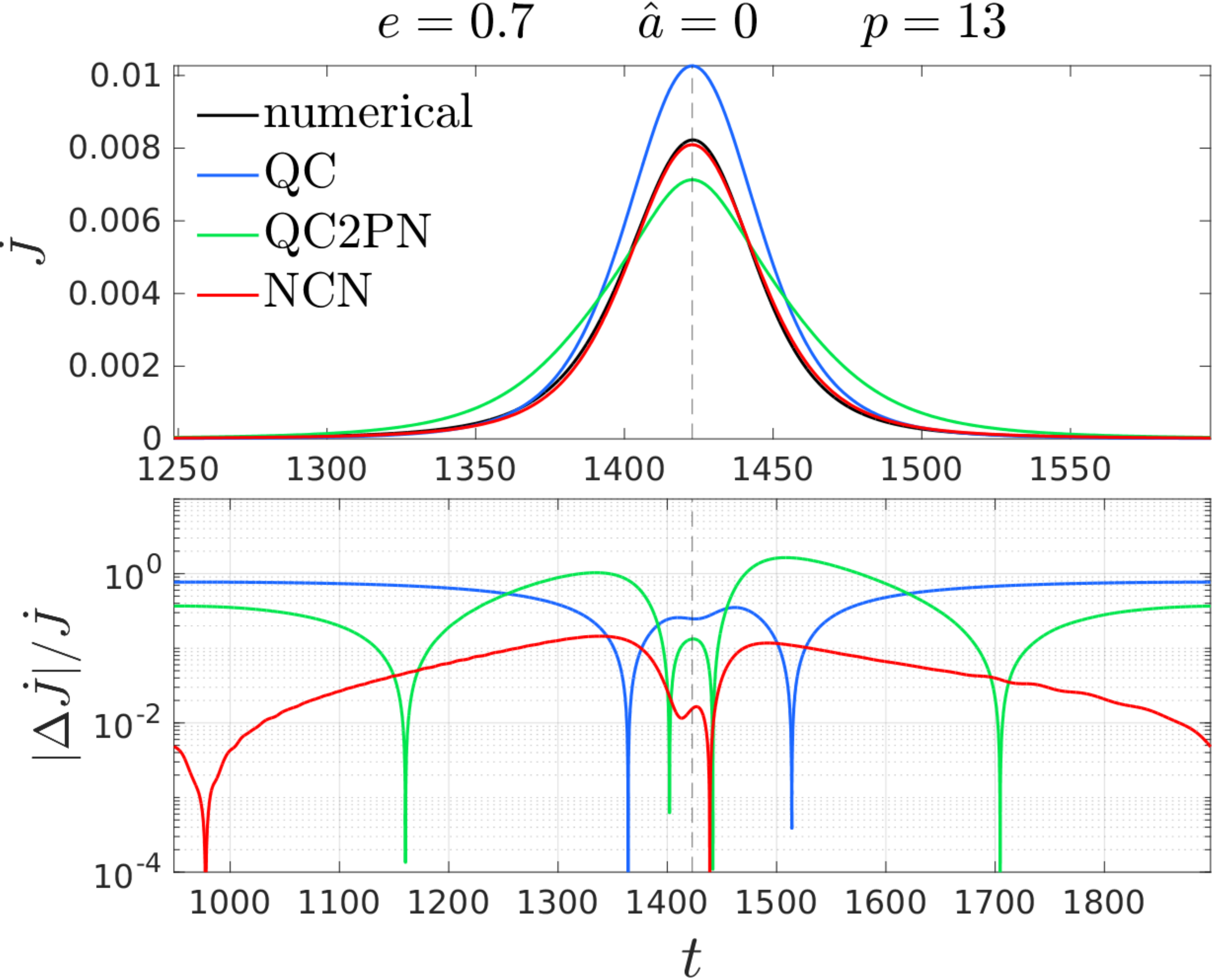}
  \hspace{0.2cm}
  \includegraphics[width=0.32\textwidth]{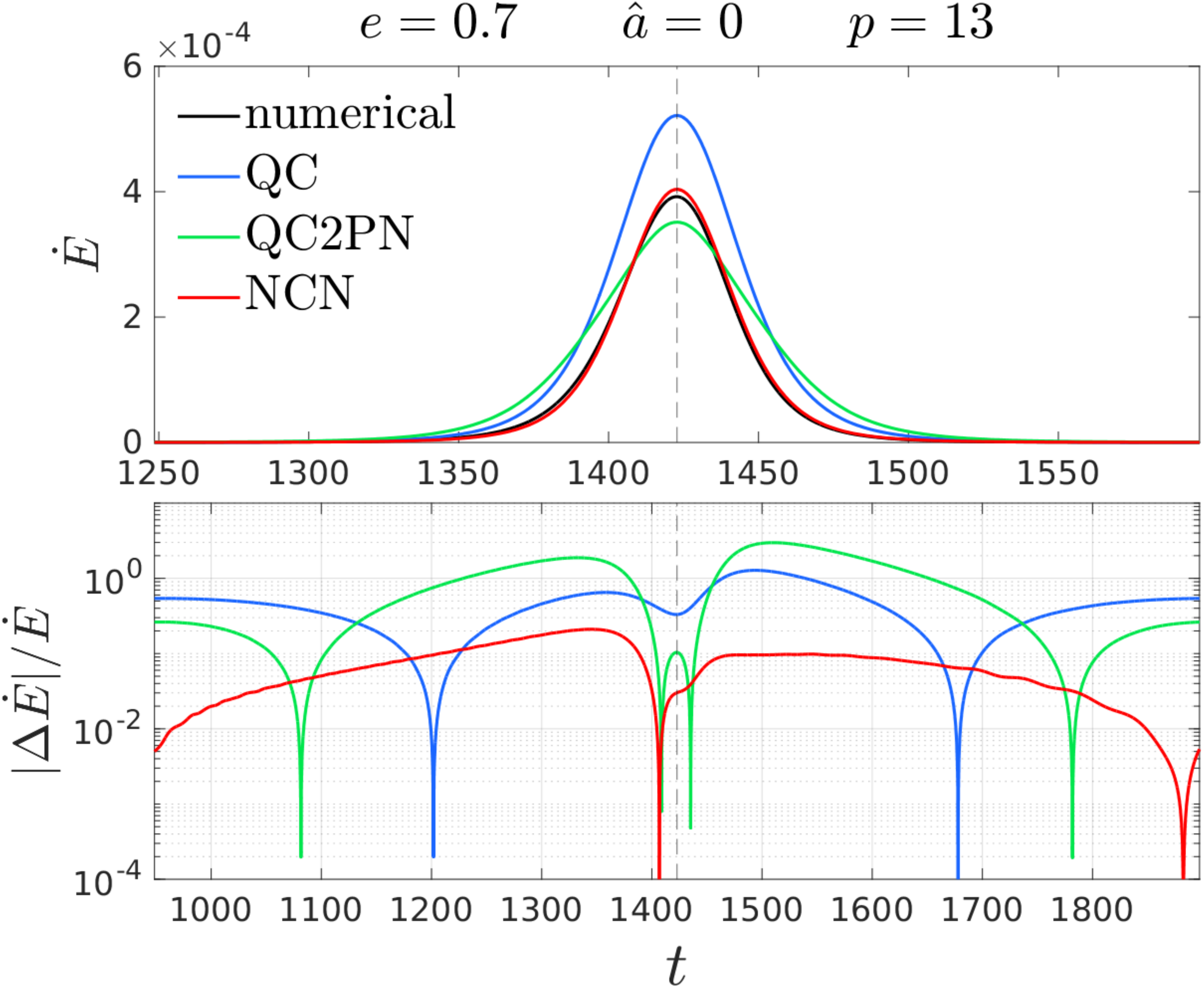}\\
   \vspace{0.5cm}
  \includegraphics[width=0.26\textwidth]{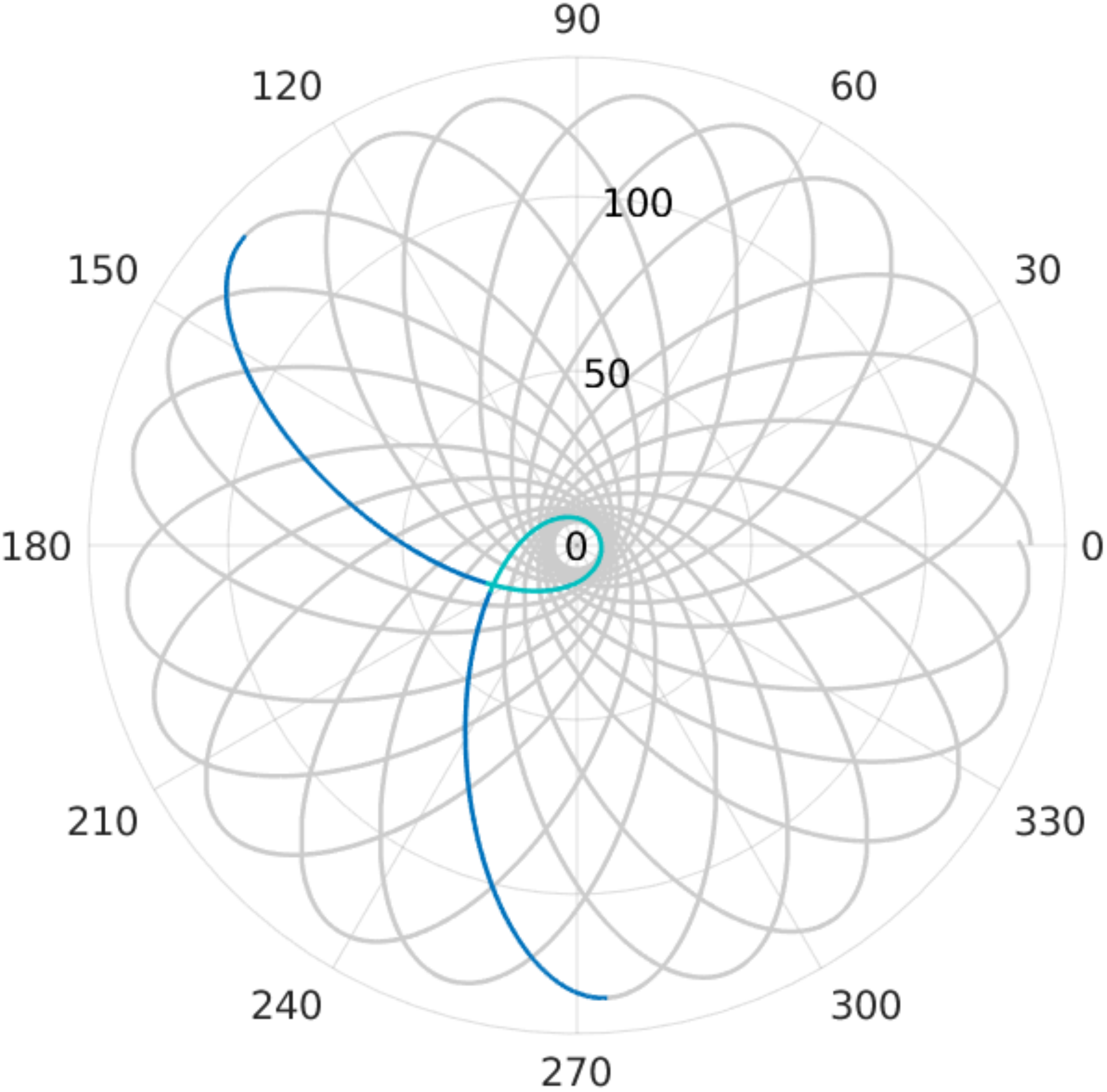}
  \hspace{0.2cm}
  \includegraphics[width=0.32\textwidth]{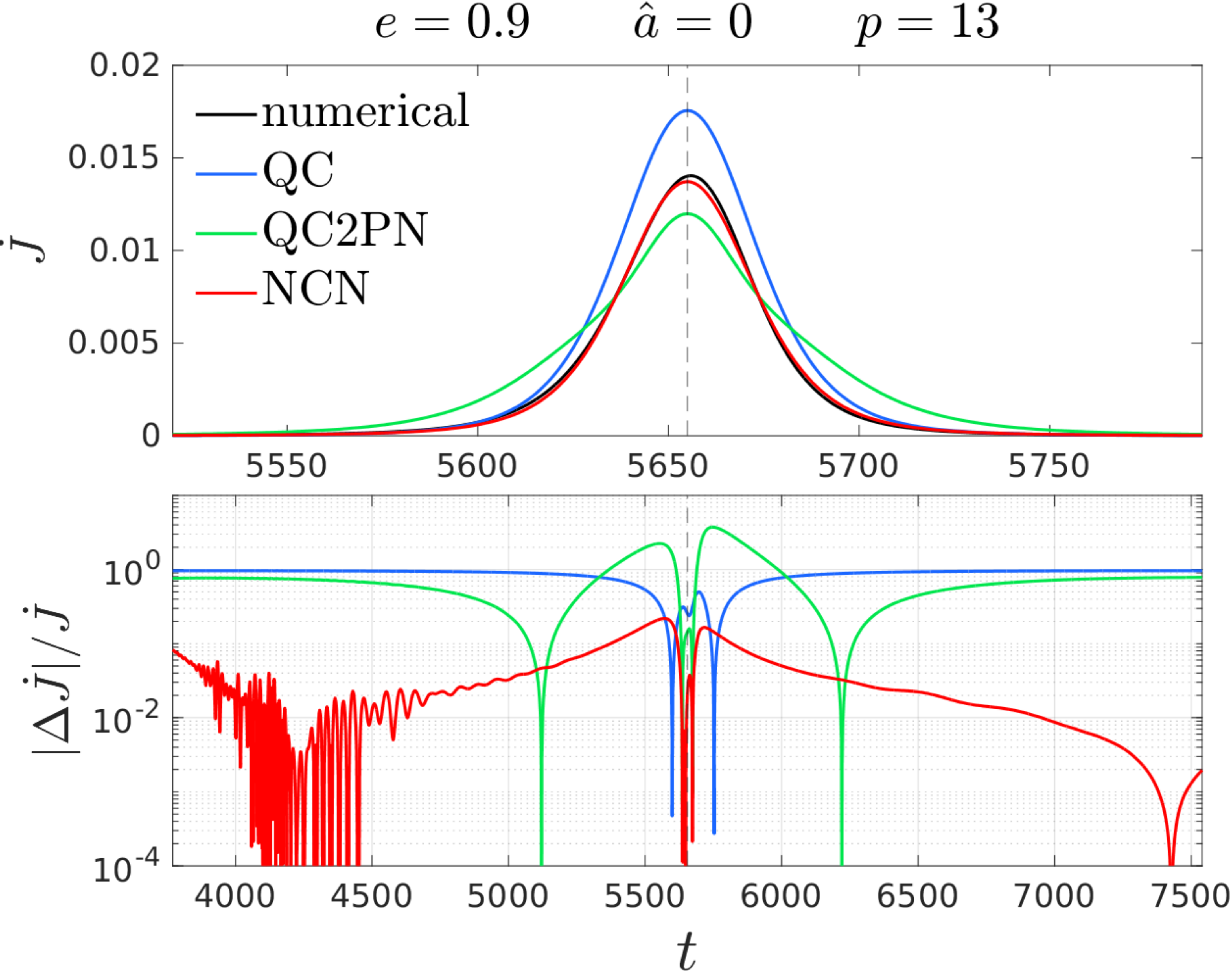}
  \hspace{0.2cm}
  \includegraphics[width=0.32\textwidth]{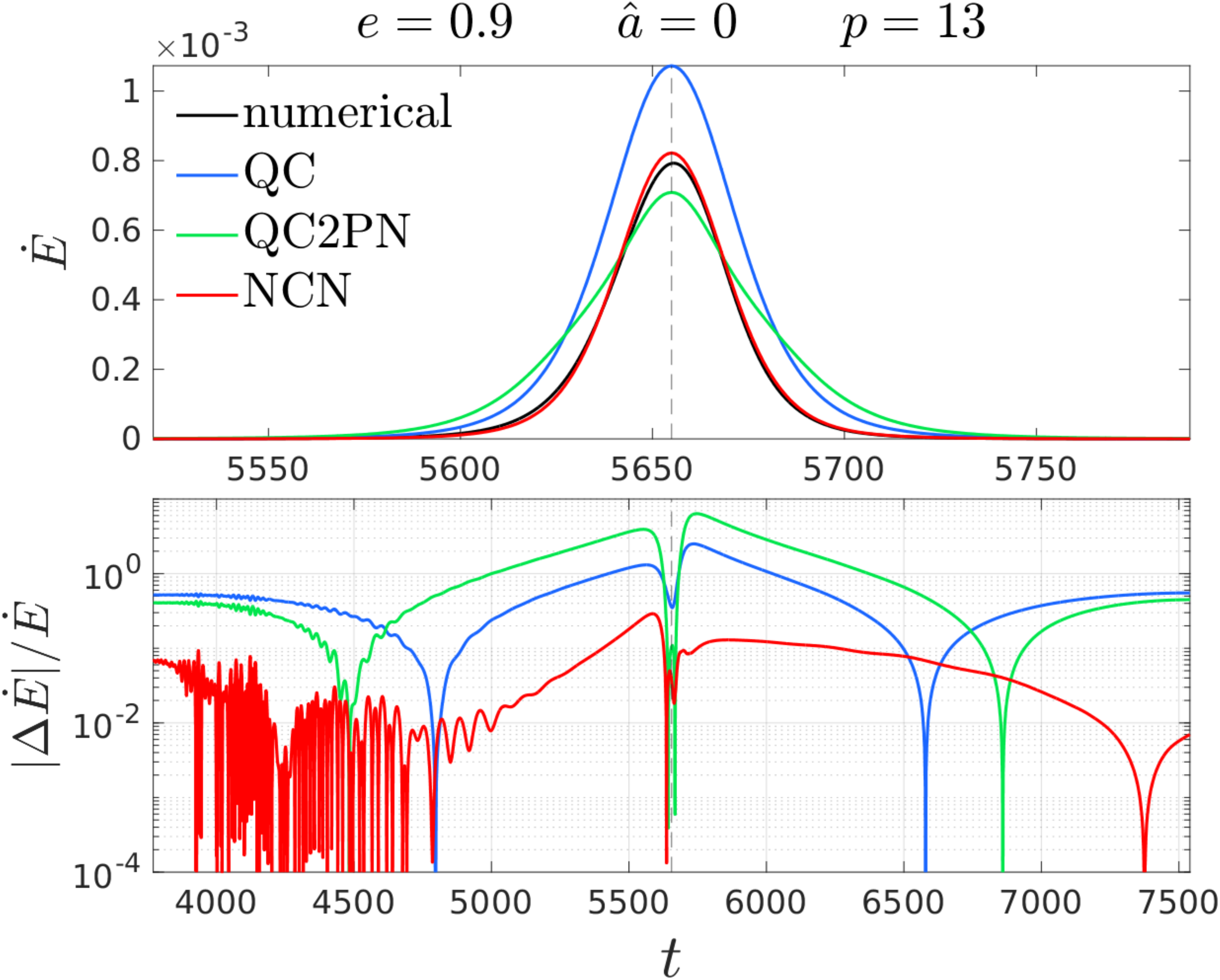}\\
  \caption{\label{fig:fluxes_nospin}
  Geodesic configurations with $\ha=0$, $p=13$ and $e=(0.1, 0.3, 0.7, 0.9)$. 
  For each one we show the trajectories highlighting one radial period
  (blue) and the corresponding fluxes. We contrast the numerical fluxes (black)
  with the three analytical fluxes considered in this work: the \NCN{} flux (red) 
  computed using Eqs.~\eqref{eq:FphiTEOB} and~\eqref{eq:FrTEOB};
  the \QC{} flux (blue) from Eq.~\eqref{eq:RRqc}, which 
  is a proxy of the {\SEOBNRe} fluxes, and the \QC2PN{} flux with 2PN noncircular 
  corrections (green) from Eqs.~\eqref{eq:FKhalil}. 
  Each subpanel also reports the analytical/numerical relative difference.
  Note that while the relative differences in the lower panels are always shown over the 
  complete radial period, the fluxes for $e\geq 0.7$ are shown on  
  smaller time intervals (highlighted in aqua-green on the trajectories) 
  in order to better highlight the burst of radiation at periastron
  passage (marked by a dashed vertical line). }
\end{figure*} 
\begin{figure*}[]
  \center
  \includegraphics[width=0.26\textwidth]{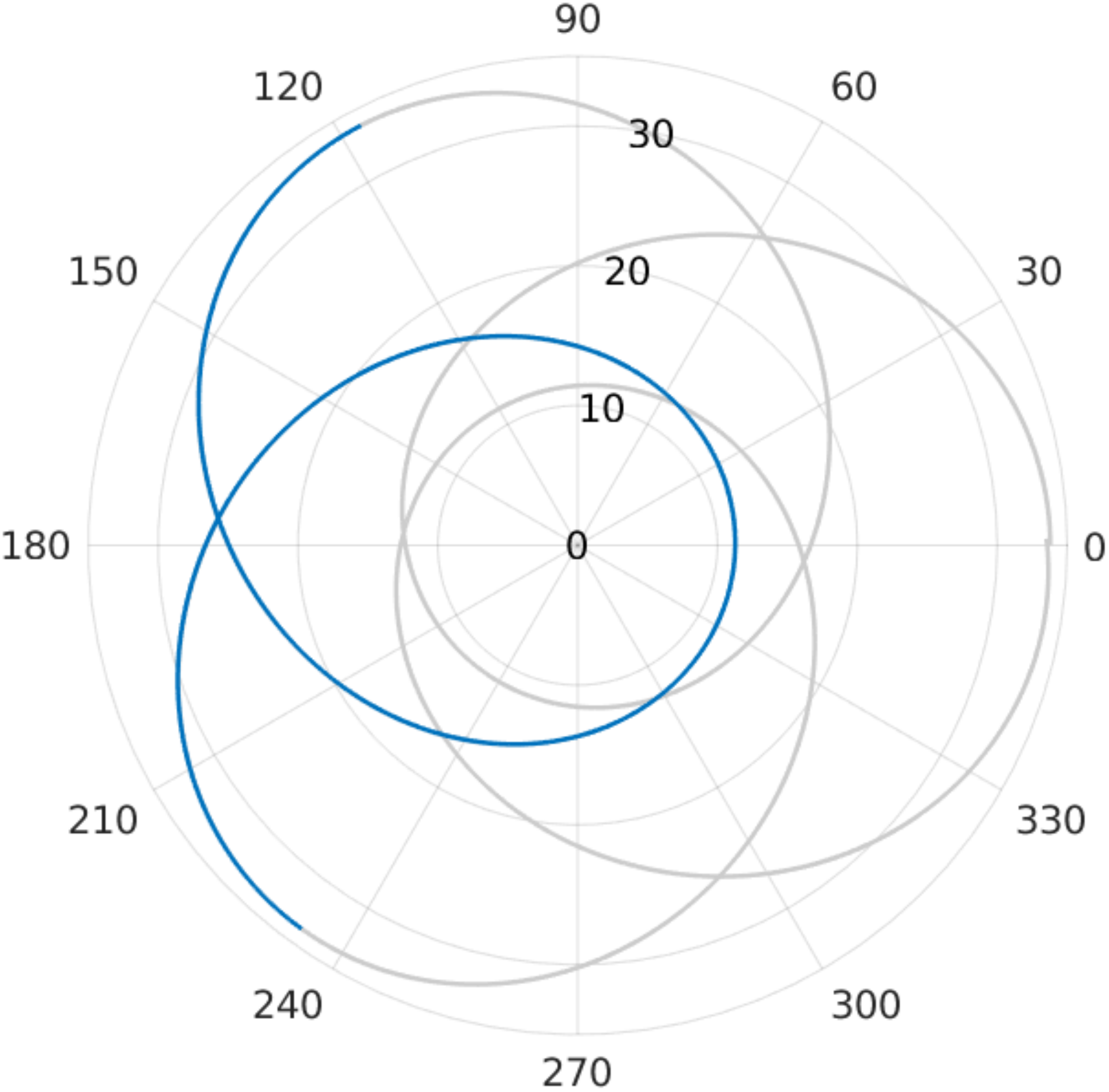}
  \hspace{0.2cm}
  \includegraphics[width=0.32\textwidth]{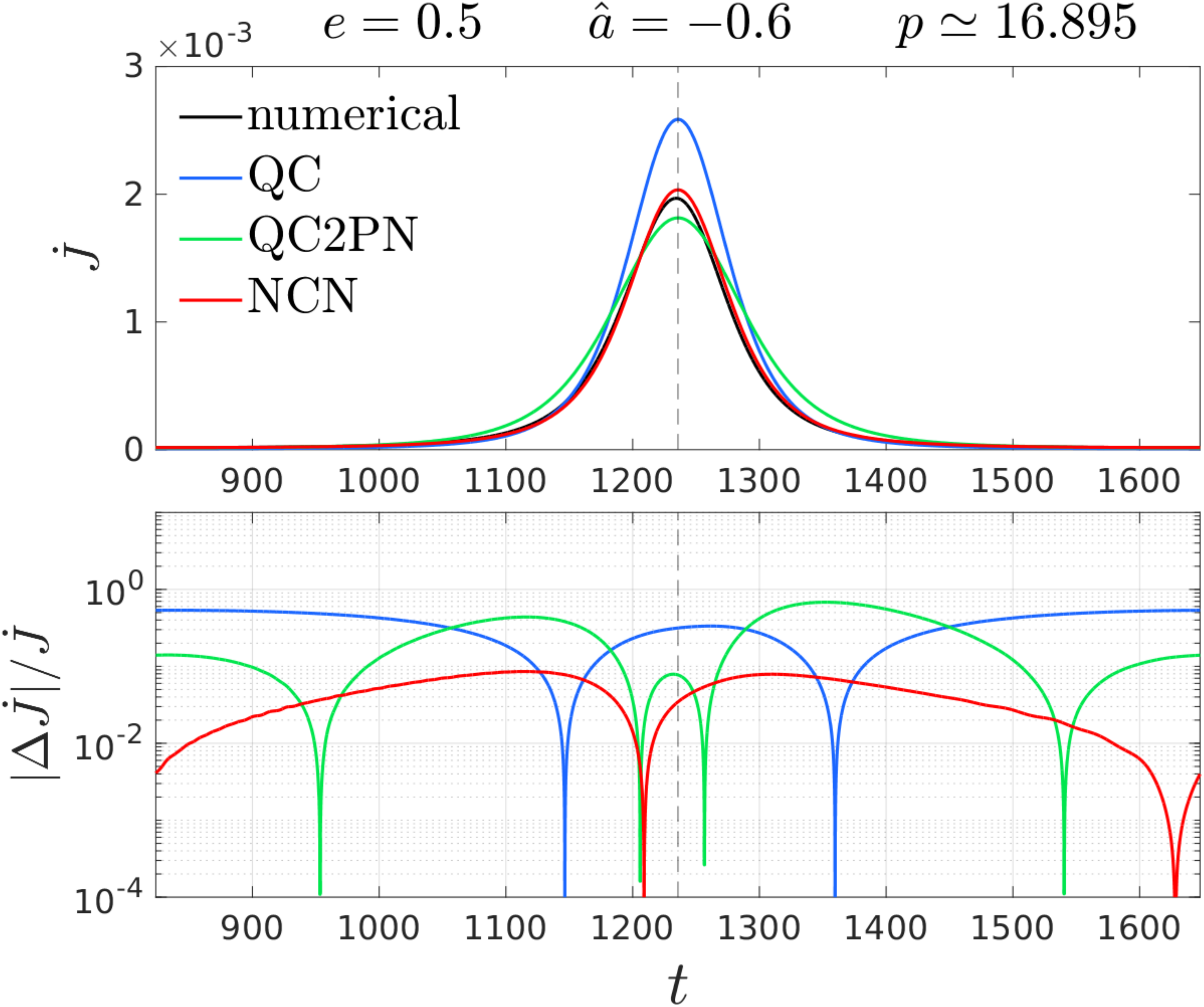}
  \hspace{0.2cm}
  \includegraphics[width=0.32\textwidth]{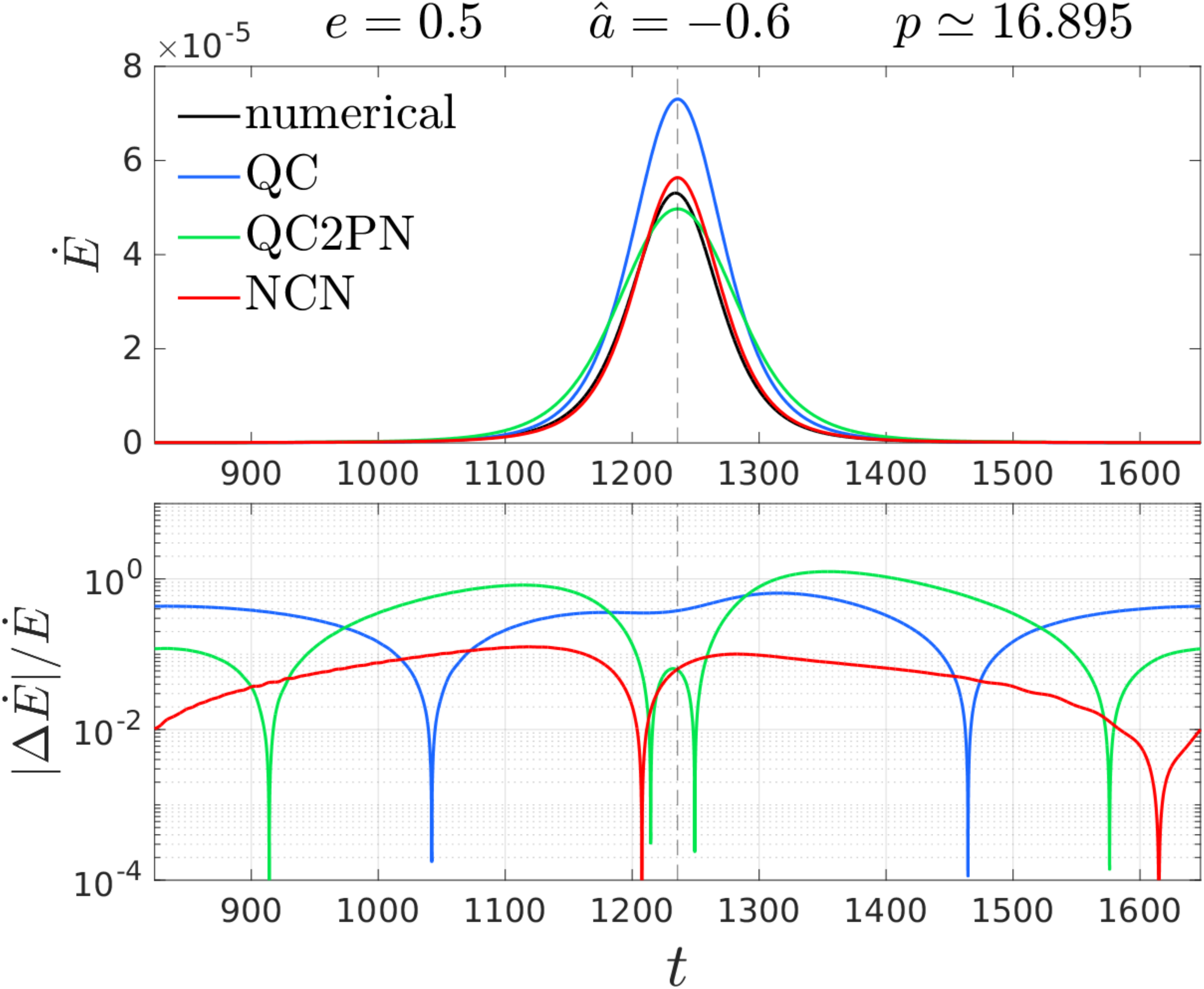}\\
  \vspace{0.5cm}
  \includegraphics[width=0.26\textwidth]{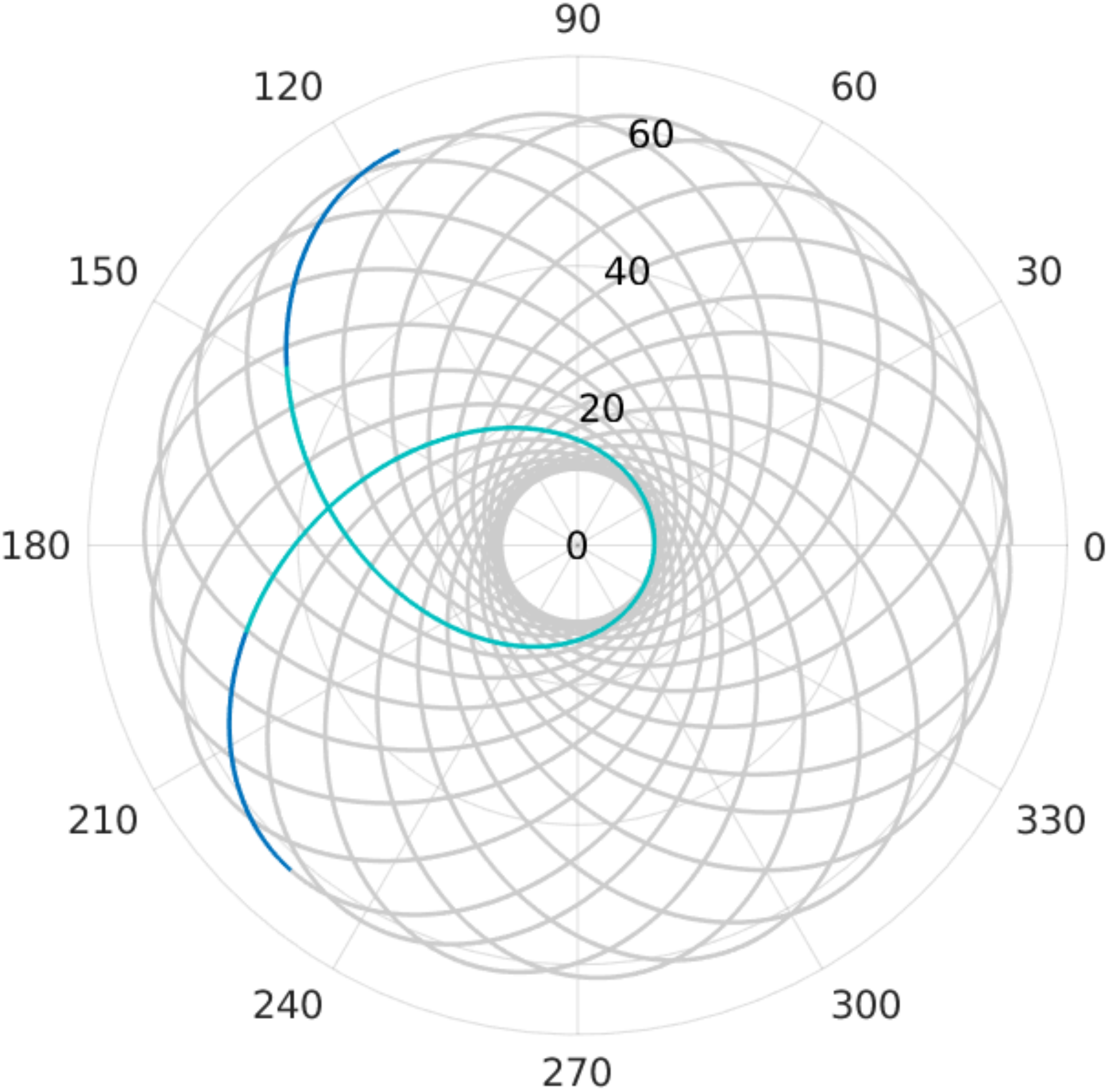}
  \hspace{0.2cm}
  \includegraphics[width=0.32\textwidth]{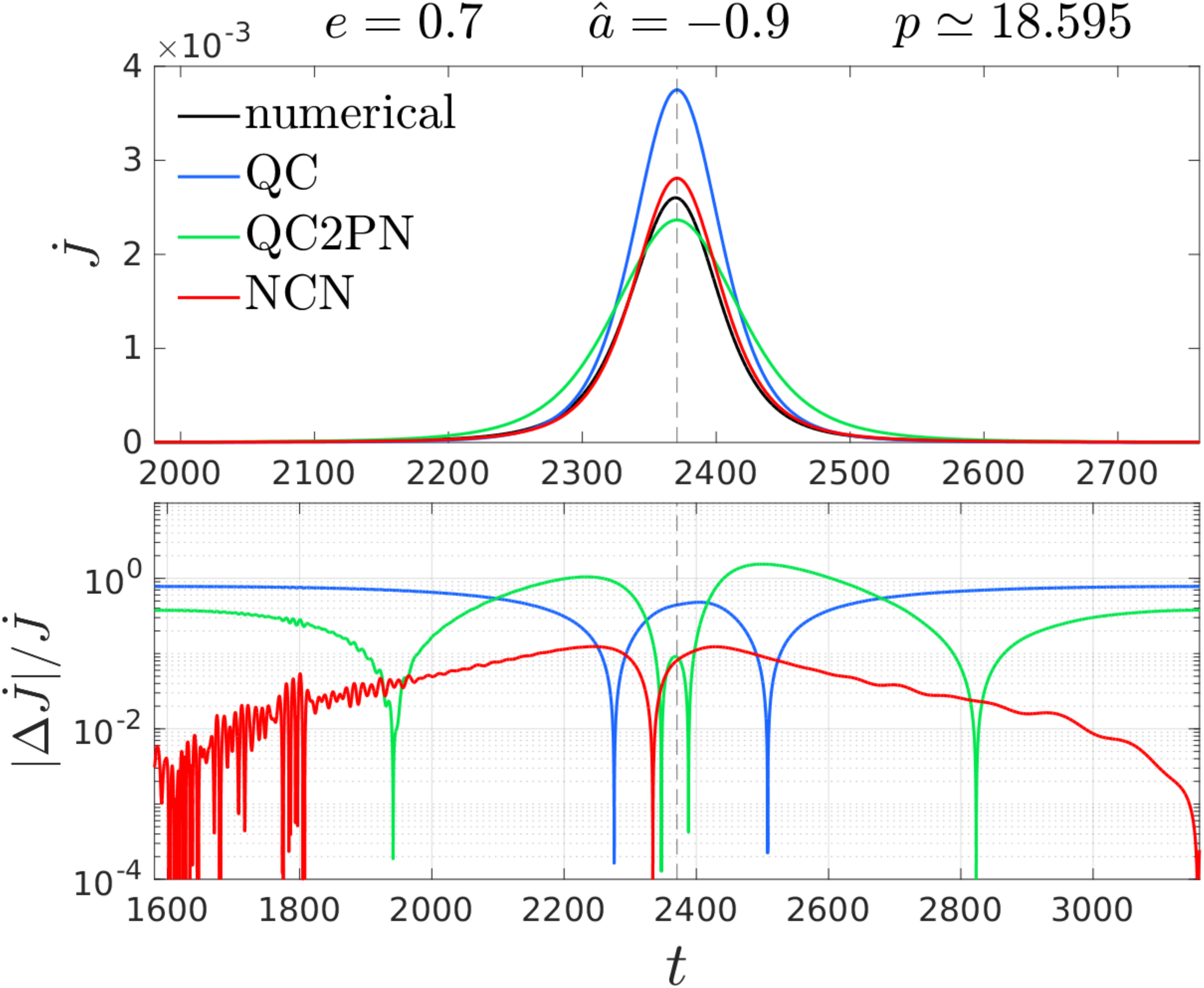}
  \hspace{0.2cm}
  \includegraphics[width=0.32\textwidth]{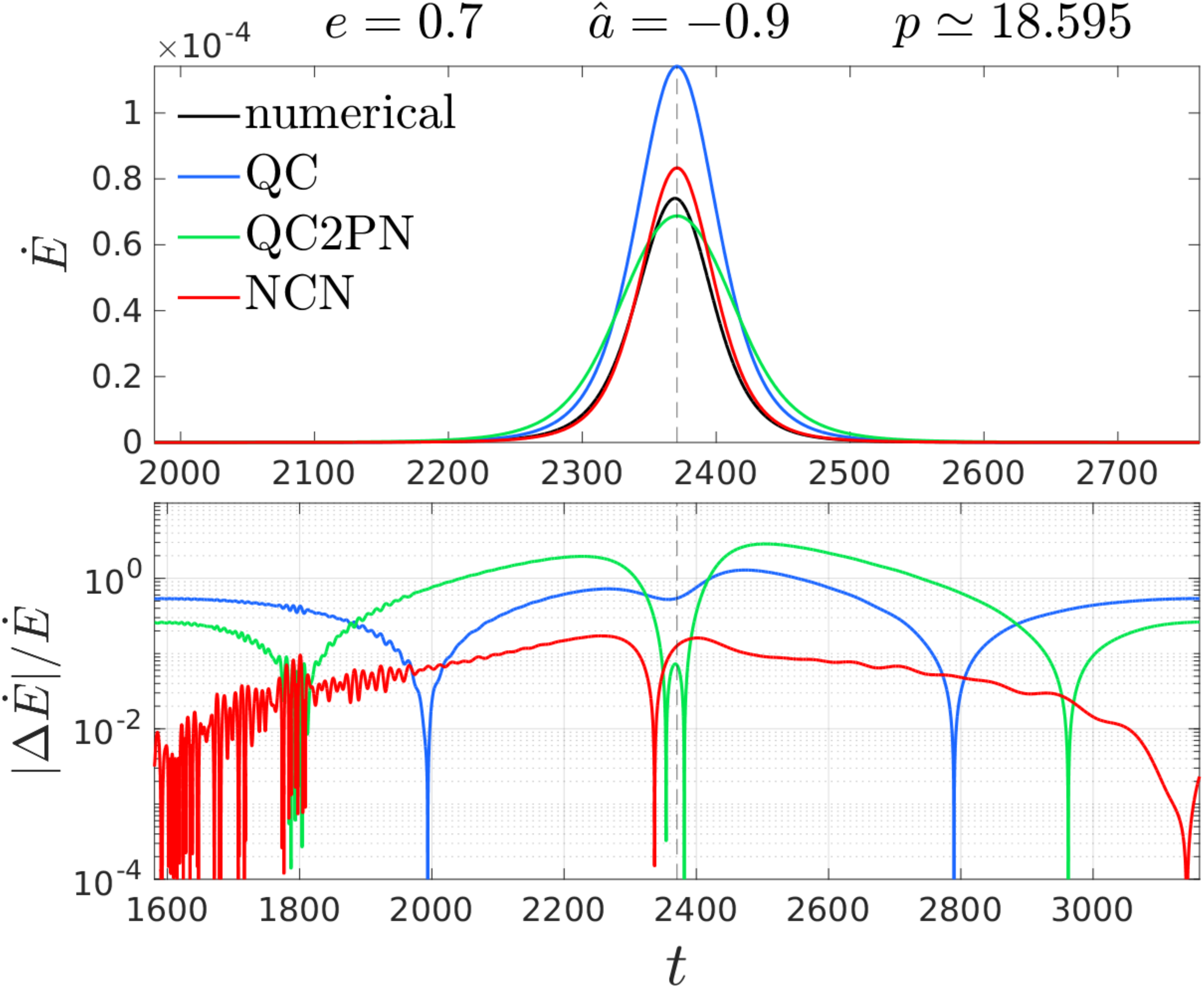}\\
  \vspace{0.5cm}
  \includegraphics[width=0.26\textwidth]{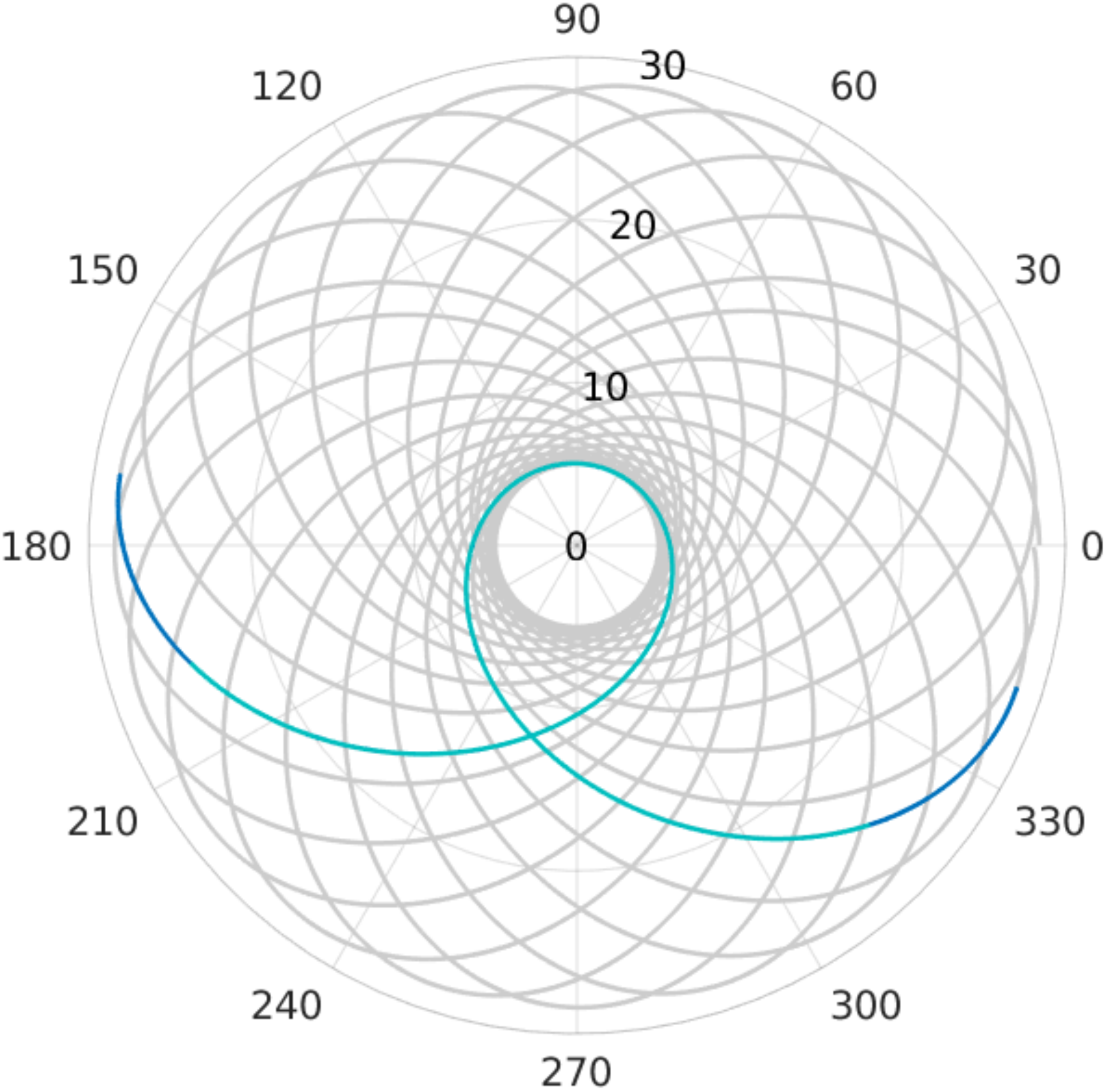}
  \hspace{0.2cm}
  \includegraphics[width=0.32\textwidth]{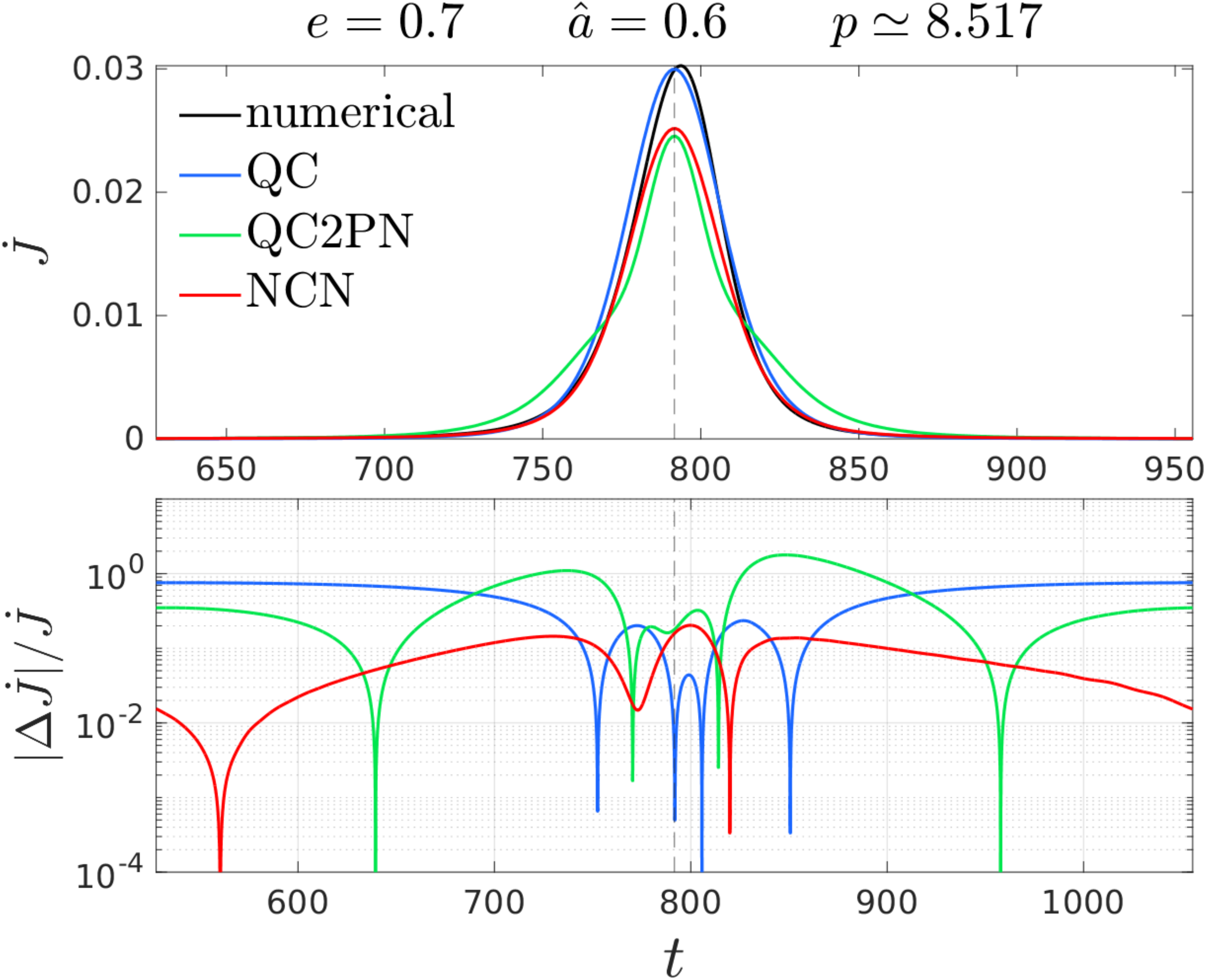}
  \hspace{0.2cm}
  \includegraphics[width=0.32\textwidth]{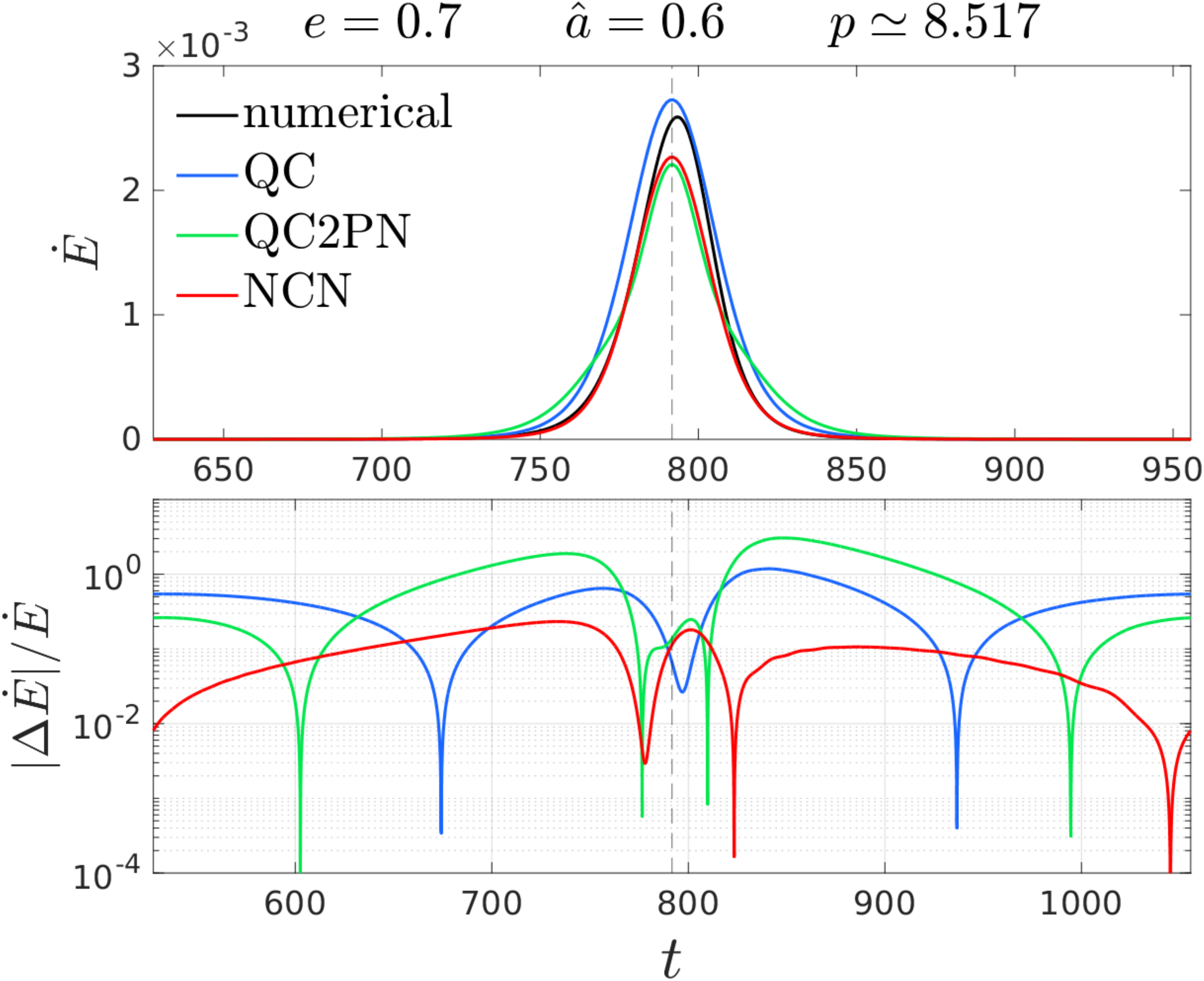}\\
   \vspace{0.5cm}
  \includegraphics[width=0.26\textwidth]{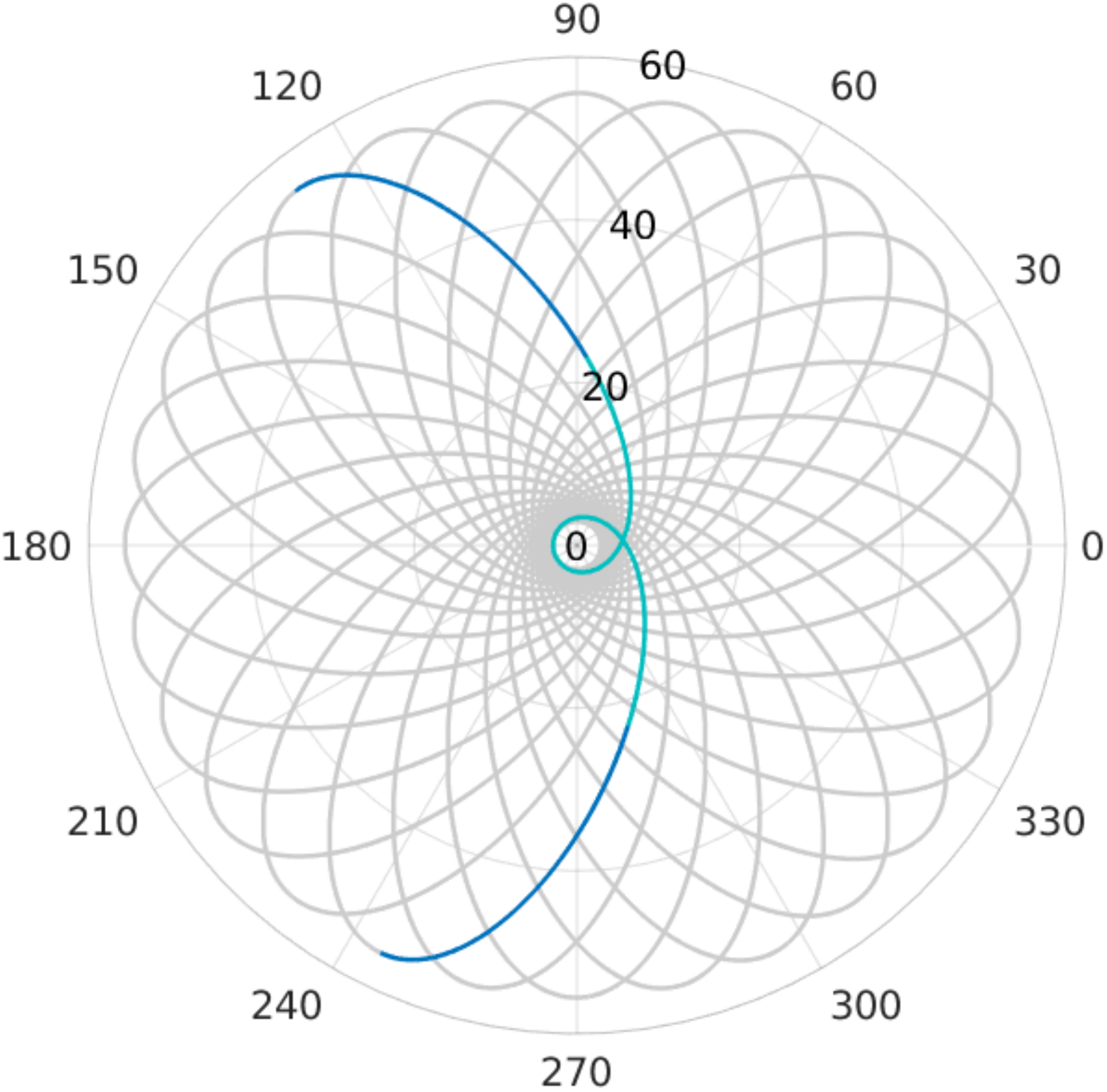}
  \hspace{0.2cm}
  \includegraphics[width=0.32\textwidth]{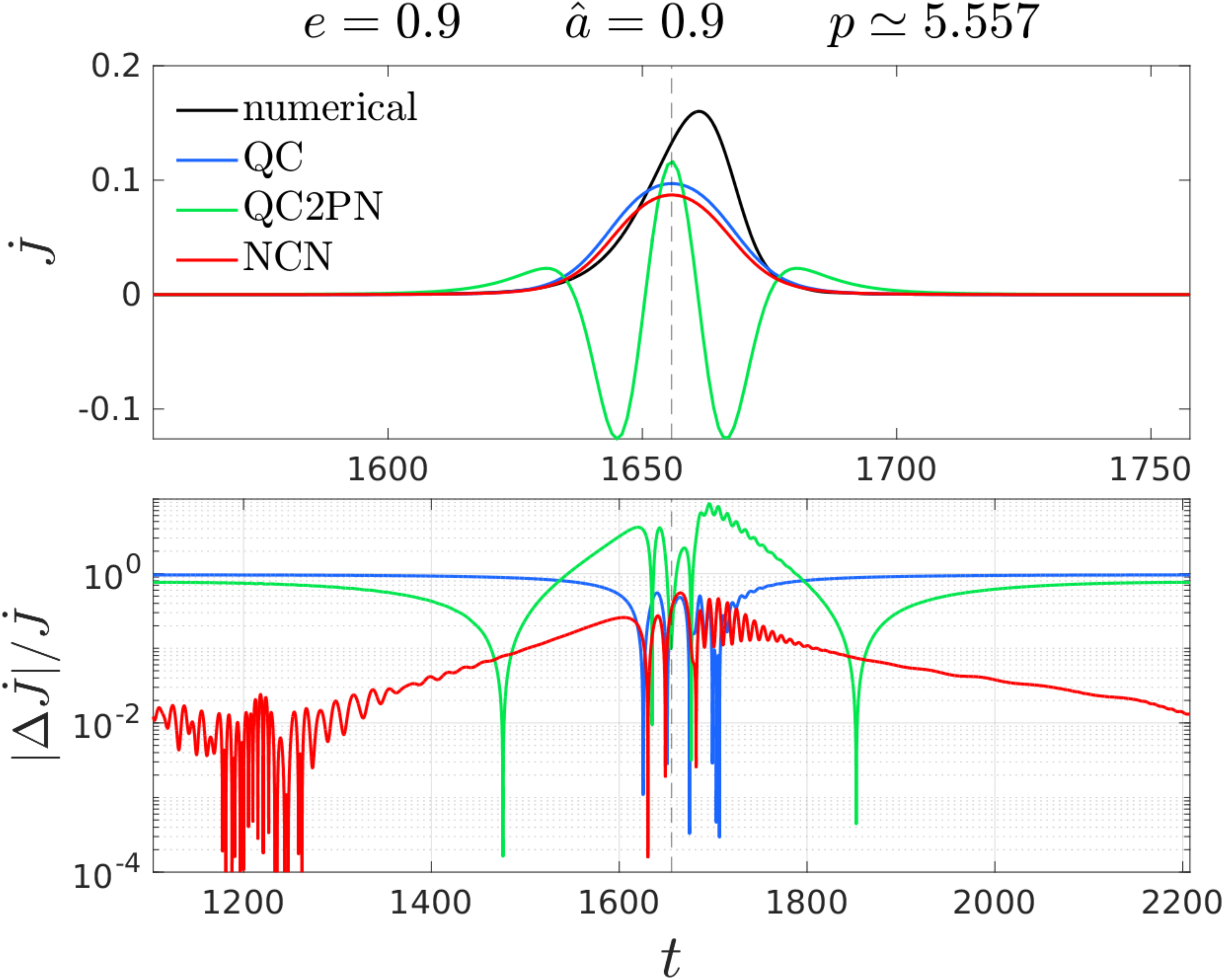}
  \hspace{0.2cm}
  \includegraphics[width=0.32\textwidth]{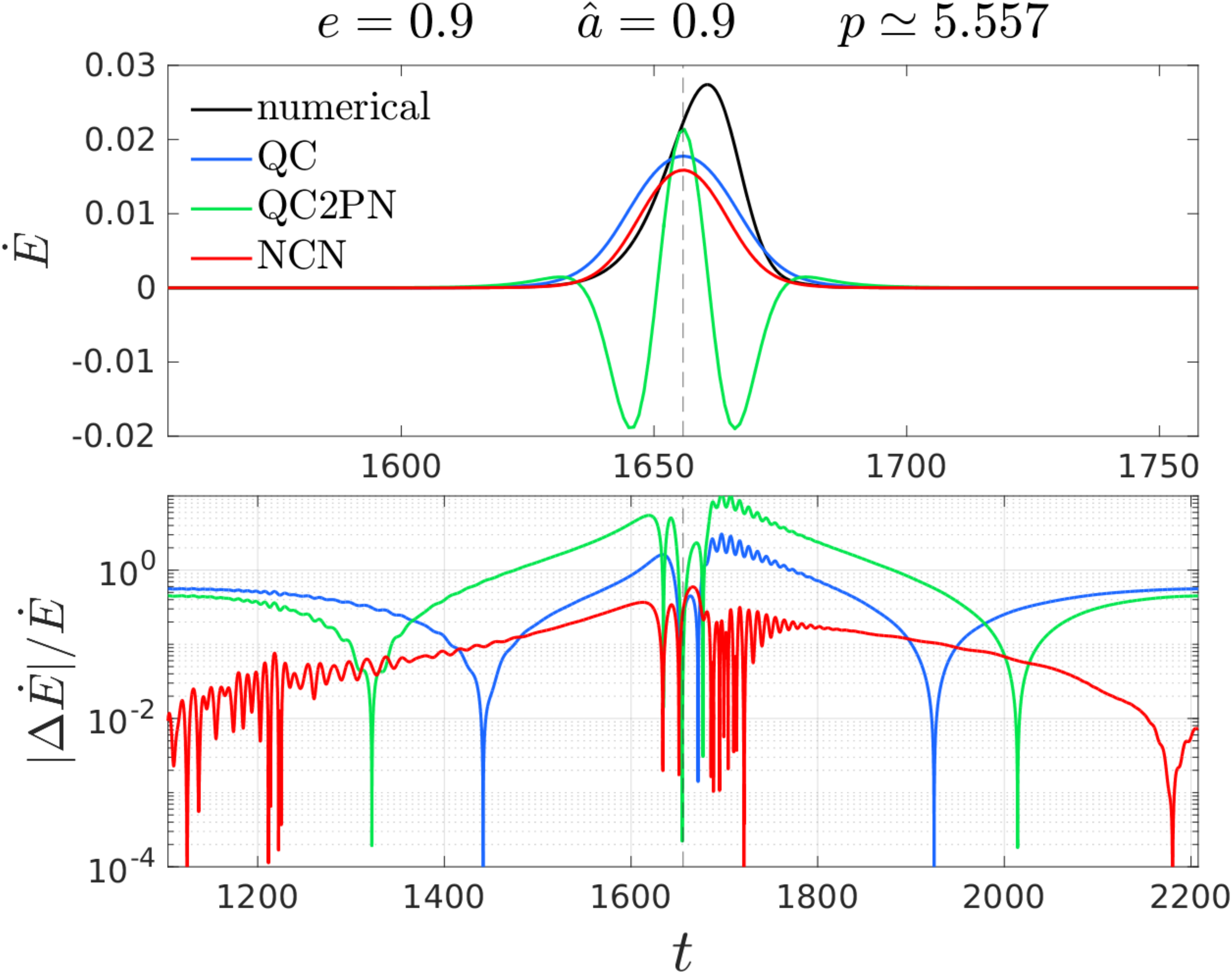}\\
  \caption{\label{fig:fluxes_spin}
  Analogous to Fig.~\ref{fig:fluxes_nospin}, but for spinning configurations with 
  different eccentricities.
  The semilatera recta are computed according to $p=13 p_s(e,\ha)/p_s(e,0)$. 
  For each configuration we show the trajectories highlighting one radial period
  (blue) and the corresponding fluxes. We contrast the numerical fluxes (black) 
  with the  three analytical fluxes considered in this work: the \NCN{} flux (red) 
  computed using Eqs.~\eqref{eq:FphiTEOB} and~\eqref{eq:FrTEOB};
  the \QC{} flux (blue) from Eq.~\eqref{eq:RRqc}, which 
  is a proxy of the {\SEOBNRe} fluxes, and the \QC2PN{} flux with 2PN noncircular 
  corrections (green) from Eqs.~\eqref{eq:FKhalil}. 
  Each subpanel also reports the analytical/numerical relative difference.
  Note that while the relative differences in the lower panels are always shown over the 
  complete radial period, the fluxes for $e\geq 0.7$ are shown on  
  smaller time intervals (highlighted in aqua-green on the trajectories) 
  in order to show better the burst of radiation at the periastron
  passage (marked by a dashed vertical line). }
\end{figure*} 
%
%=================
% Averaged differences
%=================
\begin{figure*}[]
  \center
  \includegraphics[width=0.3\textwidth,height=4.0cm]{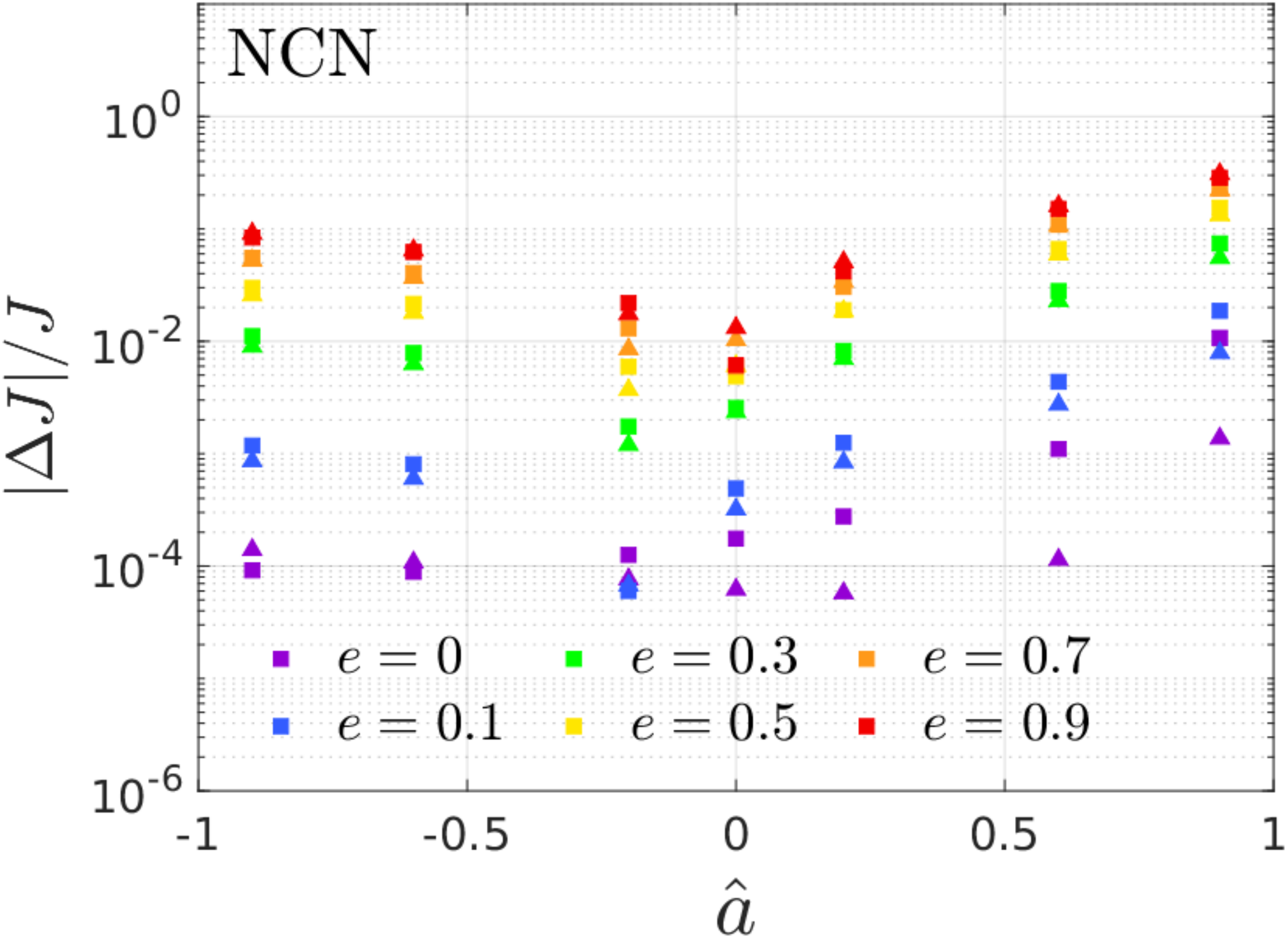}
  \hspace{0.2cm}
  \includegraphics[width=0.3\textwidth,height=4.0cm]{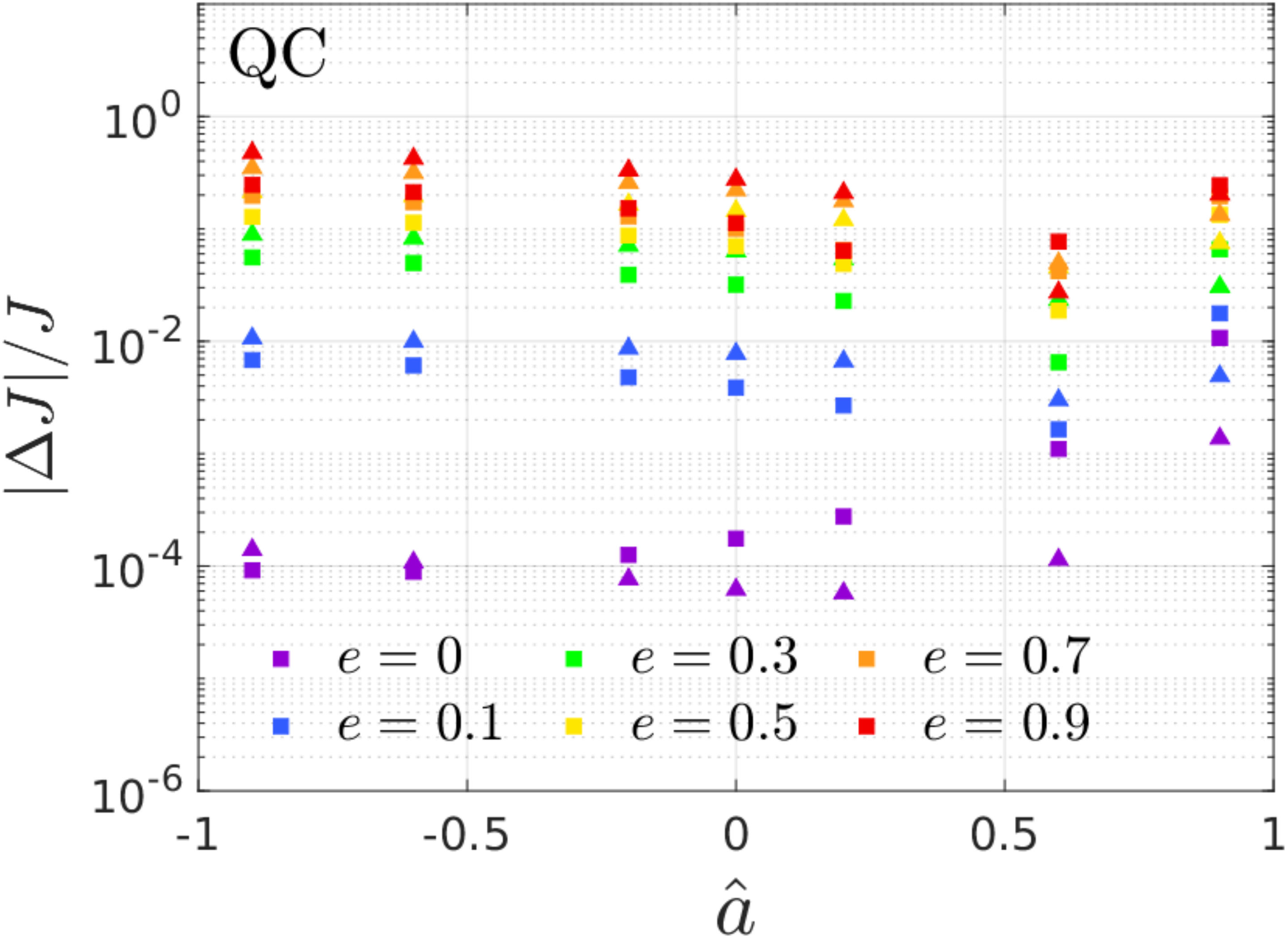}
  \hspace{0.2cm}
  \includegraphics[width=0.3\textwidth,height=4.0cm]{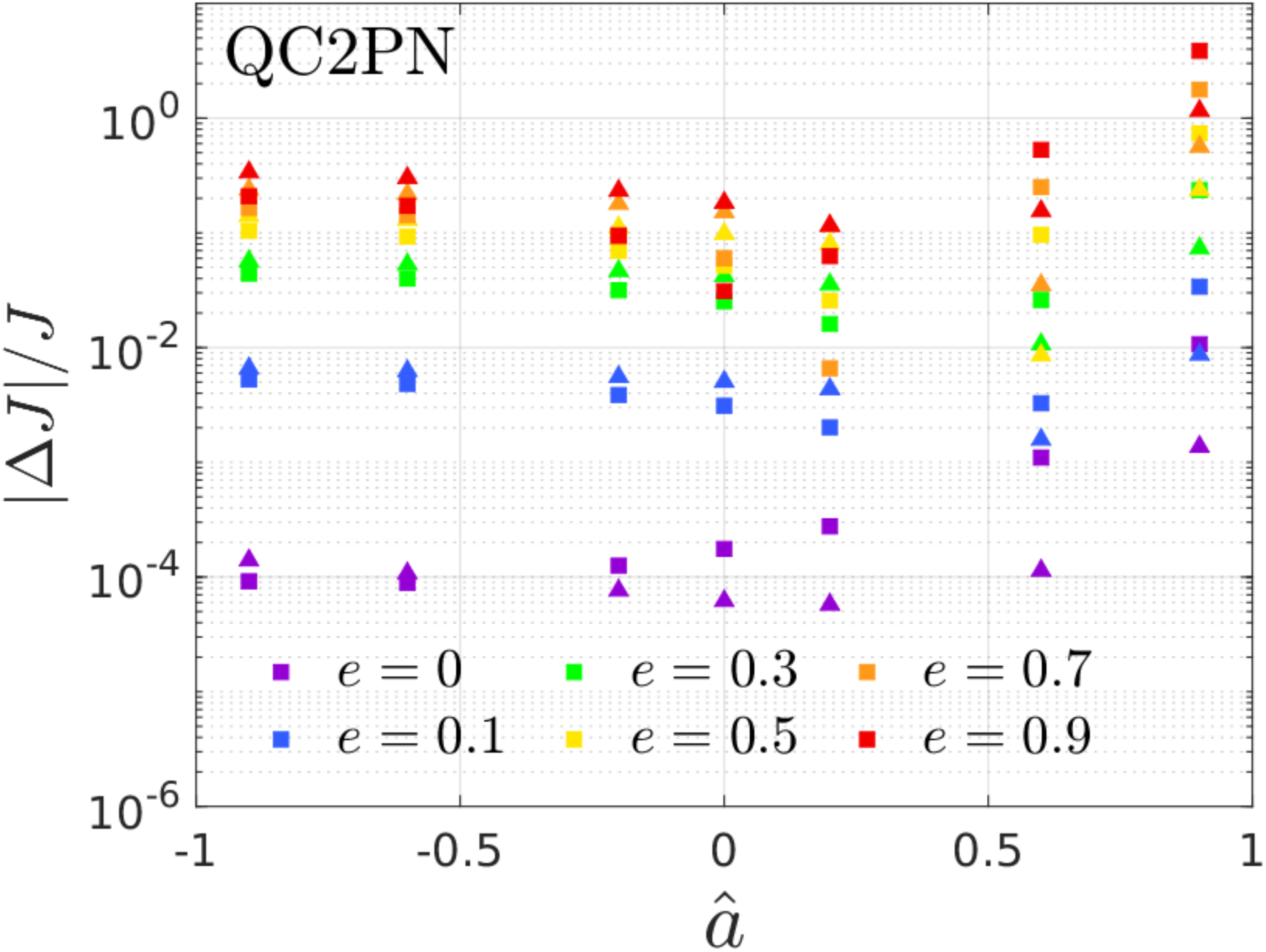}\\
  \vspace{0.2cm}
  \includegraphics[width=0.3\textwidth,height=4.0cm]{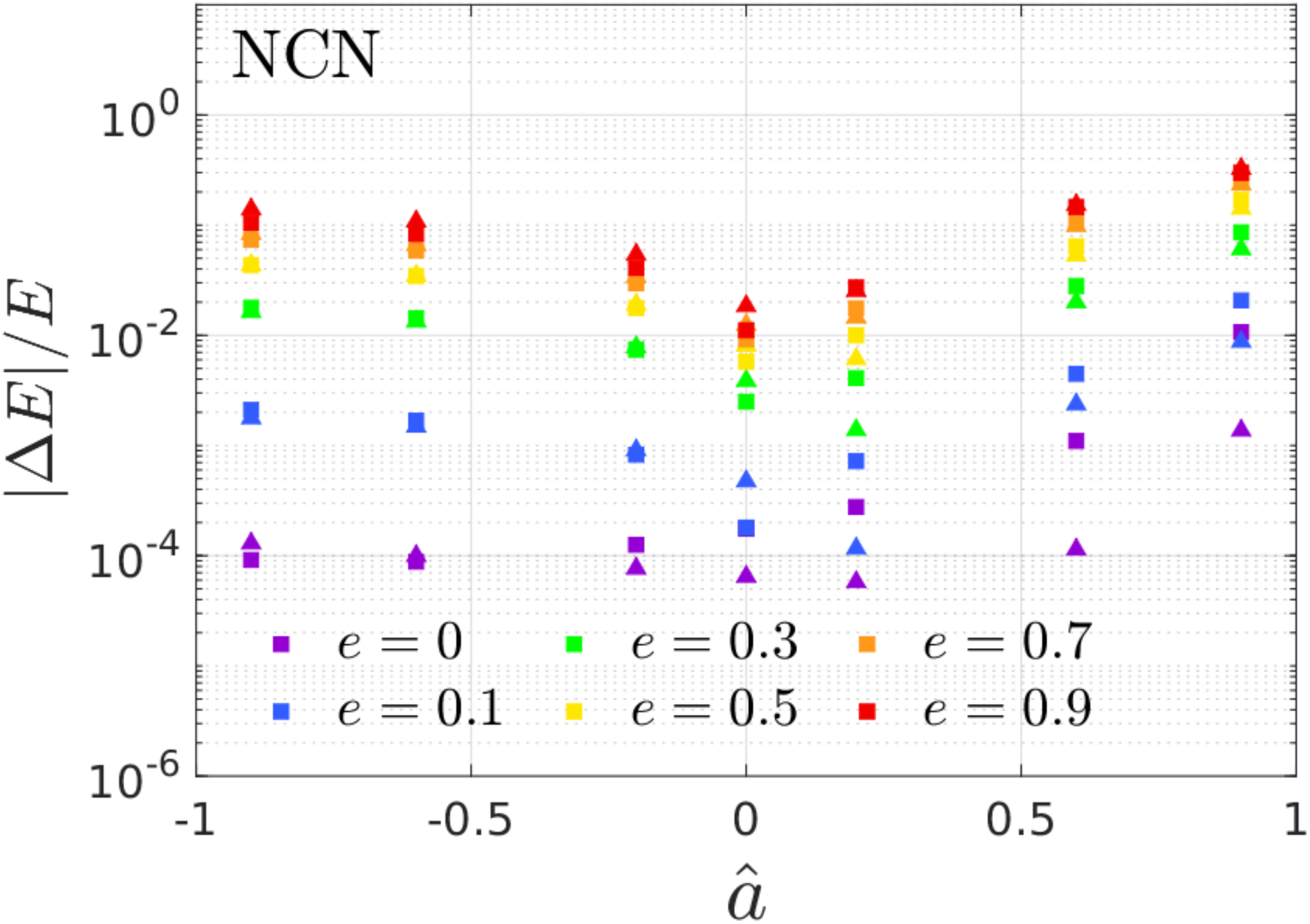}
  \hspace{0.2cm} 
  \includegraphics[width=0.3\textwidth,height=4.0cm]{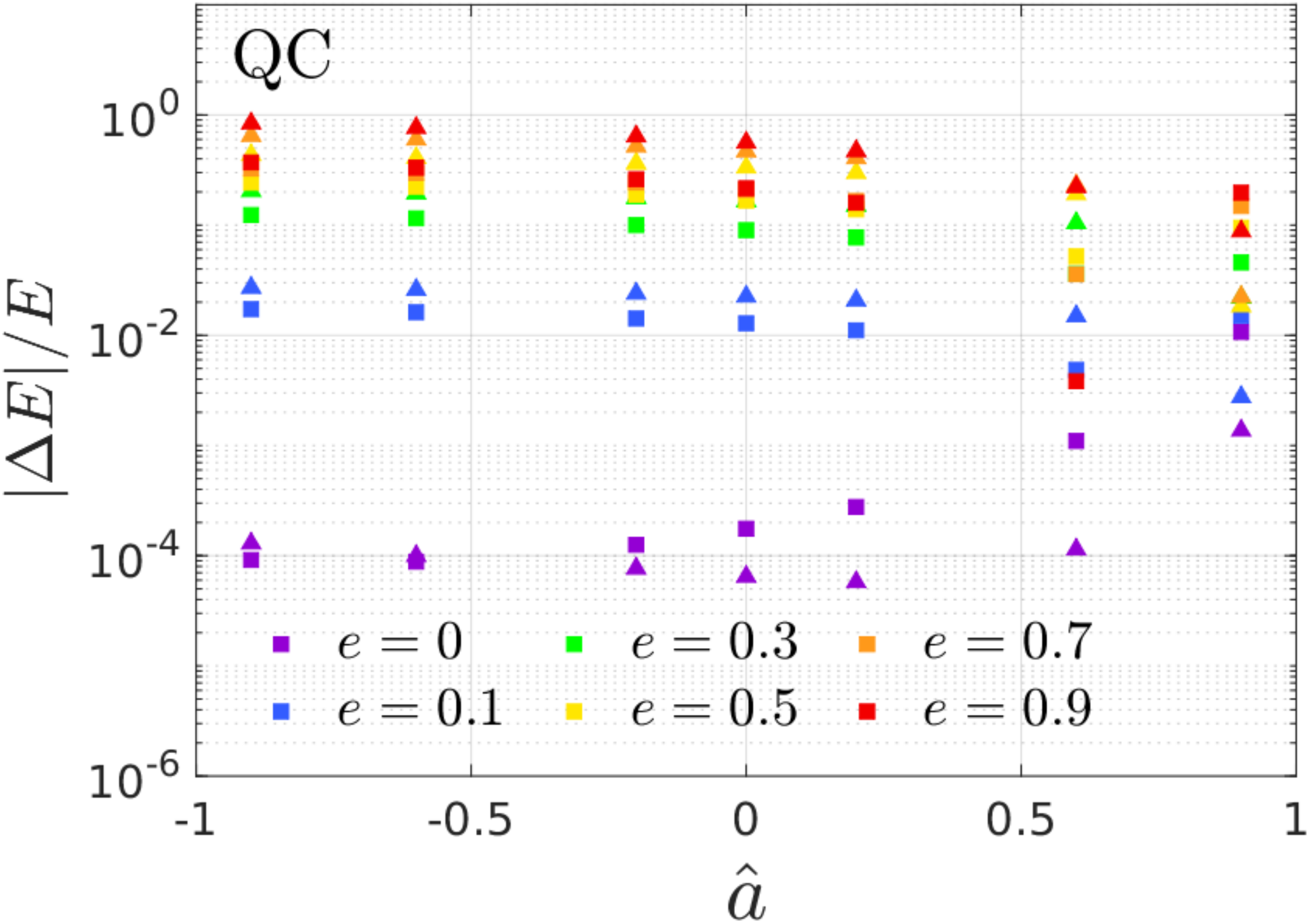}
  \hspace{0.2cm}
  \includegraphics[width=0.3\textwidth,height=4.0cm]{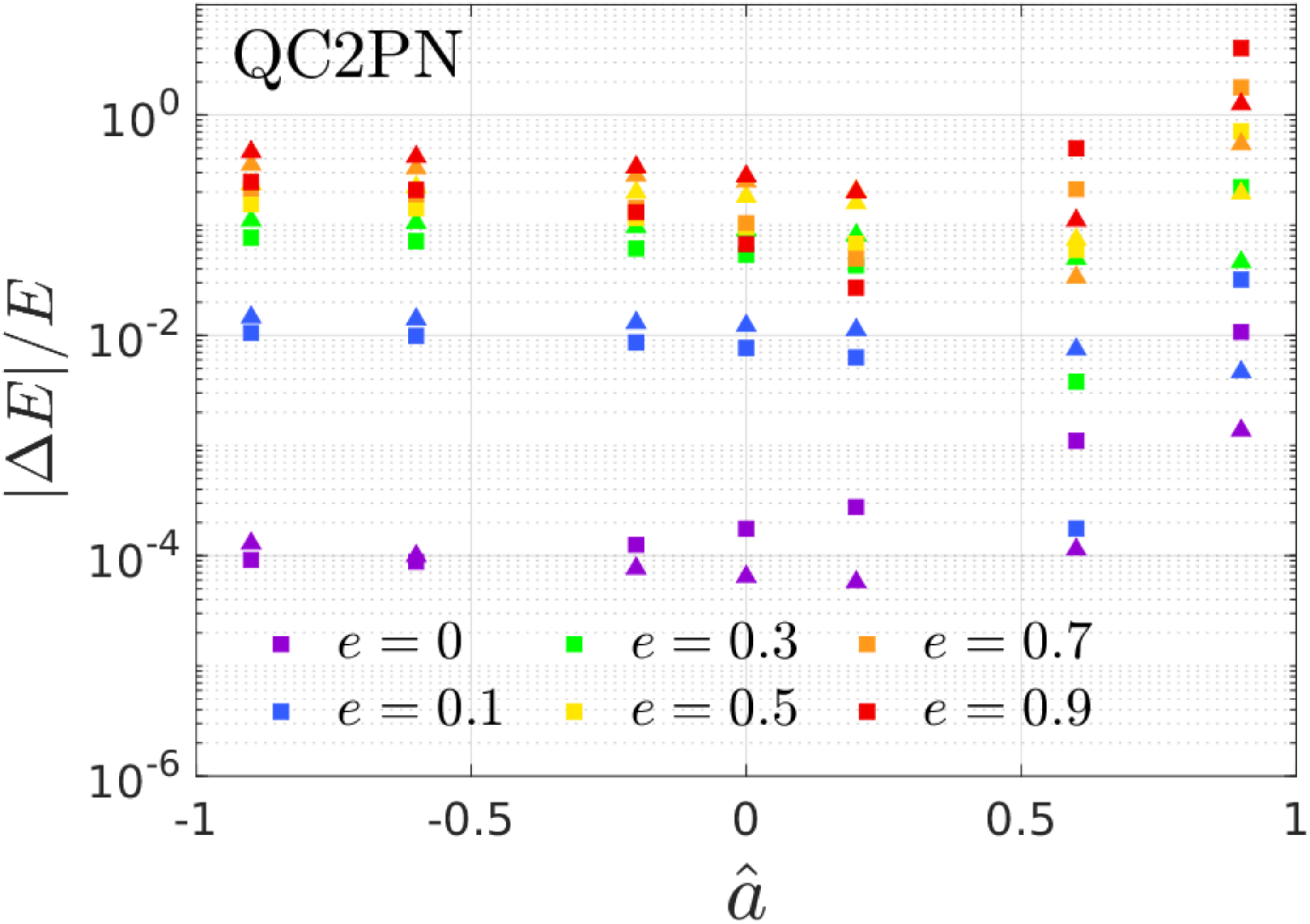}
  \caption{\label{fig:averaged_fluxes} Analytical/numerical relative differences for the 
  orbital averaged fluxes (angular momentum and energy) that correspond to eccentric configurations. 
  For each configuration the semilatus rectum is computed according to 
  $p = p_{\rm schw} p_s(e,\ha)/p_s(e,0)$ where $p_{\rm schw}=9$ (squares)
  or $p_{\rm schw}=13$ (triangles).} 
\end{figure*} 
We start our tests by computing the analytical fluxes along bound geodesics 
with eccentricities up to $e=0.9$. We report the trajectories and the instantaneous 
fluxes for some significant nonspinning cases in Fig.~\ref{fig:fluxes_nospin}, 
while in Fig.~\ref{fig:fluxes_spin} we show some spinning configurations. 
For each configuration we show the numerical fluxes (black), the \NCN fluxes (red) 
computed using the radiation reaction of Eqs.~\eqref{eq:FphiTEOB} and~\eqref{eq:FrTEOB}, 
the \QC fluxes (blue) computed using Eq.~\eqref{eq:RRqc}, and the \QC2PN fluxes (green).
Figure~\ref{fig:fluxes_nospin} illustrates that the \QC{} fluxes are the less accurate ones 
even for $e=0.1$. The discrepancies with the numerical results are more relevant for higher eccentricities. 
In particular, the \QC{} fluxes  overestimate the numerical flux at periastron in almost all the cases.
However, for some cases with high eccentricity and high spin, the \QC{} 
expressions incidentally provide the best approximation to the 
numerical fluxes at periastron, as shown in the case with 
$(e,\ha,p)=(0.7,0.6,8.517)$ of Fig.~\ref{fig:fluxes_spin}. 
To better understand this aspect, one should note that for
high spins (and high eccentricity) the periastron of the orbit
is located in stronger field, and thus the numerical fluxes 
have general-relativistic contributions that are not included in 
the EOB analytical fluxes. 
Indeed, the last panel of Fig.~\ref{fig:fluxes_spin} highlights 
that the numerical fluxes are not symmetric with respect to the periastron 
and that there are quasinormal-modes excitations of the 
Kerr black hole~\cite{10.1143/PTP.72.494,Rifat:2019fkt,Thornburg:2019ukt}. These
excitations are responsible for the oscillations in the relative differences after 
the periastron passage that can be clearly seen in the case with $(e,\ha)=(0.9,0.9)$
shown in Fig.~\ref{fig:fluxes_spin}. 
As a consequence, the analytical flux underestimates the numerical results in strong 
field, as already observed in Ref.~\cite{Albanesi:2021rby}.
Therefore considering that (i) the \QC{} fluxes overestimate the 
numerical fluxes for all the case with mild spins, and (ii) the numerical fluxes at periastron
in the strong field regime naturally overestimate analytical fluxes, we have that there
must be some configurations for which the quasicircular prescription
will provide the best approximation at periastron. However, in all 
cases with mild or high eccentricity the \QC{} fluxes are not reliable at
apastron, leading to comparatively large relative differences. 
While this is not relevant from a practical point of view since the main contribution 
of the radiation reaction to the dynamics happens at periastron, it is, however, 
an indication that the quasicircular prescription is not suited to describe eccentric 
dynamics, as {\it a priori} expected. We will further confirm this statement in the
next subsection where we will analyze the fluxes more systematically considering
the analytical/numerical differences between averaged fluxes over the whole parameter space
$(e,\ha) = [0, 0.9]\times [-0.9, 0.9]$. 
The fact that the  \QC{} radiation reaction overestimates the numerical fluxes
at periastron generally results in an unphysical acceleration of the dynamics once
the corresponding ${\cal F}_\varphi$ enters at the rhs of Hamilton's equations.
While an example in the test-mass limit has already been provided in 
Fig.~\ref{fig:overview_wave}, the relevance of this aspect for comparable-mass binaries
will be further discussed in Sec.~\ref{sec:comp_mass}.

Let us now focus on the \QC2PN{} prescription, which follows from Ref.~\cite{Khalil:2021txt} 
(green  in Figs.~\ref{fig:fluxes_nospin} and \ref{fig:fluxes_spin}). 
While for low eccentricity the corresponding fluxes are reliable, as shown
in the $e=0.1$ case of Fig.~\ref{fig:fluxes_spin}, the prescription starts
to be less accurate for mild and high eccentricities. In particular, while this 
choice is generally more accurate than the \QC{} case at apastron,
the relative differences are still larger than the \NCN{} ones.
Moreover, when the particle is almost at the periastron or shortly after
the periastron passage (i.e. when the radial momentum is close to its maximum value
along the orbit), the analytical/numerical relative differences for this prescription 
reach their maximum and are even larger than for the \QC{} case. 
This aspect can be clearly seen in all the configurations with $e\geq 0.5$ shown 
in Figs.~\ref{fig:overview},~\ref{fig:fluxes_nospin} and~\ref{fig:fluxes_spin}.
For large values of eccentricity and large spin the fluxes associated to the \QC2PN{} 
choice can show unphysical behaviors, developing multiple peaks and becoming negative. 
This can be clearly seen in the last configuration of Fig.~\ref{fig:fluxes_spin}, 
but we anticipate that a similar pathological behavior can occur even in the nonspinning case 
if the noncircularity of the orbit is high enough, as shown in the last hyperbolic scattering 
of Fig.~\ref{fig:fluxes_nospin_hyp} in Sec.~\ref{sec:hyp}.
In Fig.~\ref{fig:fluxes_spin} we consider configurations with semilatus 
rectum $p=13 p_s(e,\ha)/p_s(e,0)$, but the pathologic behavior becomes 
even more relevant if we consider smaller semilatera recta like $p=9 p_s(e,\ha)/p_s(e,0)$.

Let us finally turn to the analysis of the \NCN{} analytical fluxes, repeating for completeness
some of the analysis of Ref.~\cite{Albanesi:2021rby}.
Figure~\ref{fig:fluxes_nospin} shows that such analytical expressions are the most reliable 
among the three and this becomes especially evident for $e\geq 0.3$. Note that the \NCN{}
expressions are the only ones where the relative differences are smaller 
at apastron than at periastron, as one would naturally expect since the periastron is reached 
in a stronger gravitational field, and thus the accuracy of any PN-based expression (though resummed)
should be reduced there.

Finally, we conclude with an observation on the numerical fluxes
that could be relevant for the next subsection. 
As can be seen in the last plots of Fig.~\ref{fig:fluxes_nospin} or in 
Fig.~\ref{fig:fluxes_spin}, the relative differences of highly eccentric configuration 
are noisy for early times (i.e. near the apastron). 
This is linked to the fact that the numerical flux is still slightly contaminated by the junk
radiation\footnote{The numerical simulations do not start necessarily from $t=0$.}. 
However, the highly-eccentric fluxes have a small absolute 
magnitude near the apastron and thus this contamination does not
affect the averaged fluxes that we will consider in the next section. 
To be more quantitative, consider that for the case with 
$(e,\ha,p)=(0.9,0,13)$ in the last row of Fig.~\ref{fig:fluxes_nospin}
the angular momentum flux and the energy flux have values of
$5\cdot 10^{-8}$ and $2\cdot 10^{-11}$ at apastron, respectively.

\subsection{Orbital averaged fluxes for eccentric orbits}
To have a more systematic picture of the accuracy of the analytical
expressions discussed above, it is useful to consider the fluxes 
averaged along a radial orbit and the corresponding relative differences
with the exact quantities.
However, this method tests the reliability of the prescriptions in the neighborhood
of the periastron, where there is the main contribution to the integrated flux. 
In other words, with this method it is not possible to see that the \QC{}
and \QC2PN{} expressions  are not accurate at apastron. 
We report the analytical/numerical relative differences in absolute value plotted
against the Kerr spin parameter $\ha$ in Fig.~\ref{fig:averaged_fluxes}. 
We use squares as markers to indicate configurations with semilatus rectum 
$p=9 p_s(e,\ha)/p_s(a,0)$, while we use triangles for $p=13 p_s(e,\ha)/p_s(a,0)$.
Note that the results for the \NCN{} were already shown in Ref.~\cite{Albanesi:2021rby}.

Let us first note in Fig.~\ref{fig:averaged_fluxes} that the purple markers,
corresponding to the circular configurations, are the same in all cases
since the circular expressions are shared by the \QC{}, \QC2PN{} and \NCN{}
expressions. Their accuracy is further discussed in Appendix~\ref{appendix:hatf}. 
Already for $e=0.1$ the three prescriptions provide quite different results: both the \QC{} and \QC2PN{}
choices yield significantly larger differences than the \NCN{} one for both energy
and angular momentum. This remains true also for larger eccentricities.
On the other hand, for spins $\ha \geq 0.6$, the relative differences of the quasicircular 
prescription are similar to the ones of the {\TEOBResumS} fluxes, and even slightly better in
some cases. Nonetheless, we have already discussed in Sec.~\ref{sec:ecc_inst} 
that generally the quasicircular fluxes overestimate the numerical results,
but for high spins the numerical fluxes include contributions that are not
described by EOB models. For the last prescription shown in Fig.~\ref{fig:averaged_fluxes},
the one that includes 2PN noncircular information, for $\ha\geq 0.6$ we have
greater relative differences than for the other two prescriptions,  
but we already argued that for high spins and/or high
eccentricity this prescription shows pathological behavior. 
%It is interesting to note that the relative differences of the closer simulations
%(squares) are bigger than the farther simulations (triangles) only for 
For smaller spins instead the analytical/numerical discrepancies are bigger for 
farther simulations, as happens also for the other radiation reactions independently
of the spin considered (with few exceptions). 
This issue of the \QC2PN{} choice can even lead to negative averaged fluxes;
for example, this happens in the configurations with $\ha=0.9$, $e\geq 0.7$
and $p=9 p_s(e,\ha)/p_s(e,0)$.
%=============================
% Global difference for averaged fluxes
%=============================
\begin{figure}[t]
  \center
  \includegraphics[width=0.22\textwidth]{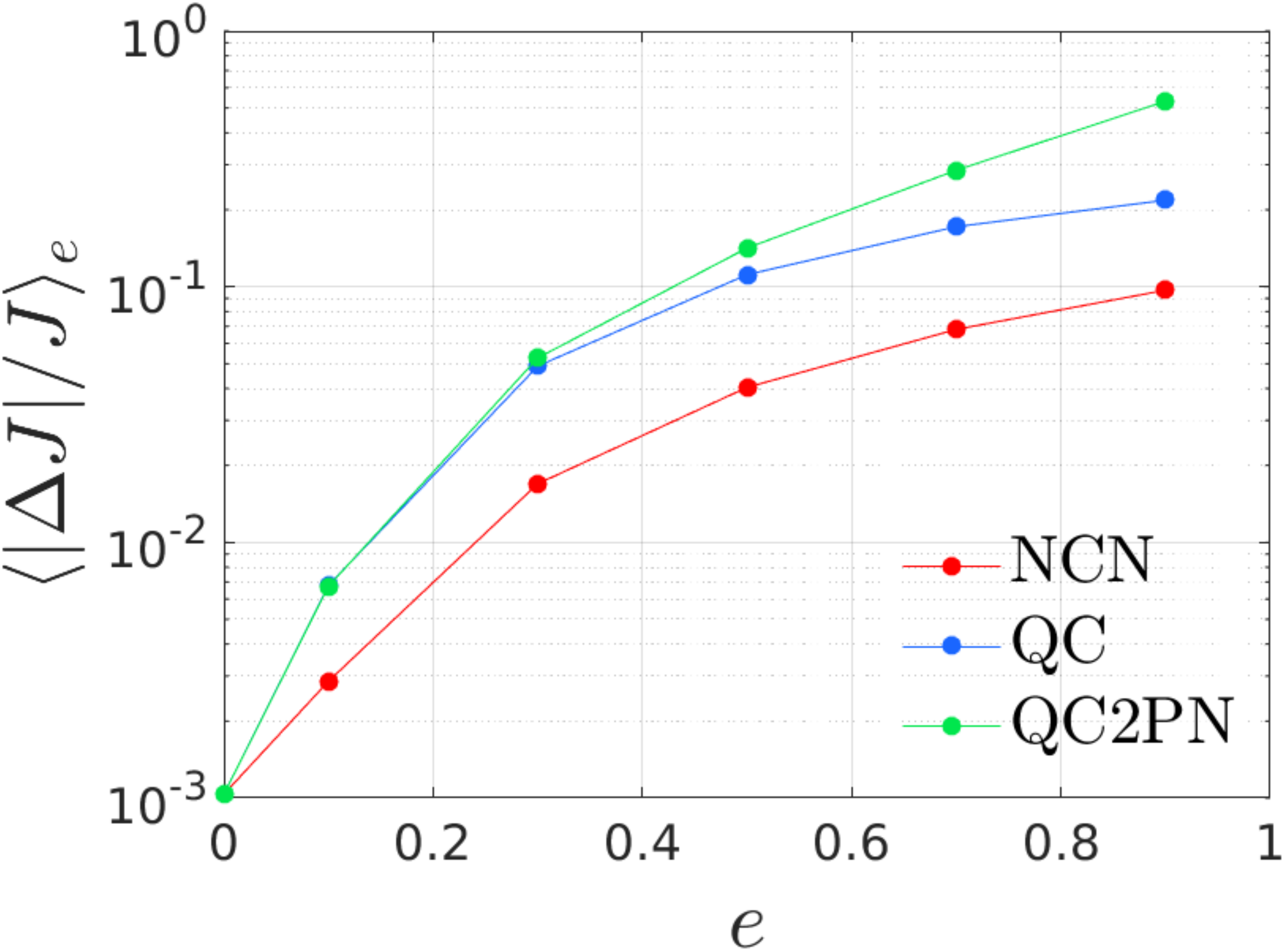}
  \vspace{0.3cm}
  \includegraphics[width=0.22\textwidth]{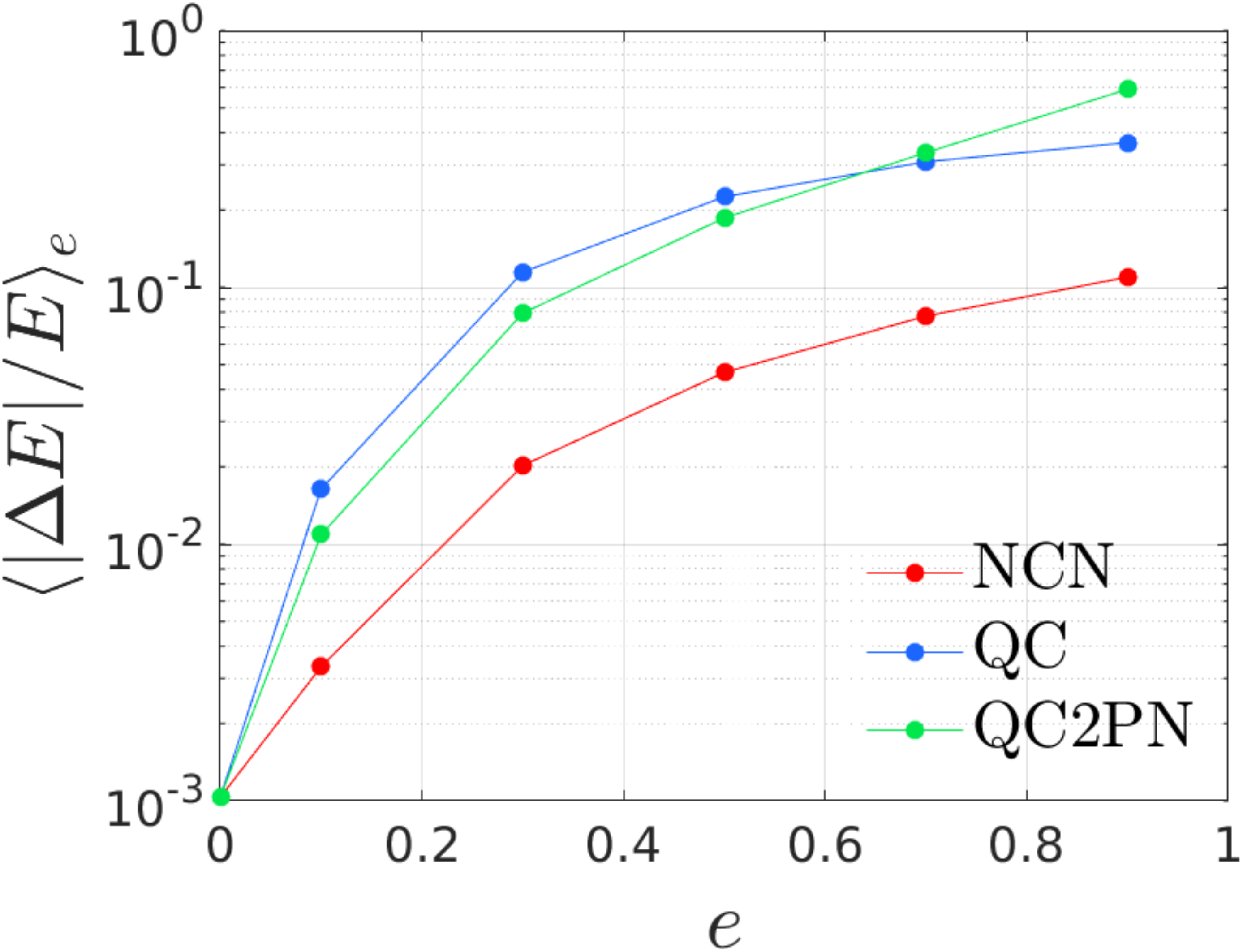}
  \caption{\label{fig:mean_averaged}
  Means of the analytical/numerical relative differences of the orbital averaged fluxes, 
  graphical representation of Table~\ref{tab:mean_differences}. 
  Each point represents the mean of the averaged fluxes of all the configurations with
  a certain value of eccentricity, that we indicate as $\langle ... \rangle_e$. 
  Note that we consider $\ha=(0, \pm 0.2, \pm 0.6, \pm 0.9)$ and two semilatera recta 
  for each pair $(e,\ha)$, so that each point is 
  the mean value over 14 averaged fluxes. As can be seen, \NCN{} is,
  on average, the most reliable radiation reaction.}
\end{figure} 

\begin{table}[]
   \caption{\label{tab:mean_differences}
   Averaged analytical/numerical relative differences 
   $\langle \delta F  \rangle = \langle |\Delta F|/F \rangle$  
   for the fluxes averaged over a radial orbit for the three analytical possibilities,
   \NCN{}, \QC{} and \QC2PN{}.
   We consider the means over the simulations with same eccentricity, 
   the average for the nonspinning configurations, and the average for all the configurations.
   The values are reported in percentage. The relative differences averaged over
   the circular configurations are
   $\langle \delta J \rangle_c = \langle \delta E \rangle_c = 0.10\%$.
   The averages at fixed eccentricity are shown in Fig.~\ref{fig:mean_averaged}.}
\begin{center}
\begin{ruledtabular}
\begin{tabular}{c | c c | c c | c c}
 $[\%]$ & \multicolumn{2}{c|}{NCN} & \multicolumn{2}{c|}{QC} & 
 \multicolumn{2}{c}{QC2PN} \\
\hline
 & $\langle \delta J \rangle $ & $\langle \delta E \rangle $ 
 & $\langle \delta J \rangle $ & $\langle \delta E \rangle $
 & $\langle \delta J \rangle $ & $\langle \delta E \rangle $ \\
\hline
$e = 0.1$ & 0.29 &  0.33  &  0.68 &  1.64  &  0.67 &  1.09 \\
$e = 0.3$ & 1.70 &  2.02  &  4.89 & 11.46  &  5.26 &  7.93 \\
$e = 0.5$ & 4.04 &  4.66  & 11.15 & 22.53  & 14.17 & 18.65 \\
$e = 0.7$ & 6.81 &  7.73  & 17.17 & 30.84  & 28.54 & 33.47 \\
$e = 0.9$ & 9.69 & 10.97  & 21.80 & 36.60  & 53.15 & 59.08 \\
\hline
$\ha = 0$ & 0.44 &  0.60  &  8.60 & 18.72  &  5.43 &  9.49 \\
\hline
   all    & 3.77 &  4.30  &  9.30 & 17.20  & 16.98 & 20.05 \\
\end{tabular}
\end{ruledtabular}
\end{center}
\end{table}

Finally, we can consider all the simulations for a certain value of eccentricity and we can 
obtain the mean of the corresponding averaged fluxes. 
We indicate this averaged at fixed eccentricity as 
$\langle ... \rangle_e$. In Fig.~\ref{fig:mean_averaged} we report these averages 
for the fluxes considered so far, showing once again that the NCN is
the most reliable for every value of eccentricity.

To conclude, also the systematic study of the reliability of the three prescriptions
considered so far suggests that the standard radiation reaction of {\TEOBResumS} is 
the most accurate, both for low and high eccentricity. 

\subsection{Hyperbolic orbits}
\label{sec:hyp}
%============
% Hyperblic orbits
%============
\begin{figure*}[]
  \center
  \includegraphics[width=0.26\textwidth]{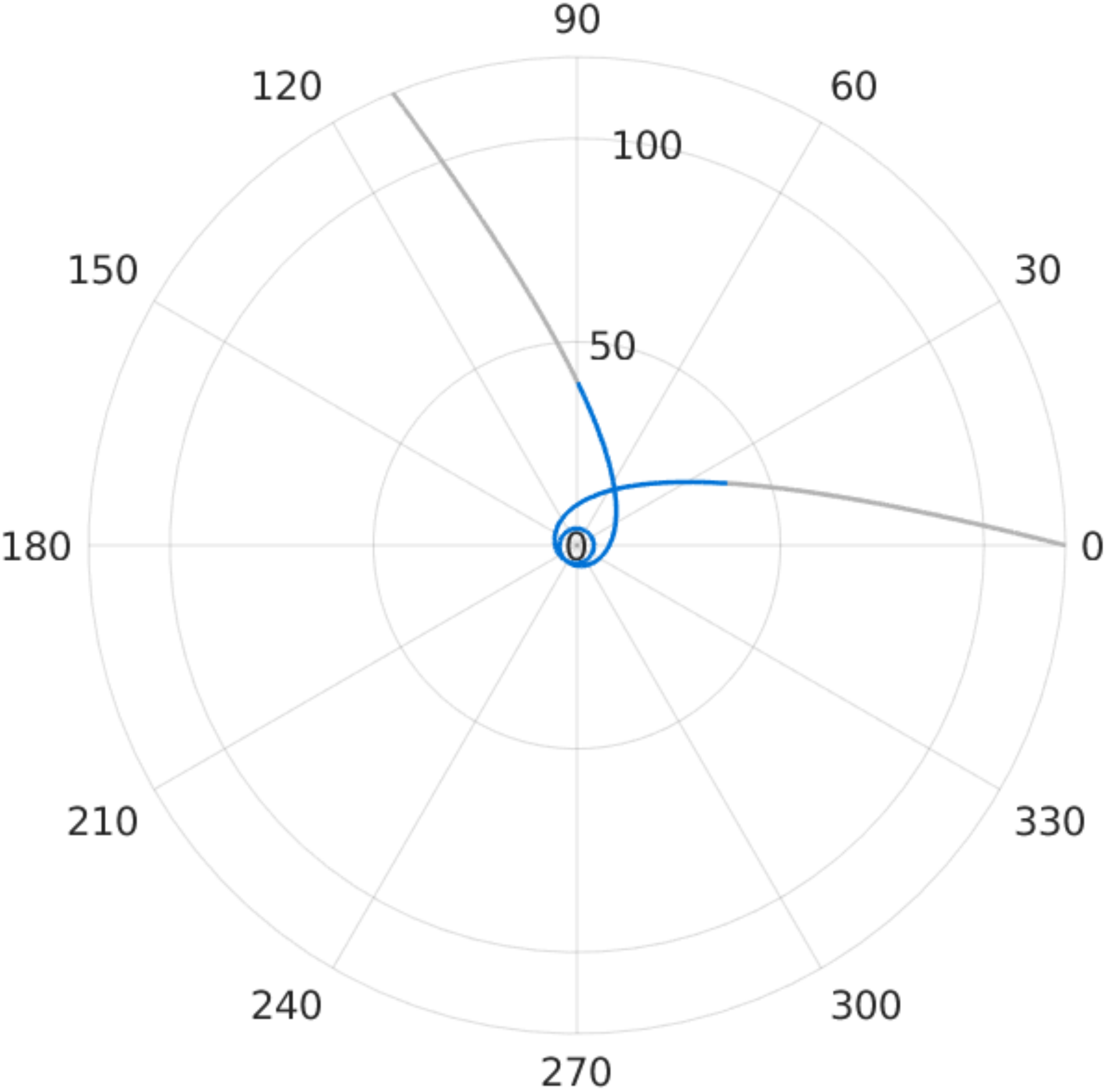}
  \hspace{0.2cm}
  \includegraphics[width=0.32\textwidth]{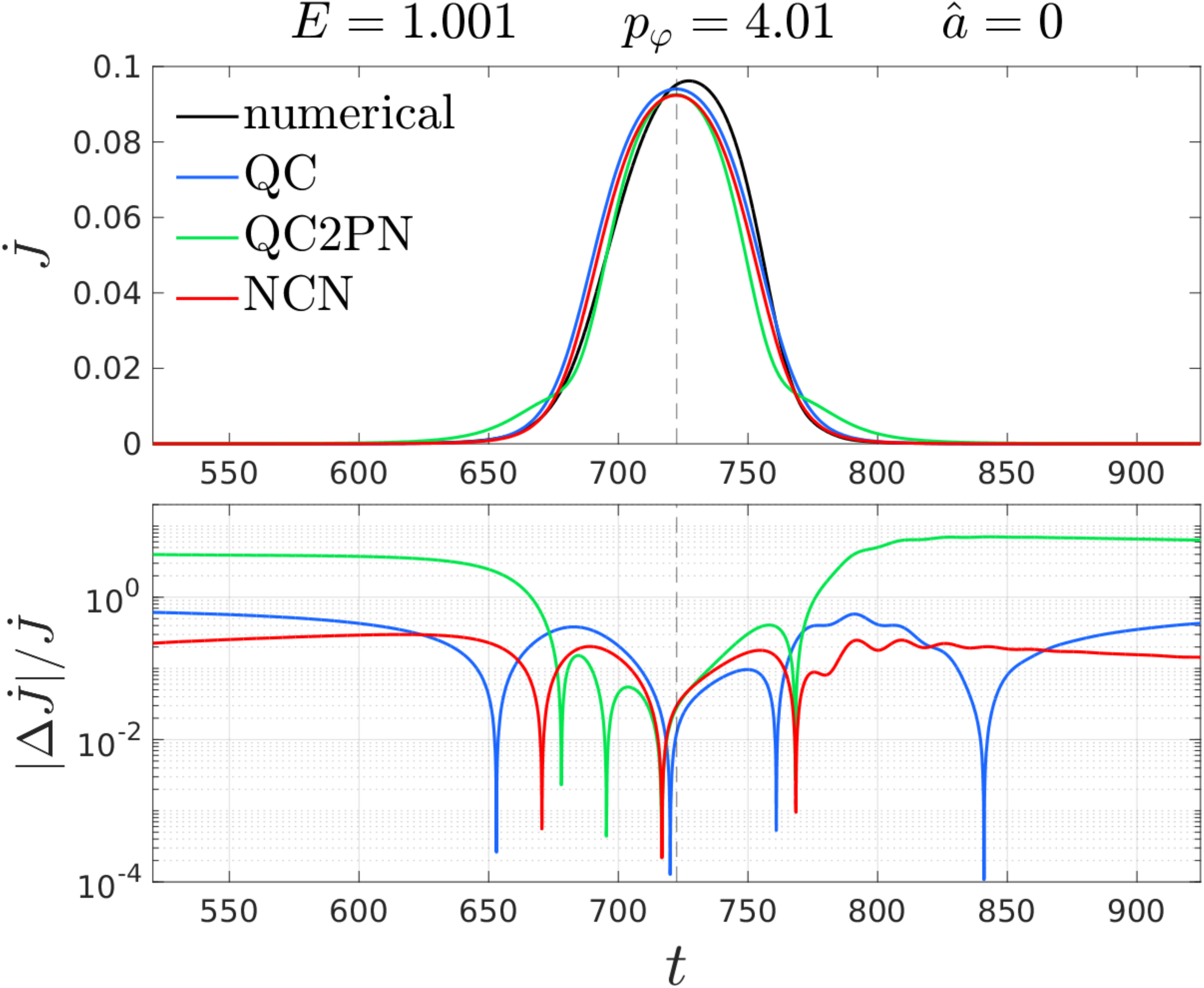}
  \hspace{0.2cm}
  \includegraphics[width=0.32\textwidth]{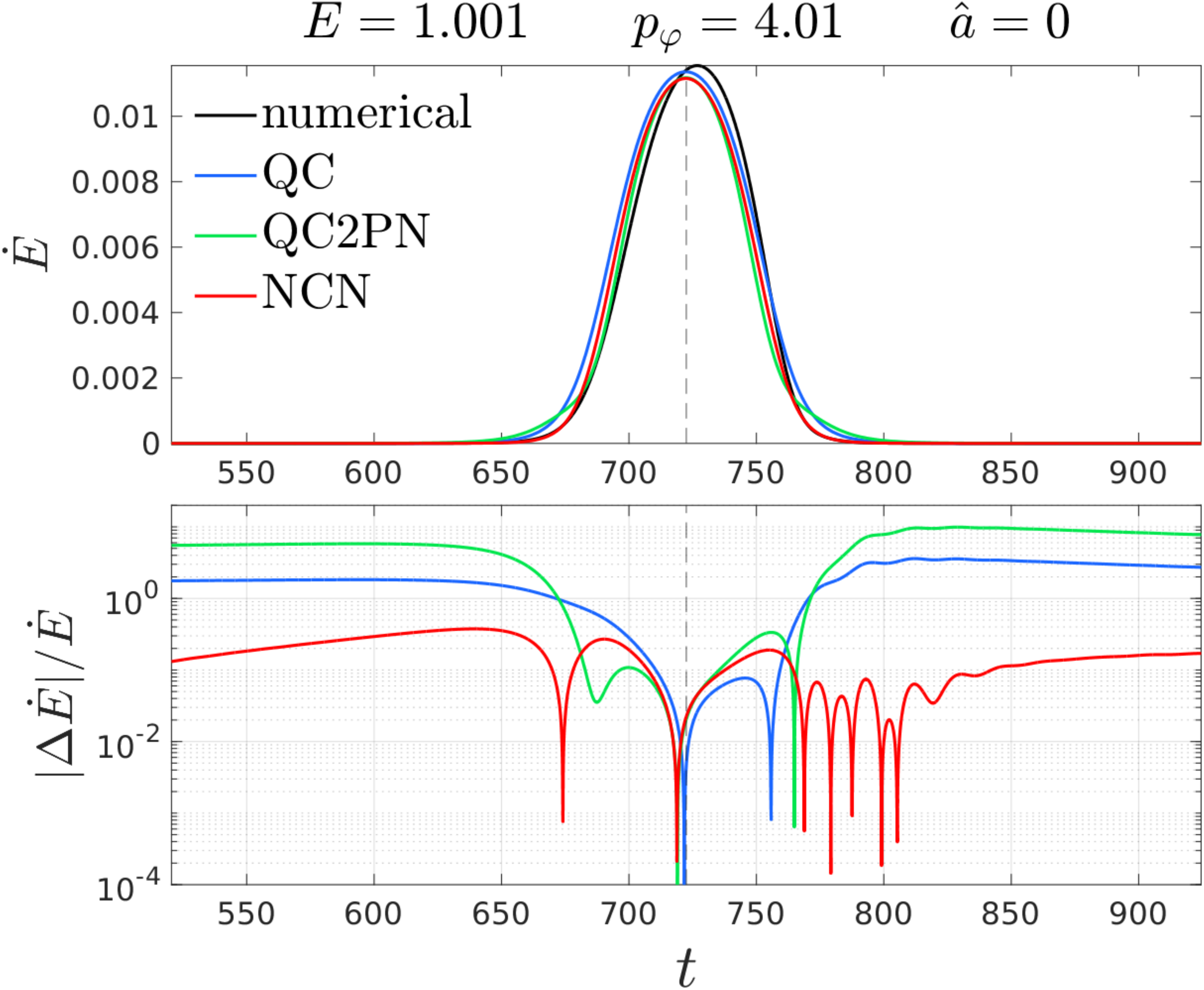}\\
  \vspace{0.5cm}
  \includegraphics[width=0.26\textwidth]{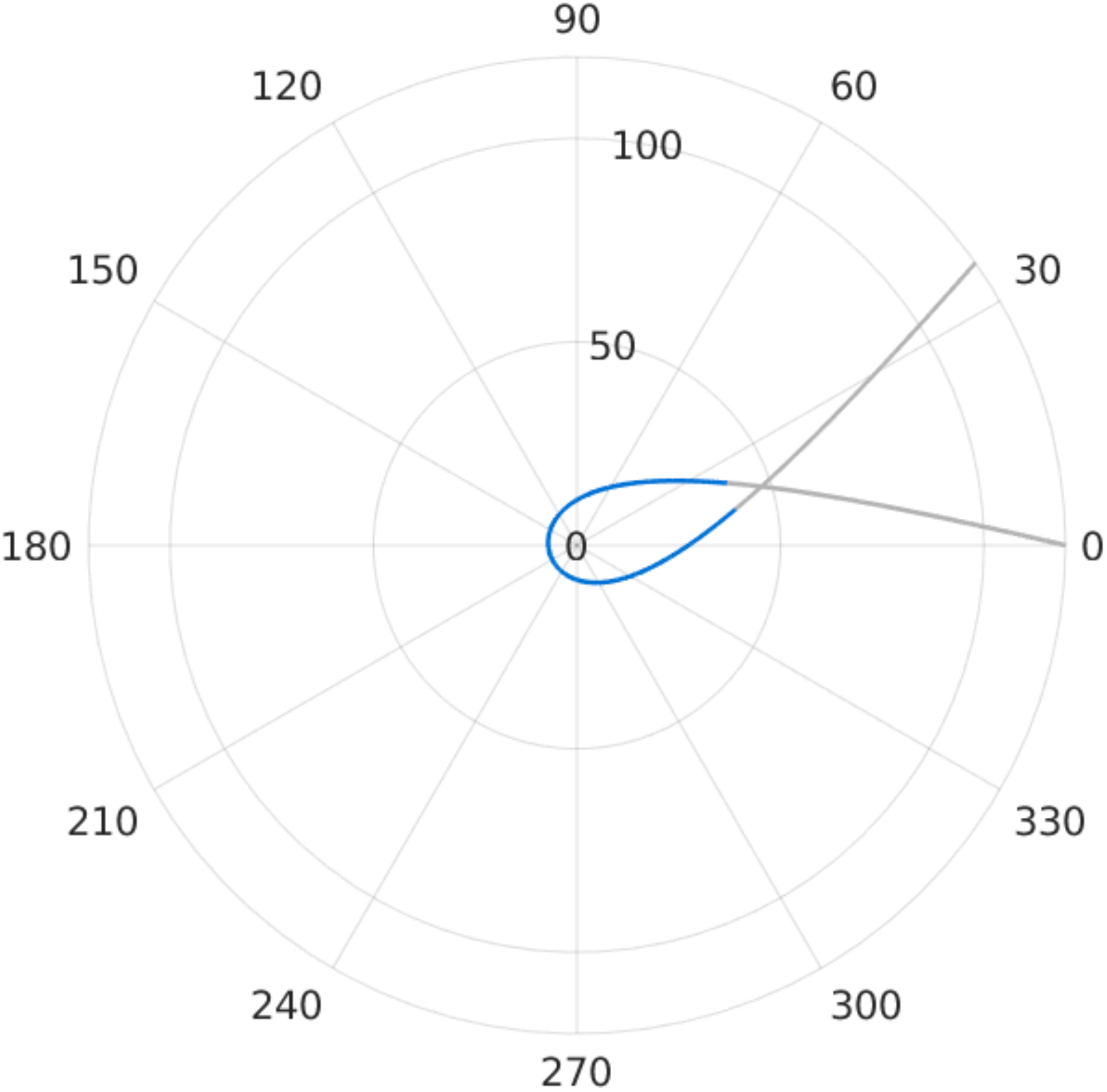}
  \hspace{0.2cm}
  \includegraphics[width=0.32\textwidth]{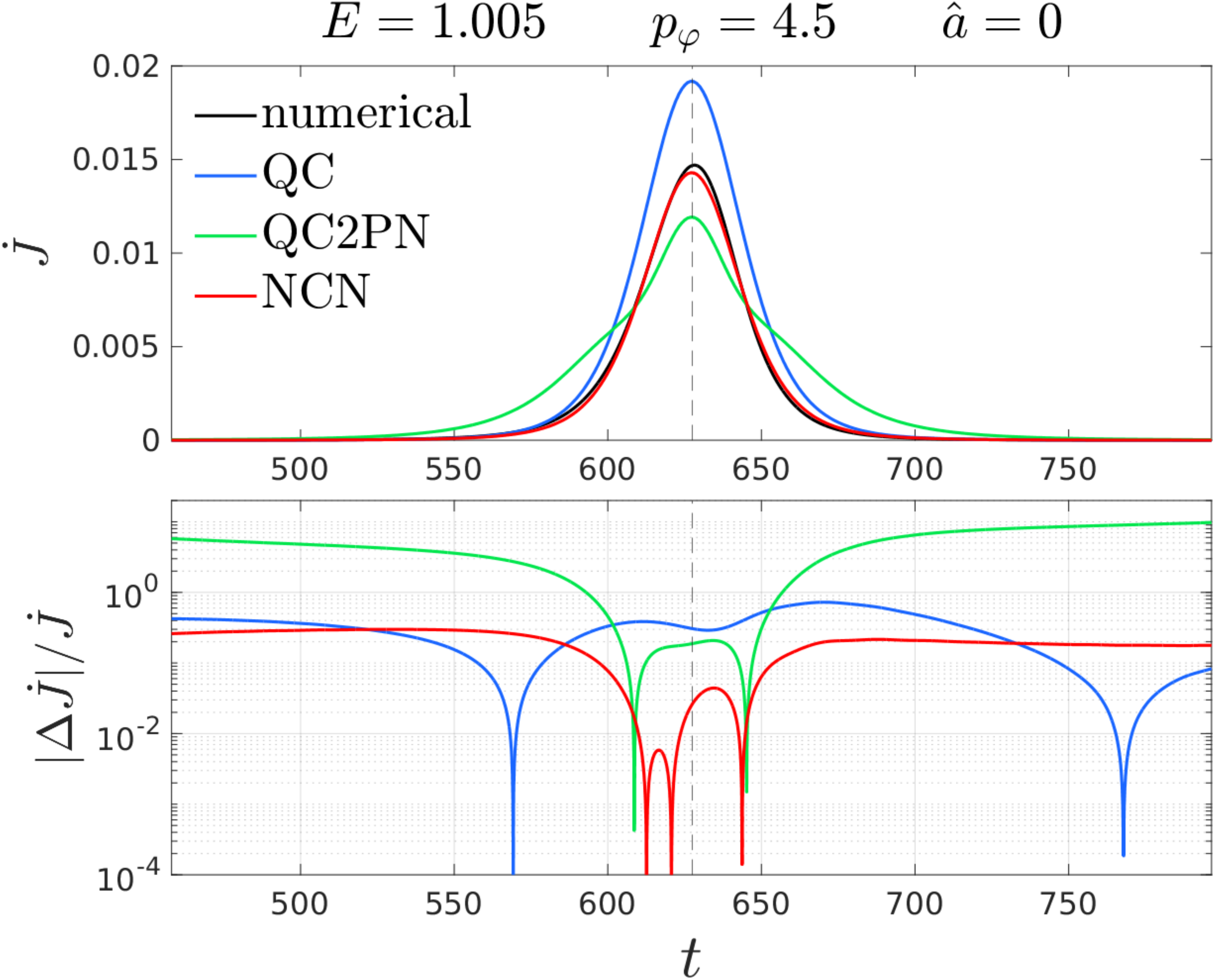}
  \hspace{0.2cm}
  \includegraphics[width=0.32\textwidth]{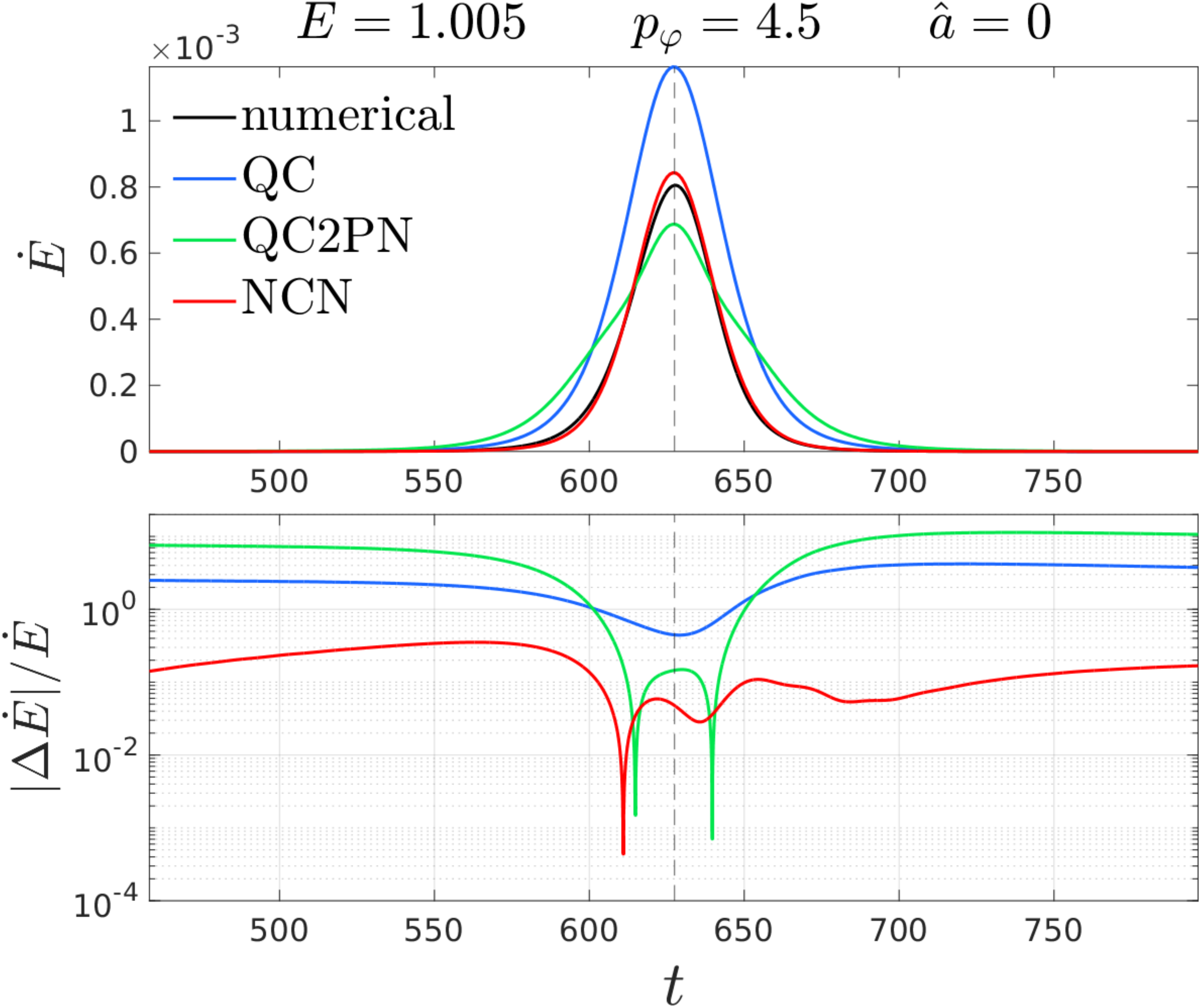}\\
  \vspace{0.5cm}
  \includegraphics[width=0.26\textwidth]{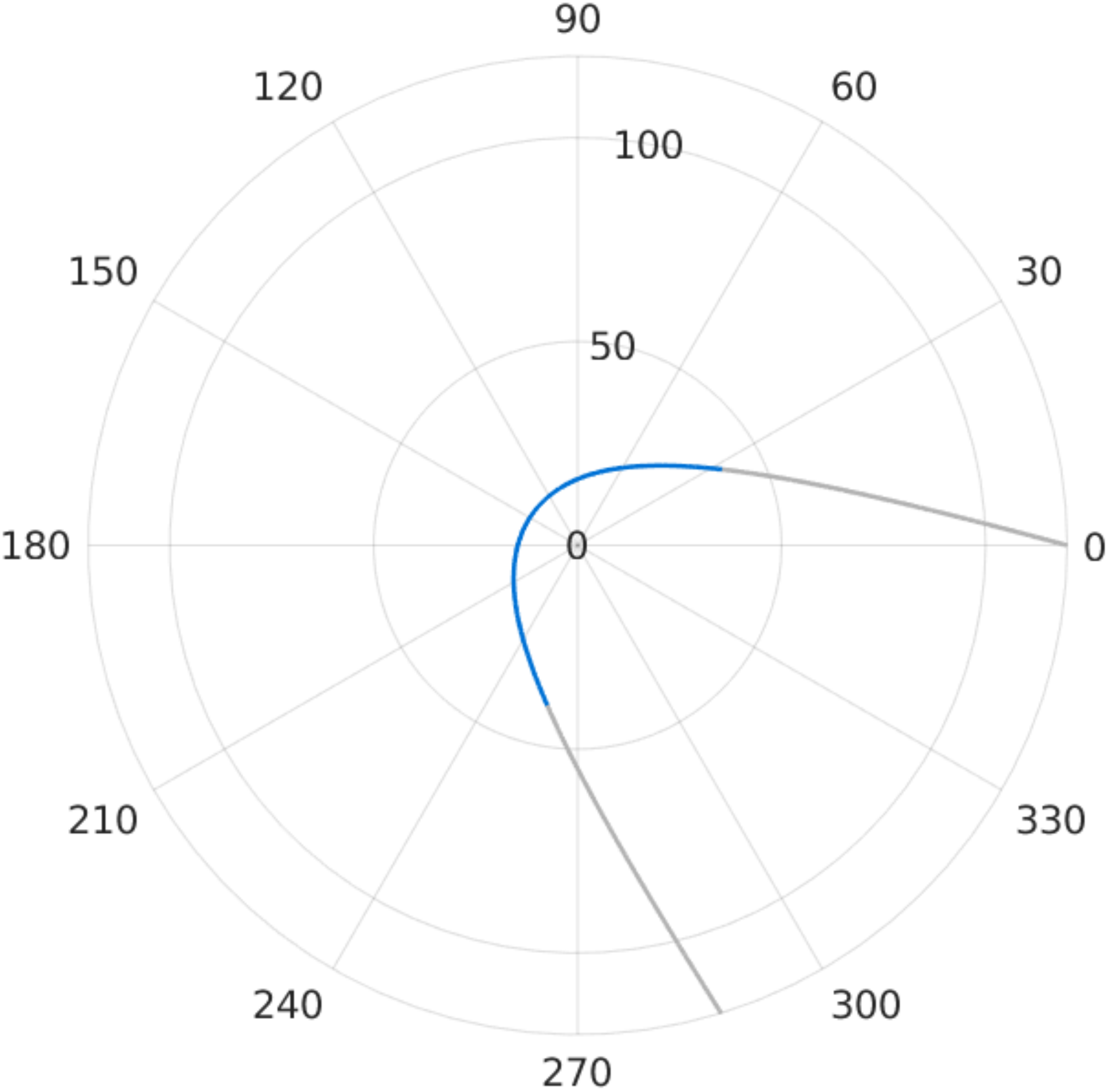}
  \hspace{0.2cm}
  \includegraphics[width=0.32\textwidth]{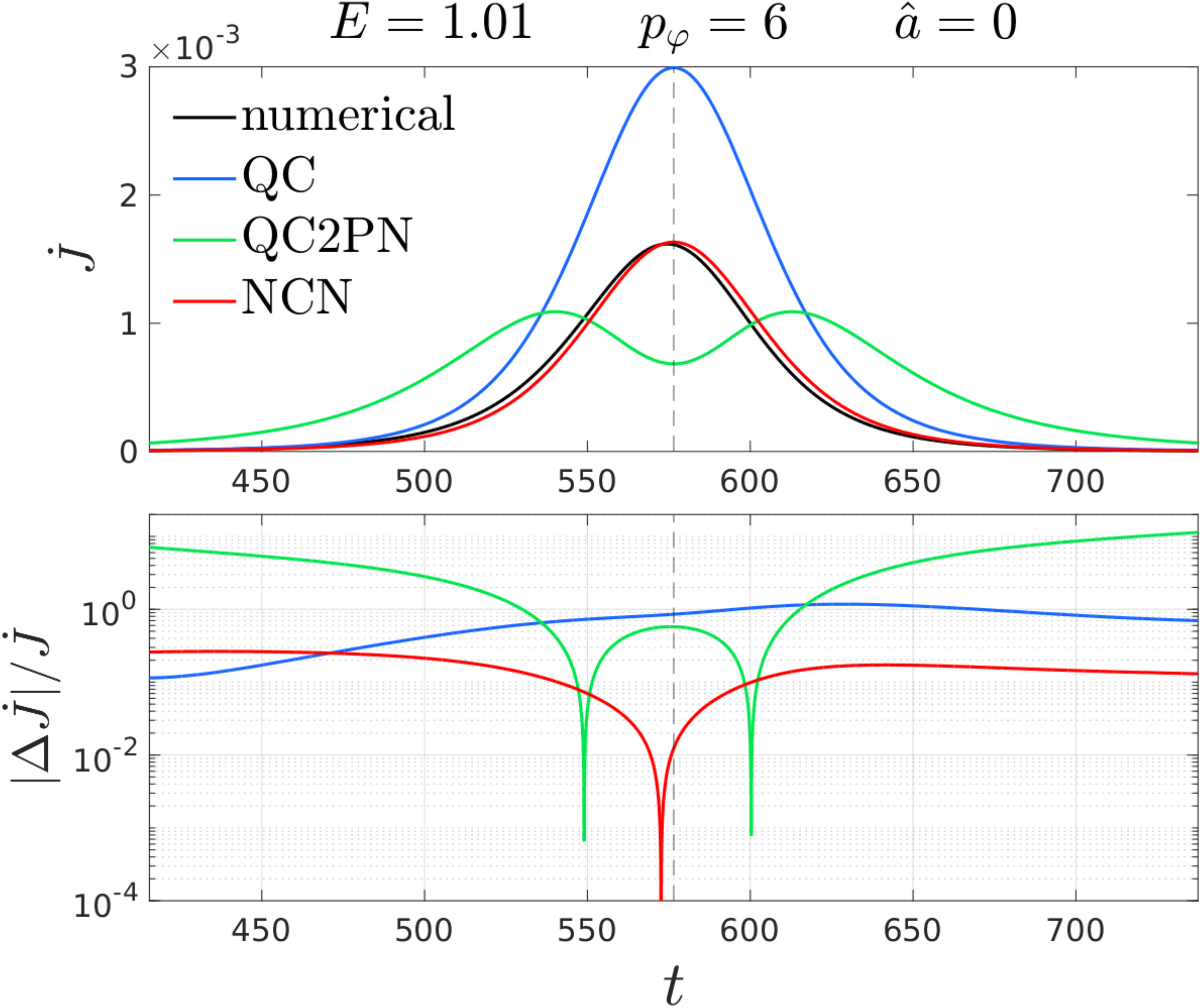}
  \hspace{0.2cm}
  \includegraphics[width=0.32\textwidth]{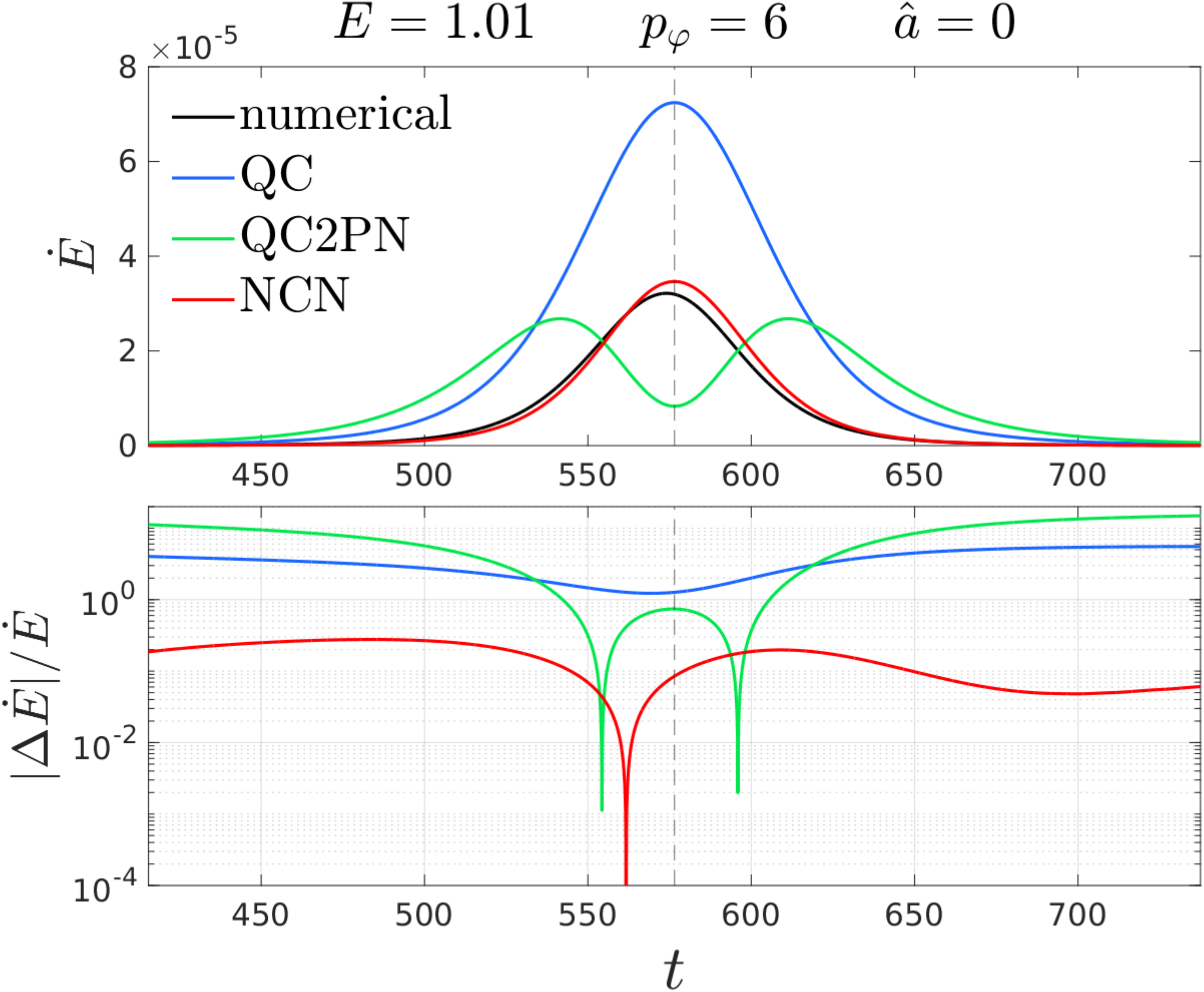}\\
  \caption{\label{fig:fluxes_nospin_hyp}
  Hyperbolic geodesics in Schwarzschild spacetime. 
  In the trajectories we highlight in blue the part that corresponds to
  $r\leq 40$; the fluxes are computed on this interval.
  We contrast the numerical fluxes (black)
  with the three analytical fluxes considered in this work: the \NCN{} flux (red) 
  computed using Eqs.~\eqref{eq:FphiTEOB} and~\eqref{eq:FrTEOB};
  the \QC{} flux (blue) from Eq.~\eqref{eq:RRqc}, that 
  is a proxy of the {\SEOBNRe} fluxes, and the \QC2PN{} flux with 2PN noncircular 
  corrections (green) from Eqs.~\eqref{eq:FKhalil}. 
  Each subpanel also reports the analytical/numerical relative difference.
  For each analytical flux we also show the relative difference with the numerical result.
  The dashed vertical line marks the closest passage to the central black hole.}
\end{figure*} 
We now turn our attention to unbound orbits. We consider
three hyperbolic geodesic scatterings in Schwarzschild spacetime whose 
orbits are shown in Fig.~\ref{fig:fluxes_nospin_hyp} together with the corresponding 
analytical/numerical angular momentum and energy fluxes.
In all the three cases considered, the initial separation is $r=120$, 
but we show only the time interval that corresponds to $r\leq 40$. 
As already discussed in Sec.~\ref{sec:EOB}, our definitions of eccentricity
and semilatus rectum are not valid for unbound motion; thus, we use the energy $E$ and
the angular momentum $p_\varphi$, which are constants of motion, to characterize our orbits. 
The selected values are also reported in Fig.~\ref{fig:fluxes_nospin_hyp} for each case.

The first configuration considered exhibits a strong zoom-whirl behavior. 
While during the circular whirl at periastron all the three 
prescriptions provide similar results, the differences between them are clearly visible
in the zoom part. Here, the less reliable prescription is the one using the \QC2PN{}
flux, as it gives differences well beyond the $100\%$ either before and
after the whirl phase. This issue is even more evident when inspecting 
the other two configurations, which have larger energies and larger angular momenta. 
In particular, for $(E,p_\varphi)=(1.01,6)$ the \QC2PN{} fluxes even develop multiple peaks.
The \QC{} fluxes are more solid, but the relative differences 
with the numerical results are still very large and they always overestimate 
the numerical results at periastron, especially for the energy fluxes. 
Finally, the fluxes computed using the \NCN{} choice are, once again, 
the most reliable and robust, showing smaller quantitative discrepancies 
with the numerical results for all the three configurations considered. 

%========================
\section{comparable-mass case}
%========================
\label{sec:comp_mass}
%==================
% 1363: effect on phasing
%==================
\begin{figure}[t]
  \center
  \includegraphics[width=0.5\textwidth]{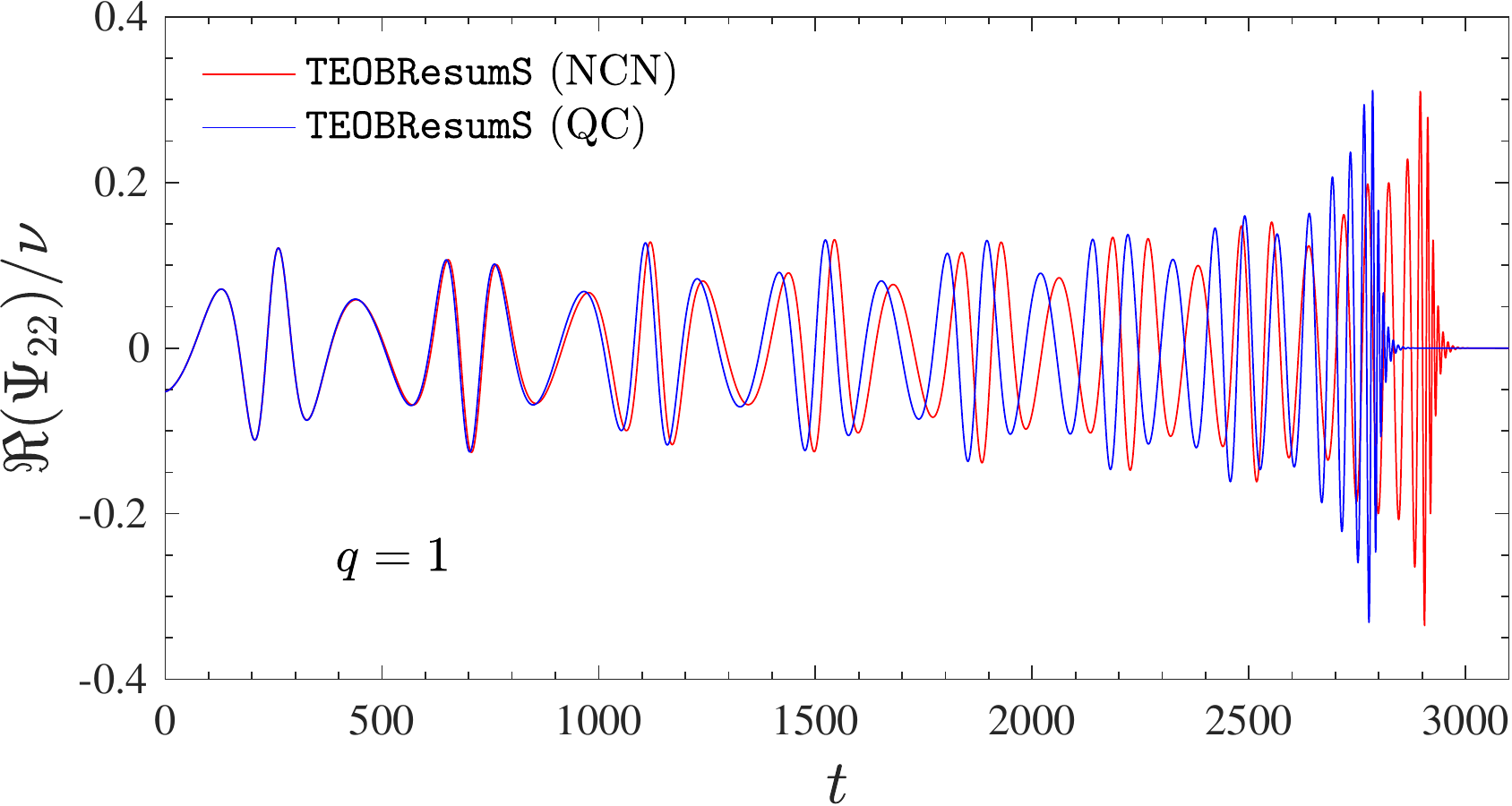}
  \caption{\label{fig:psi22_eob}Equal-mass case ($q=1$), initial EOB eccentricity $\sim 0.2$, 
  \TEOBResumS{}~\cite{Placidi:2021rkh} evolution: effect of using the circular
   $\hat{\cal F}_\varphi$ (\QC{}) instead of the full one (\NCN{}) with the Newtonian noncircular correction.
   Both dynamics are started with the same initial data. This BBH configuration corresponds to 
   the SXS:BBH:1363 configuration discussed in Table~IV of Ref.~\cite{Placidi:2021rkh}. 
   The qualitative behavior is analogous to the test-mass case shown in Fig.~\ref{fig:overview_wave}.
  } 
\end{figure} 

We investigate to which extent our findings in the test-mass case
carry over to comparable-mass binaries.
We concentrate on a single case study of an equal-mass, nonspinning
binary black hole (BBH) configuration with initial (nominal) EOB eccentricity $\sim 0.2$, and
use the SXS:BBH:1363 dataset of the SXS catalog~\cite{SXS:catalog}. 
This is configuration $9$ in Table~IV of Ref.~\cite{Placidi:2021rkh} used in both our
previous work and in Ref.~\cite{Ramos-Buades:2021adz}. For the details about the 
setup of initial data, we refer the reader to Refs.~\cite{Chiaramello:2020ehz,Nagar:2021gss,Nagar:2021xnh,Placidi:2021rkh}
%A more detailed comparison with numerical relativity simulations is
%left for the future, when more and more reliable waveforms will be available to us.

On the EOB side, we consider the latest development of the eccentric version of \TEOBResumS{},
as introduced  in Ref.~\cite{Nagar:2021xnh}, since it delivers the highest EOB/NR agreement (including
scattering angle measures).  The only change with respect to Ref.~\cite{Nagar:2021xnh} is that we adopt
the 2PN-accurate, factorized and resummed, waveform of Ref.~\cite{Placidi:2021rkh}. 
Using this baseline model, we employ two different radiation reactions:
(i) the complete (\NCN) version of $\hat{\cal F}_\varphi$, which
includes the leading-order noncircular correction in the $\ell=m=2$
mode and (ii) the quasicircular (\QC{}) $\hat{\cal F}_\varphi$ but keeping ${\cal F}_r\neq 0$. 
In both cases, the dynamics starts at the apastron\footnote{See Ref.~\cite{Ramos-Buades:2021adz}
for an improved initial data setup that, involving the eccentric anomaly, 
allows to start the dynamics at a different point of the orbit. } with the same
apastron frequency, $\omega_a^{\rm EOB}$, and EOB eccentricity at this frequency, 
$e_{\omega_a}^{\rm EOB}$, $(e_{\omega_a}^{\rm EOB},\omega_a^{\rm EOB})=(0.30479,0.01908)$.
Figure~\ref{fig:psi22_eob} compares the two waveforms obtained from the
two EOB dynamics. Analogously to the test-mass case of
Fig.~\ref{fig:overview_wave}, the \QC{} inspiral is shorter when
compared to the \NCN{} because of the larger amount of radiation emitted at 
each periastron passage. The previous results using the exact test-mass data 
suggest that the \QC{} acceleration is unphysical.
This is key to understand the EOB/NR comparison in the equal-mass case.

\begin{figure*}[t]
  \center
  \includegraphics[width=0.47\textwidth]{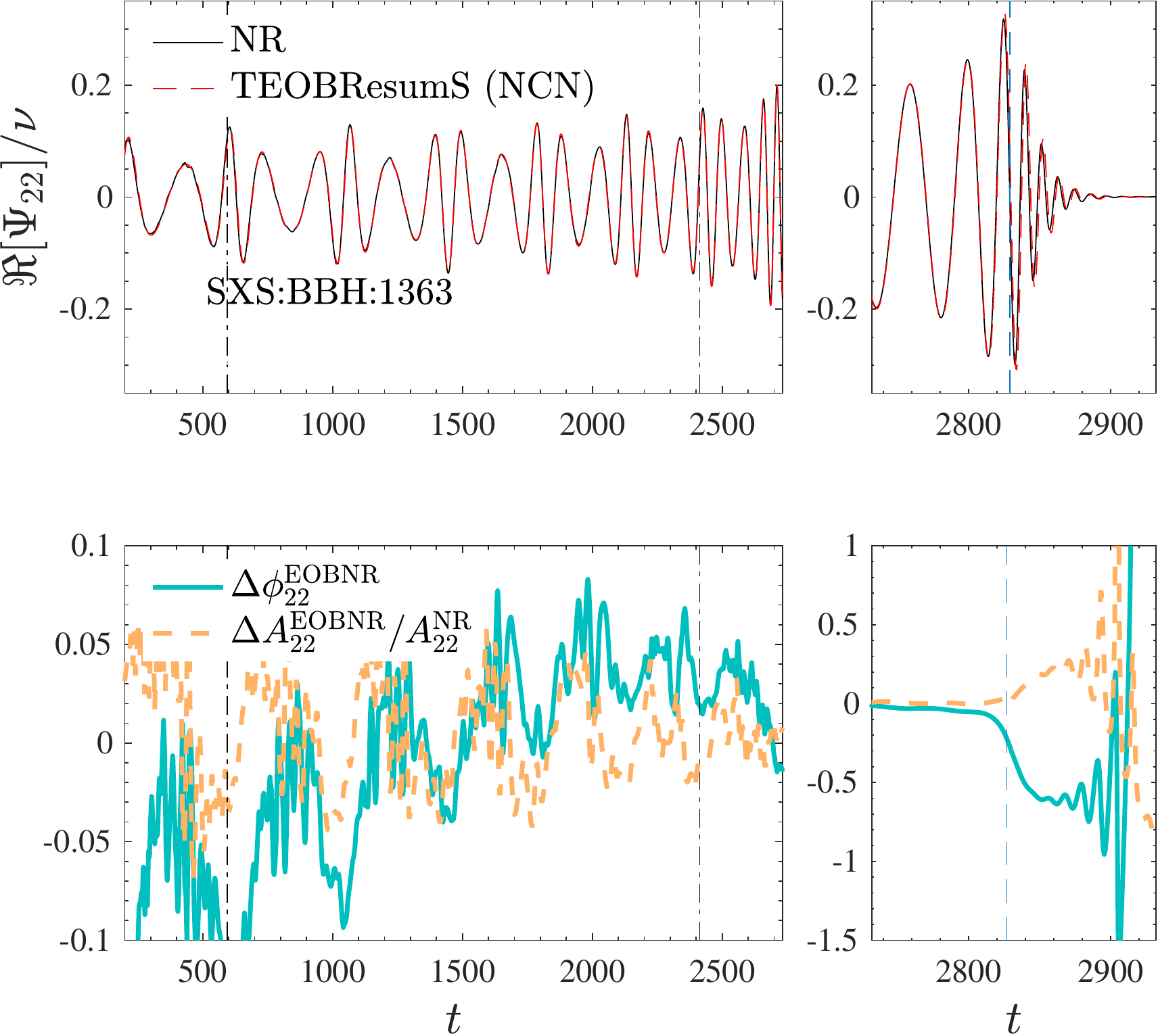}
  \hspace{5mm}
  \includegraphics[width=0.47\textwidth]{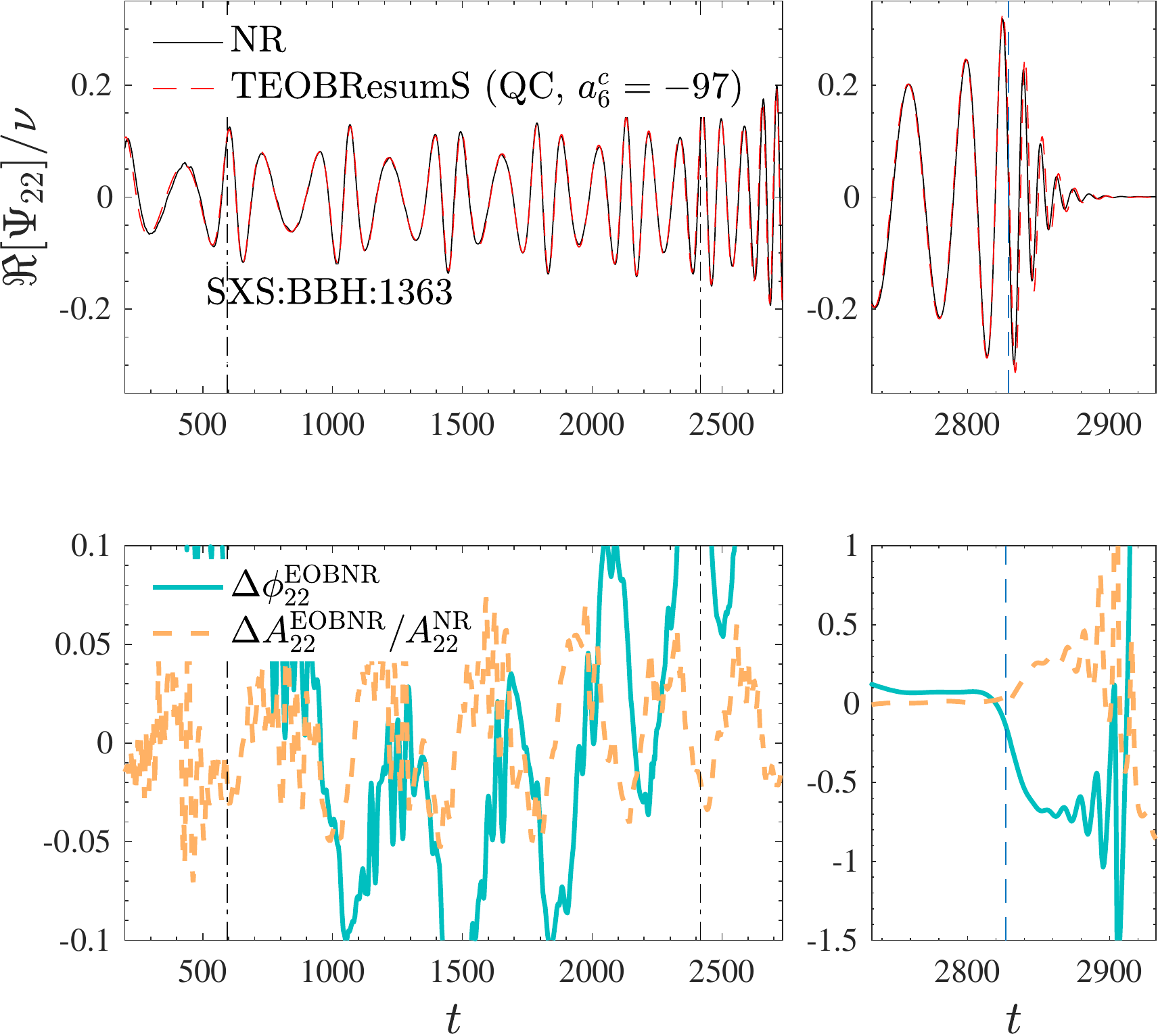}
  \caption{\label{fig:eobnr}EOB/NR phasing comparison for
    SXS:BBH:1363.
    Left: \TEOBResumS{} waveform model of
    Refs.~\cite{Nagar:2021xnh,Placidi:2021rkh} using the complete
    \NCN{} expression for $\hat{\cal{F}}_\varphi$ with initial
    parameters $(e_{\omega_a}^{\rm EOB},\omega_a^{\rm
      EOB})=(0.30479,0.01908)$ [see Fig.~12
      of~\cite{Placidi:2021rkh}].
    Right: Modified \TEOBResumS{} with the {\it quasicircular} (\QC{}) 
    $\hat{\cal F}_\varphi$, without leading noncircular terms, and
    $a_6^c=-97$. The initial parameters are  $(e_{\omega_a}^{\rm
      EOB},\omega_a^{\rm EOB})=(0.315,0.018239)$.
    The vertical dot-dashed lines indicate the alignment region.
    On such a
    short-length waveform the absence of noncircular corrections in
    $\hat{\cal{F}}_\varphi$ can be (partially) reabsorbed in the initial
    conditions. However, the stronger orbital circularization yielded 
    by the \QC{} model clearly shows up in the last three orbits.
    } 
\end{figure*} 
%==================
% EOB/NR unfaithfulness
%==================
\begin{figure}[t]
  \center
  \includegraphics[width=0.42\textwidth]{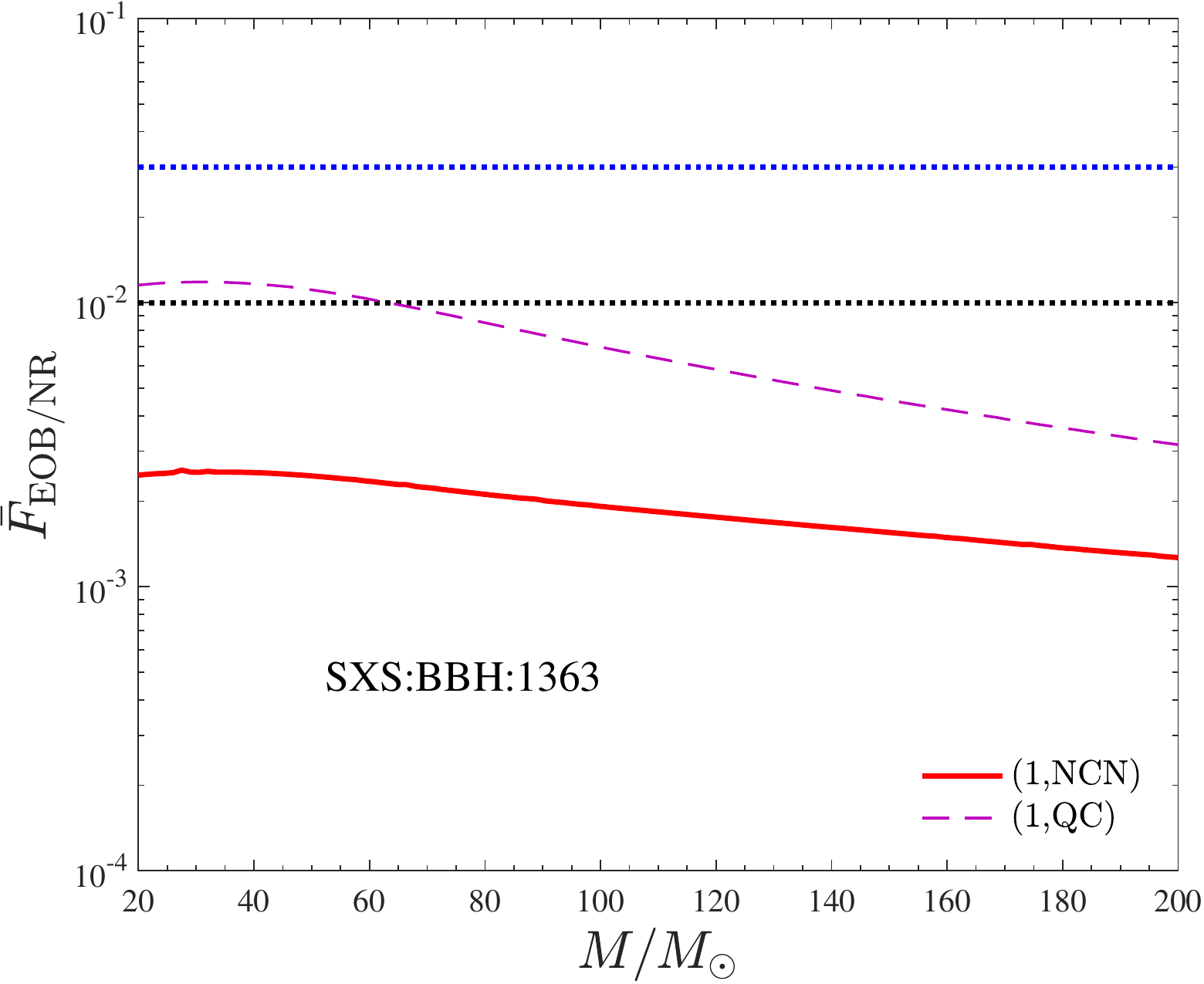}
   \caption{\label{fig:barF}EOB/NR unfaithfulness computation with the Advanced LIGO noise using the standard \NCN{} \TEOBResumS{} model
   of Ref.~\cite{Placidi:2021rkh} or the \QC{} version discussed here. The faster circularization of the \QC{} model results 
   in $\bar{F}_{\rm EOB/NR}$ slightly above the $1\%$ threshold (see Fig.~\ref{fig:omg_comp}).} 
\end{figure} 
%=========

An EOB/NR phasing analysis for SXS:BBH:1363 with the {\TEOBResumS{}} \NCN{} 
radiation reaction was recently considered in Ref.~\cite{Placidi:2021rkh} (see
Fig.~12 there). This result is reported here in the left panel of
Fig.~\ref{fig:eobnr} to ease the reader: the agreement is excellent with
$\sim -0.22$~rad accumulated up to merger, a phase difference within the 
NR uncertainty ($\delta\phi^{\rm NR}_{\rm mrg}=0.58$~rad, see Table~IV of~\cite{Placidi:2021rkh}).
The corresponding EOB/NR maximum unfaithfulness using Advanced LIGO
noise was $\bar{F}^{\rm max}_{\rm EOB/NR}=0.25\%$ (see Ref.~\cite{Placidi:2021rkh} and Fig.~\ref{fig:barF} below).
The left panel of Fig.~\ref{fig:eobnr} highlights that not only is the EOB/NR phase difference rather small 
(always between $[-0.1,0.05]$~rad), but it also {\it decreases with time}, as one is 
{\it a priori}
expecting from two orbital dynamics that progressively circularize in the same way.
An EOB/NR phasing for SXS:BBH:1363 with the \QC{} radiation reaction
was also recently reported in \cite{Ramos-Buades:2021adz}, using the
$\hat{\cal F}_\varphi$ implementation of \SEOBNRe{}.
Although no phase differences are explicitly shown, 
the authors report EOB/NR unfaithfulness ${\sim} 0.1\%$.
The findings of Ref.~\cite{Ramos-Buades:2021adz} seem to suggest that
the \QC{} azimuthal force yields a faithful
representation of the eccentric inspiral dynamics and waveform, at
least for mild eccentricities, comparable to the \NCN{}. This would be
in stark contrast with the differences between \NCN{}-\QC{} and NR 
observed in Fig.~\ref{fig:psi22_eob} and discussed above.
We now use the flexibility of \TEOBResumS{} to demonstrate that 
the conclusion is incorrect and, at the same time, to highlight subtle
aspects in EOB/NR comparisons using short, eccentric waveforms.

As a first step, we construct an EOB \QC{} model consistent with the main
features of \SEOBNRe{}. This is done by: (i) switching off the noncircular
correction to $\hat{\cal F}_\varphi$ and
(ii) redetermining the effective 5PN parameter
$a_6^c$ using the \QC{} flux.~\footnote{%
  This step is necessary since the effective 5PN parameter
  $a_6^c$ 
  %was informed in Ref.~\cite{Nagar:2021xnh} to quasicircular NR simulations
  was determined in Ref.~\cite{Nagar:2021xnh} by comparison with quasicircular NR simulations
  with the full \NCN{} flux and will impact only the late plunge
  and merger part of the waveform.
}
We consider only the $q=1$ case,
and we  find that $a_6^c=-97$ (instead of $a_6^c=-93.0366$, Eq.~(41) of Ref.~\cite{Nagar:2021xnh})  
is a good choice for our purposes. 
In the second step, we proceed with EOB/NR phasing with this {\it flexed} \QC{} model.
Following Ref.~\cite{Nagar:2021xnh} we retune the initial
dynamical parameters $(e_{\omega_a}^{\rm EOB},\omega_a^{\rm EOB})$
to minimize as much as possible the EOB/NR
phase difference during the eccentric inspiral~\footnote{
  Differently from Ref.~\cite{Ramos-Buades:2021adz} we perform this
  step manually, by tuning the parameters iteratively, without any
  automatized procedure. Note also that Ref.~\cite{Ramos-Buades:2021adz} 
  does not look at the phase difference, but at the EOB/NR  unfaithfulness,
  which in our case is checked {\it a posteriori}. Our results might be improved 
  but are sufficient for our purposes.
}.
Using $(e_{\omega_a}^{\rm EOB},\omega_a^{\rm EOB})=(0.30479,0.01908)$,
we obtain an EOB/NR phase difference of ${\sim} -0.14$~rad accumulated
to merger, as shown in the right panel of Fig.~\ref{fig:eobnr}. 
Although less good than for the \NCN{} model, this is a close performance. 
In summary, this demonstrates that by suitably tuning the initial data parameters 
it is possible to obtain small phasing errors and faithful waveforms even with 
a significantly less accurate \QC{} radiation reaction prescription.
However, we crucially note that the phase difference is now {\it increasing with time}
in the last few orbits, with a clear quasi-linear trend starting around $t\sim 1500$. 
This is qualitatively very different from the \NCN{} model, although {\it globally} 
the performances of the two waveforms are approximately comparable.

\begin{figure}[t]
  \center
  \includegraphics[width=0.42\textwidth]{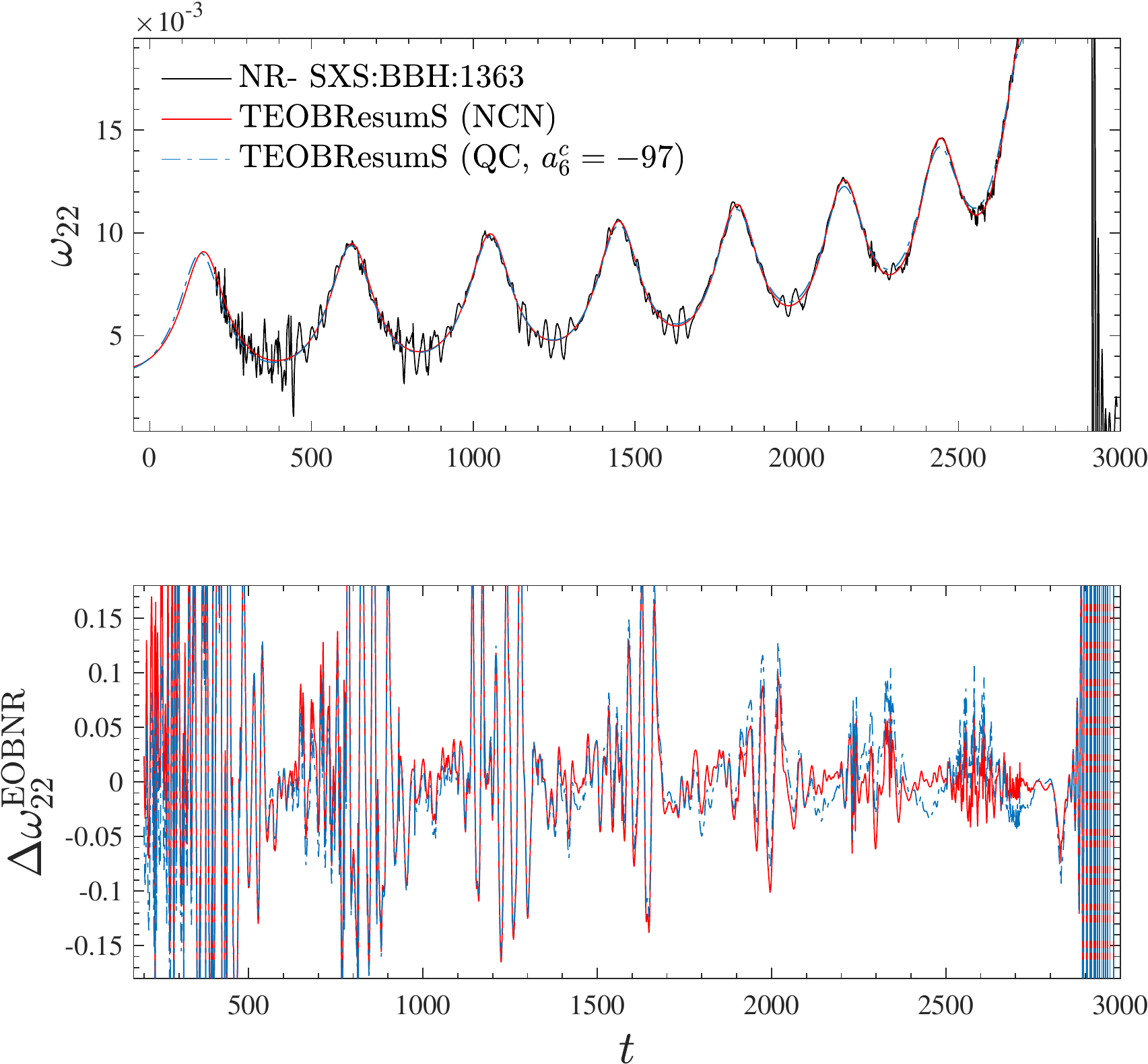}
   \caption{\label{fig:omg_comp}Time domain frequency evolution. Top panel: comparison between NR and the \NCN{} or \QC{} version of \TEOBResumS{}.
   Bottom panel: fractional frequency difference with NR. The picture confirms that the \NCN{} and \QC{} initial conditions are approximately coherent among 
   themselves and with the NR curve. The stronger circularization yielded by the \QC{} $\hat{\cal F}_\varphi$ entails a noticeable EOB/NR disagreement toward the late inspiral.} 
\end{figure}

Following Ref.~\cite{Nagar:2021xnh}, we calculate the EOB/NR
unfaithfulness with LIGO noise, and present it in
Fig.~\ref{fig:barF}. Although the choice of $(e_{\omega_a}^{\rm
  EOB},\omega_a^{\rm EOB})$ in the \QC{} model might be further
optimized, the \NCN{} model shows better matches than the \QC{}
model. It does not seem possible to fully cancel the difference
between \NCN{} and \QC{} as the latter is related to the faster
circularization of the orbit yielded by the \QC{} flux.
This fact is evident from Fig.~\ref{fig:omg_comp}, in which we compare the
time evolution of the instantaneous GW frequency of the two models.
Even if the \NCN{} and \QC{} choices for  $(e_{\omega_a}^{\rm EOB},\omega_a^{\rm EOB})$
give frequency oscillations that are very compatible among themselves and
with the NR during the first orbits (say up to $t\sim 1500$ in Fig.~\ref{fig:omg_comp})
the circularization of \QC{} is faster: the amplitude of the frequency oscillation decreases, 
with fractional difference with NR (bottom panel) that is more than twice larger than
the \NCN{} case. 

This analysis indicates that the \QC{} expression of
$\hat{\cal F}_\varphi$ can introduce large systematics and it is thus {\it not suited} 
to  construct faithful and robust EOB template waveforms for eccentric binaries.
Crucially, standard phasing and faithfulness diagnostic can fail
to capture these systematic errors, especially if a short NR waveform is
employed. 
The faster circularization of the orbit given by the \QC{} choice can
be easily overlooked in the minimization of the initial data
parameters thus leading to large phase errors on longer signals.
  
%==================
% Frequency comparison
%==================
%=========

%========================================================================
\section{2PN corrections in factorized and resummed form}
\label{sec:2PN_resum}
%========================================================================
Reference~\cite{Placidi:2021rkh} introduced high PN noncircular terms in
the Newtonian-factorized waveform as a multiplicative correction to the generic 
Newtonian prefactor.
After suitable resummations, this yielded an improved analytical/numerical waveform 
agreement with respect to the use of only the generic Newtonian prefactor, especially for the phase. 
The aim of this section is to
explore the performance of an analogous procedure applied to the flux. For simplicity, and since in
any case our $\hat{\cal F}_\varphi$ is noncircular flexed only in the $\ell=m=2$ mode, we limit here
our analysis only to this mode. A more complete investigation is postponed to future work.

\subsection{2PN-accurate quadrupolar flux}
In Ref.~\cite{Placidi:2021rkh} we recovered the full multipolar fluxes at 2PN accuracy in EOB coordinates
as a consistency check of our waveform calculation. In particular, starting from the instantaneous part of the waveform we could
explicitly recover the 2PN-accurate flux in EOB coordinates as obtained in Ref.~\cite{Bini:2012ji}.
For our current factorization purposes, we show here explicitly the $\ell=m=2$ flux multipole up to 2PN order,
including {\it both} the instantaneous and tail parts,
\begin{widetext}
	\begin{align}
	\label{eq:F2PN22}
		F^{\rm 2PN}_{22} & = p_{r_*}^2 p_\varphi u^3-2 p_\varphi^3 u^5-2 p_\varphi u^4+ \frac{1}{c^2}\Bigg\{  p_\varphi u^5 \Bigg[\frac{689}{42}-\frac{33 \nu}{7}+\left(\frac{109}{7}-\frac{61 \nu }{7}\right) p_\varphi^2 u+\left(\frac{17}{42}-\frac{12 \nu }{7}\right) p_\varphi^4 u^2\Bigg] \cr
		&- p_{r_*}^2 p_\varphi u^4\Bigg[ \frac{148}{21}-\frac{36 \nu }{7}+p_\varphi^2 u \left(\frac{7}{3}-\frac{21 \nu }{2}\right)\Bigg] -p_{r_*}^4 p_\varphi u^3 \left(\frac{145}{42}+\frac{9 \nu }{14}\right) \Bigg\} + \frac{\pi}{c^3} \Bigg[ u^5  \Bigg(\frac{9}{128}-\frac{173 p_\varphi^2 u}{10}\cr
		&+\frac{79309 p_\varphi^4 u^2}{5760}-\frac{889 p_\varphi^6 u^3}{32}+\frac{9157 p_\varphi^8 u^4}{384}-\frac{697 p_\varphi^{10} u^5}{72}+\frac{47 p_\varphi^{12} u^6}{640}+\frac{643 p_\varphi^{14} u^7}{480}-\frac{97 p_\varphi^{16} u^8}{288}\Bigg) \cr
		&+p_{r_*}^2 u^4 \Bigg(\frac{215}{192}+\frac{91109 p_\varphi^2 u}{5760}-\frac{811 p_\varphi^4 u^2}{24}+\frac{14659 p_\varphi^6 u^3}{192}-\frac{51487 p_\varphi^8 u^4}{576}+\frac{22975 p_\varphi^{10} u^5}{384}-\frac{3489 p_\varphi^{12} u^6}{160}\cr
		&+\frac{485 p_\varphi^{14} u^7}{144}\Bigg) -p_{r_*}^4 u^3 \Bigg( \frac{1219}{1152}-\frac{1471 p_\varphi^2 u}{576}+\frac{449 p_\varphi^4 u^2}{128}-\frac{3677 p_\varphi^6 u^3}{576}+\frac{769 p_\varphi^8 u^4}{144}-\frac{49 p_\varphi^{10} u^5}{32} \Bigg)   \Bigg]\cr
		&- \frac{1}{c^4} \Bigg\{ p_\varphi u^6 \Bigg[\frac{8852}{189}-\frac{12335 \nu }{378}+\frac{268 \nu ^2}{27}+p_\varphi^2 u \left(\frac{21001}{882}-\frac{52147 \nu }{588}+\frac{7661 \nu ^2}{294}\right)+p_\varphi^4 u^2 \Bigg(\frac{59}{882}-\frac{60901 \nu }{1764} \cr
		&+\frac{7663 \nu ^2}{1764}\Bigg)+p_\varphi^6 u^3 \left(\frac{851}{882}-\frac{4987 \nu }{882}-\frac{3295 \nu ^2}{1764}\right)\Bigg] - p_{r_*}^2 p_\varphi u^5 \Bigg[ \frac{10277}{882}-\frac{60065 \nu }{1764}+\frac{7799 \nu ^2}{441}\cr
		&+p_\varphi^2 u \left(-\frac{6465}{392}-\frac{48385 \nu }{441}+\frac{22237 \nu ^2}{441}\right)+p_\varphi^4 u^2 \left(\frac{709}{84}-\frac{685 \nu }{24}+\frac{463 \nu ^2}{168}\right) \Bigg]- p_{r_*}^4 p_\varphi u^4 \Bigg[ \frac{6527}{441}+\frac{6809 \nu }{1764}\cr
		&-\frac{14921 \nu ^2}{1764}+p_\varphi^2 u \left(\frac{2923}{588}-\frac{4889 \nu }{588}-\frac{6317 \nu ^2}{294}\right) \Bigg] \Bigg\},
	\end{align}
\end{widetext}
where for the sake of simplicity we considered an expansion around 0 of $p_{r_*}$ up to order $\mathcal{O}(p_{r_*}^4)$.
For completeness, Appendix \ref{app:2PNfluxes} also reports the other flux multipoles relevant at 2PN order.
\subsection{Factorization and resummation}
\label{sec:fac_resum}
Our aim here is to add 2PN-accurate noncircular corrections to the flux contribution of the 
mode $\ell=m=2$. 
This is achieved by dressing the first term of Eq.~\eqref{eq:hatfnc22}
with an additional correcting factor  $\hat{F}^{\rm 2PN_{nc}}_{\varphi,22}$ derived from 
the full noncircular 2PN flux $\hat{F}^{\rm 2PN}_{22}$ written in Eq.~\eqref{eq:F2PN22}.
Following the same reasoning implemented for the waveform in Ref.~\cite{Placidi:2021rkh}, the straightforward procedure we use is the following:  
\begin{itemize}
\item[(i)] starting from the Taylor expanded flux $F^{\rm 2PN}_{22}$, we factorize the full Newtonian contribution $F^N_{22 }\NP22$, and we use in the latter  the 2PN-accurate EOB equations of motion to replace the time derivatives and expand the residual up to $\mathcal{O}(1/c^4)$,
\item[(ii)] we single out the circular part $\hat{F}^{\rm 2PN_{c}}_{22}$ of the Newton-factorized flux by simply taking the limit $p_{r_{*}}\to 0$,
\item[(iii)] we factorize the circular part computed in the previous step and compute the desired noncircular correction $\hat{F}^{\rm 2PN_{nc}}_{22}$.
\end{itemize} 
In formulas we have
\begin{align}
	\label{eq:hatF2PNc}
	\hat{F}^{\rm 2PN_{c}}_{22}  \equiv  \lim_{p_{r_*} \to 0} \, T_{\rm 2PN} \Bigg[\dfrac{F^{\rm 2PN}_{22}}{    \left( F^N_{22 }\NP22 \right)_{\text{EOMs}}} \Bigg] ,\\
	\label{eq:hatF2PNnc}
	\hat{F}^{\rm 2PN_{nc}}_{22}  \equiv  T_{\rm 2PN} \Bigg[\dfrac{F^{\rm 2PN}_{22}}{    \left( F^N_{22 }\NP22 \right)_{\text{EOMs}}	\hat{F}^{\rm 2PN_{c}}_{22}} \Bigg] ,
\end{align}
where the operator $T_{\rm 2PN}$ performs a Taylor expansion up to the 2PN order and the subscript ``EOMs" manifests the replacement of the time derivatives in the Newtonian flux with the corresponding EOB equations of motion. The resulting noncircular factor \eqref{eq:hatF2PNnc} comes out naturally split in an instantaneous and a tail part which appear at different PN orders. For this reason one can readily factorize it further in an instantaneous and a tail factor,
\begin{equation}
\label{eq:Fphi2PNnc}
	\hat{F}^{\rm 2PN_{nc}}_{22}  = \hat{F}^{\rm 2PN_{nc, inst}}_{22}  \hat{F}^{\rm 2PN_{nc, tail}}_{22}, 
\end{equation}
which explicitly read
%\begin{widetext}
	\begin{align}
		 &\hat{F}^{\rm 2PN_{nc, inst}}_{22}  = 1 +\frac{1}{c^2}\Bigg[\frac{p_{r_*}^2}{\big(1+p_\varphi^2 u\big)^2} \Bigg( \frac{281}{168} + \frac{31 \nu }{28}\Bigg) \hat{f}^{\rm 1PN}_{p_{r_*}^2} \nonumber\\
		 & + \frac{p_{r_*}^4}{u \big(1+p_\varphi^2 u\big)^3} \Bigg(\frac{5-6 \nu }{16 }\Bigg) 
		\hat{f}^{\rm 1PN}_{p_{r_*}^4}  \Bigg] \\
		&+\frac{1}{c^4} \Bigg[\frac{p_{r_*}^2 u}{\big(1+p_\varphi^2 u\big)^3} \Bigg( \frac{159697 }{42336}-\frac{2081 \nu  }{10584}+ \frac{20345 \nu ^2 }{10584}\Bigg)
        \hat{f}^{\rm 2PN}_{p_{r_*}^2} \cr
		&+\frac{p_{r_*}^4}{p_\varphi \big(1+p_\varphi^2 u\big)^4} 
		\Bigg(\frac{225067}{84672}+\frac{18119 \nu }{10584}
		+\frac{6893 \nu ^2}{21168}\Bigg) \hat{f}^{\rm 2PN}_{p_{r_*}^4}  \Bigg],\nonumber\\
		&\hat{F}^{\rm 2PN_{nc, tail}}_{22}  = 1+\frac{\pi}{c^3}\Bigg[ -\frac{p_{r_*}^2}{p_\varphi \big(1+p_\varphi^2 u\big)^2}\frac{887}{1536} \hat{t}^{\rm 1.5PN}_{p_{r_*}^2}\nonumber\\
		&+\frac{p_{r_*}^4}{p_\varphi u \big(1+p_\varphi^2 u\big)^3}\frac{2215}{9216}
		\hat{t}^{\rm 1.5PN}_{p_{r_*}^2} \Bigg].
	\end{align}
%\end{widetext}
%
\begin{figure*}[t]
  \center
  \includegraphics[width=0.26\textwidth]{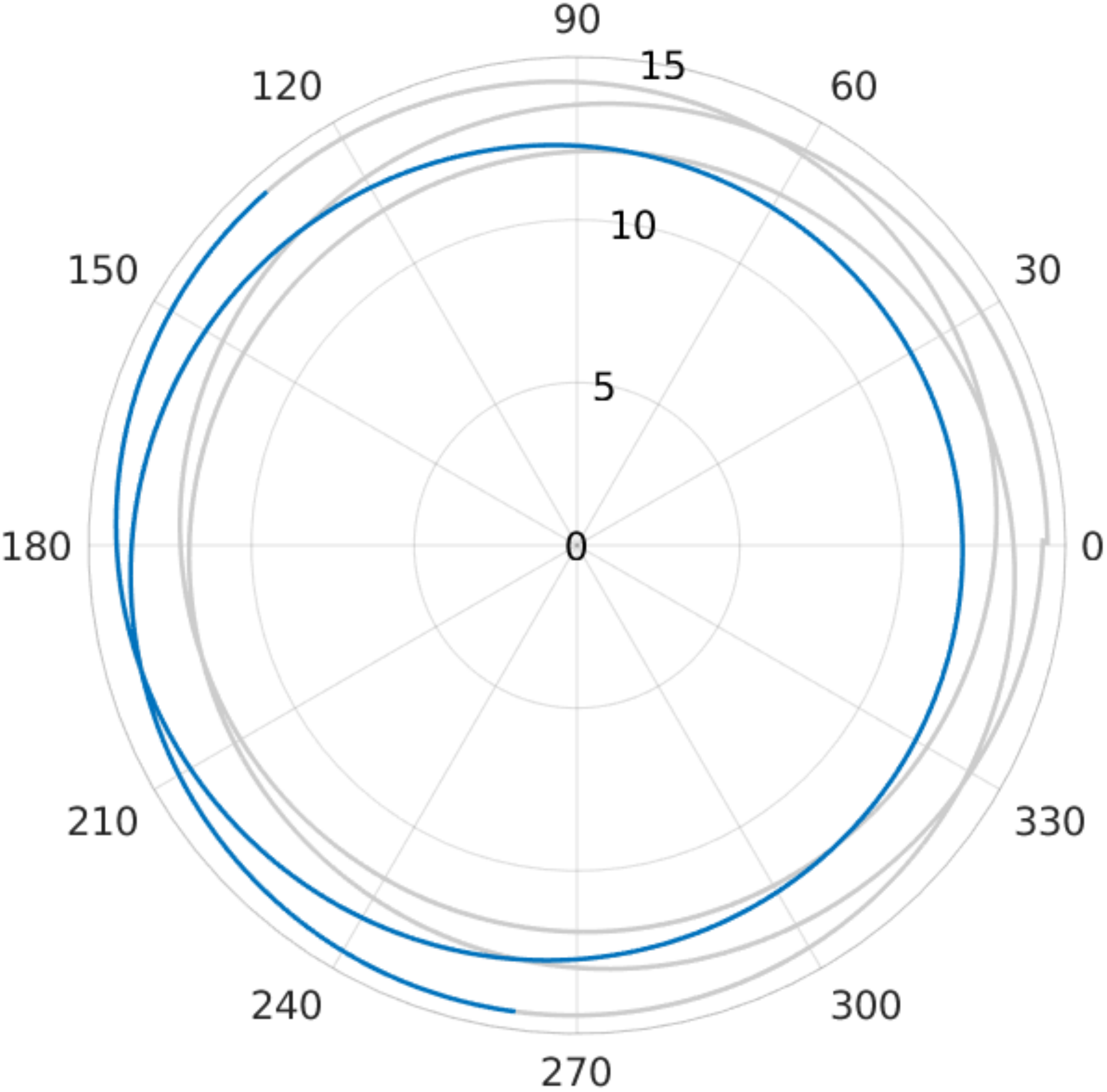}
  \hspace{0.2cm}
  \includegraphics[width=0.32\textwidth]{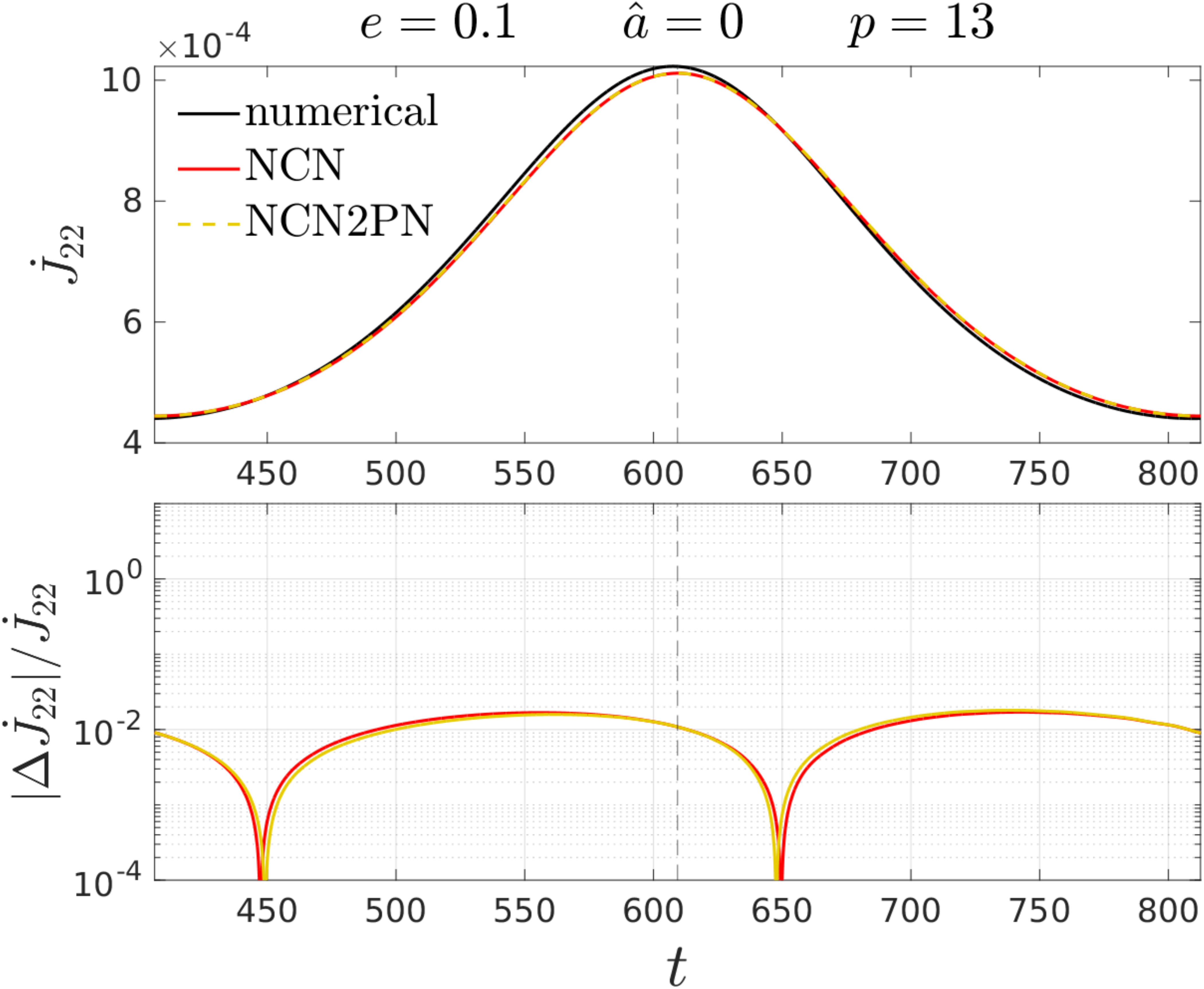}
  \hspace{0.2cm}
  \includegraphics[width=0.32\textwidth]{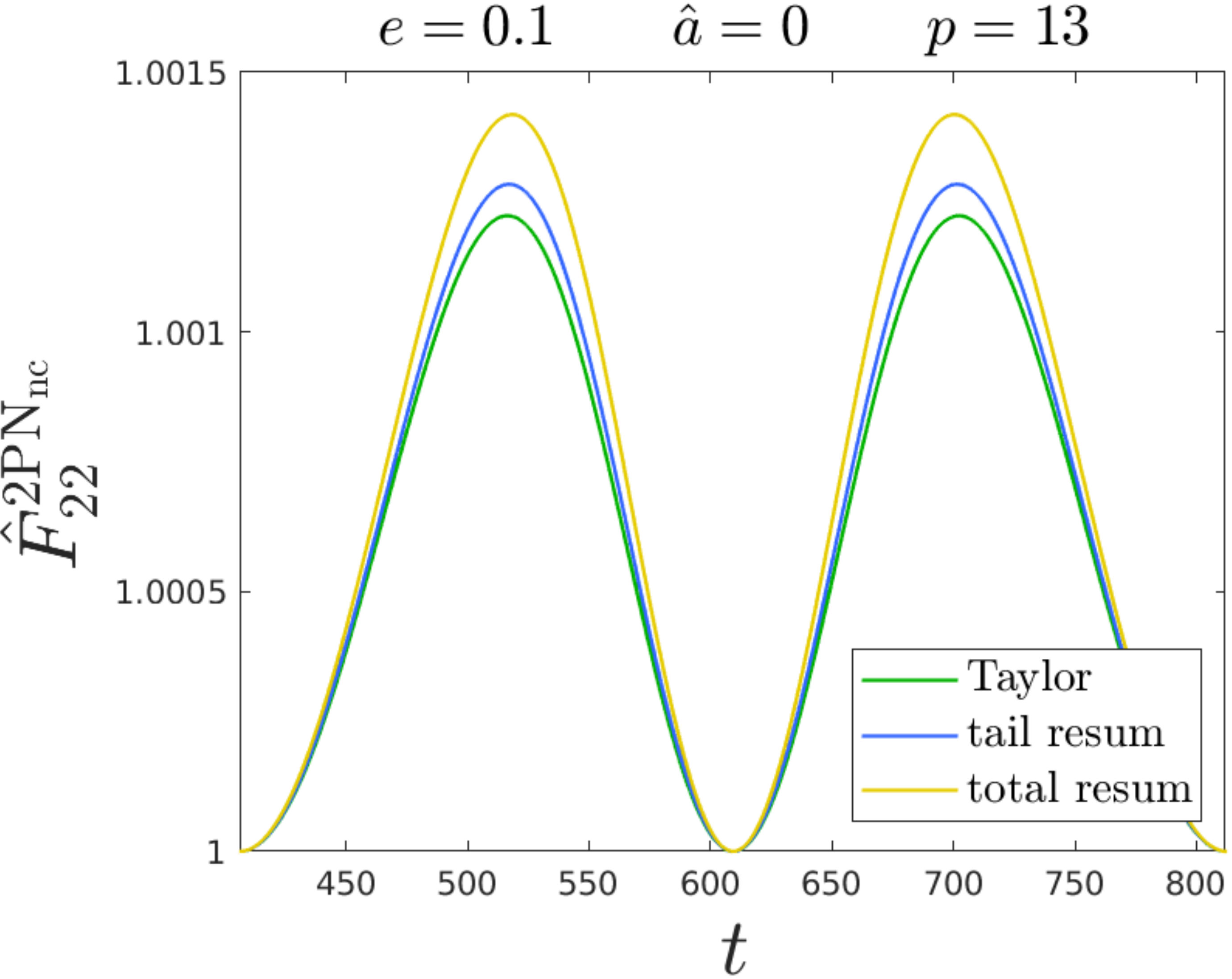}\\ 
  \vspace{0.5cm}
  \includegraphics[width=0.26\textwidth]{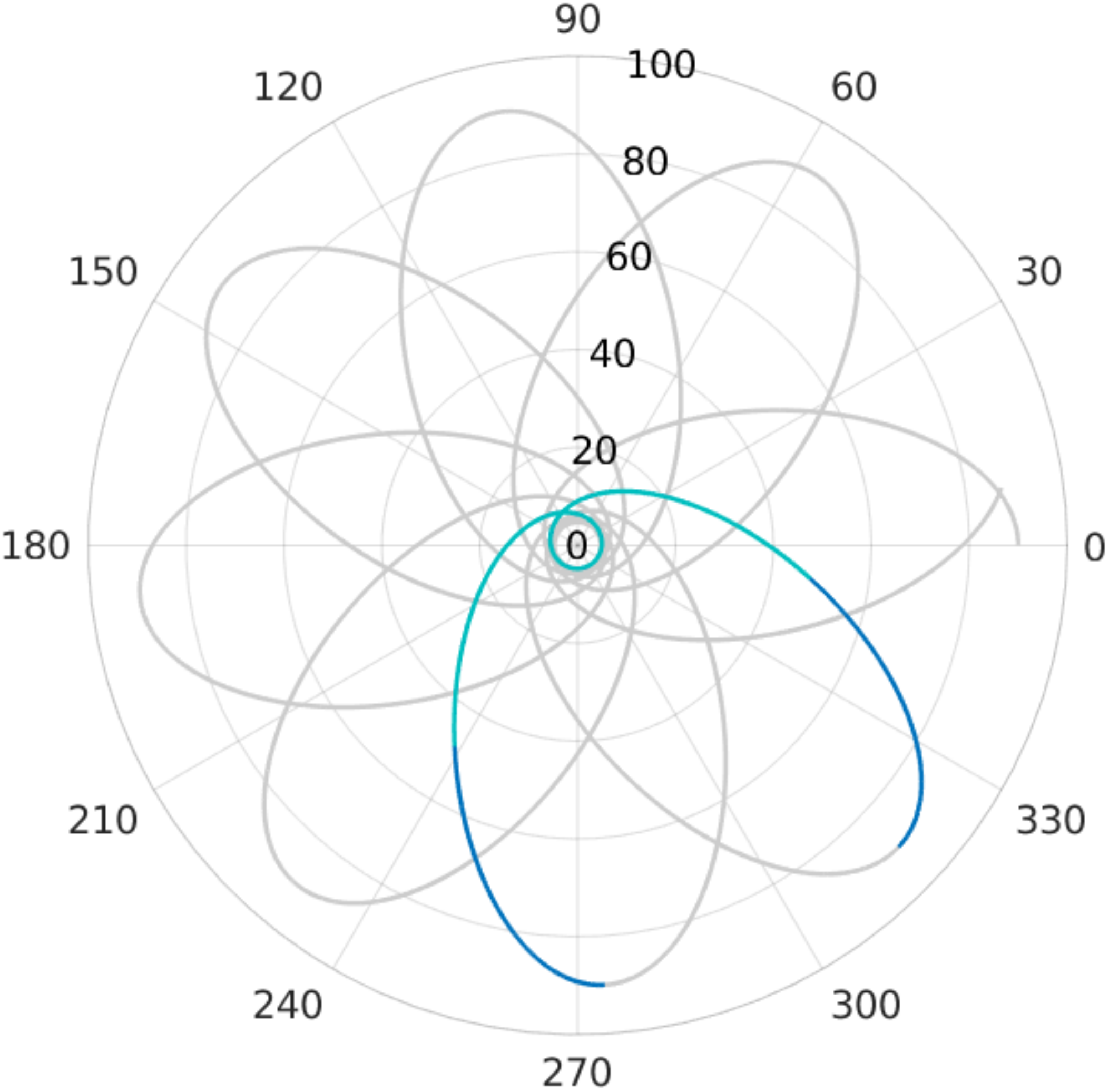}
  \hspace{0.2cm}
  \includegraphics[width=0.32\textwidth]{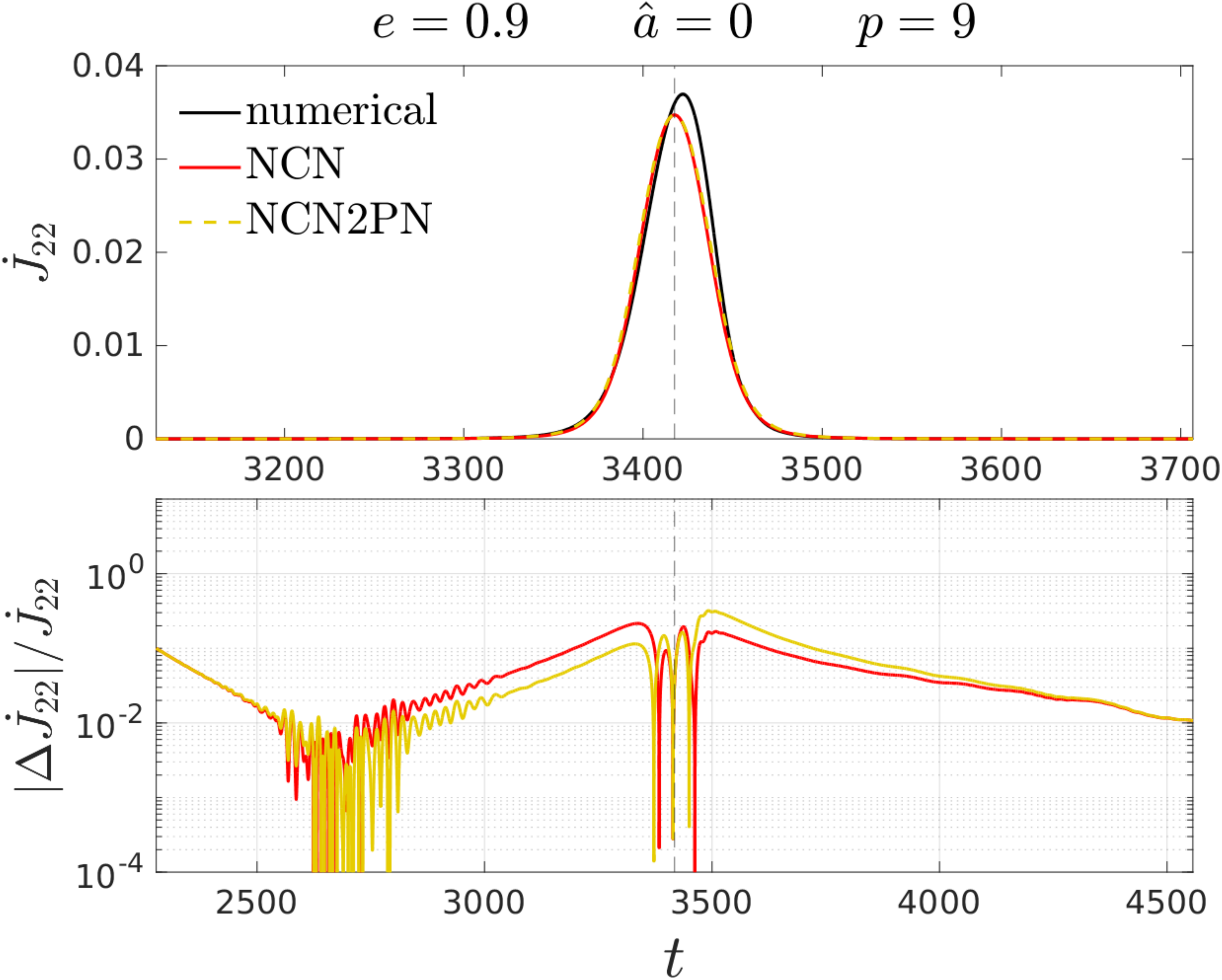}
  \hspace{0.2cm}
  \includegraphics[width=0.32\textwidth]{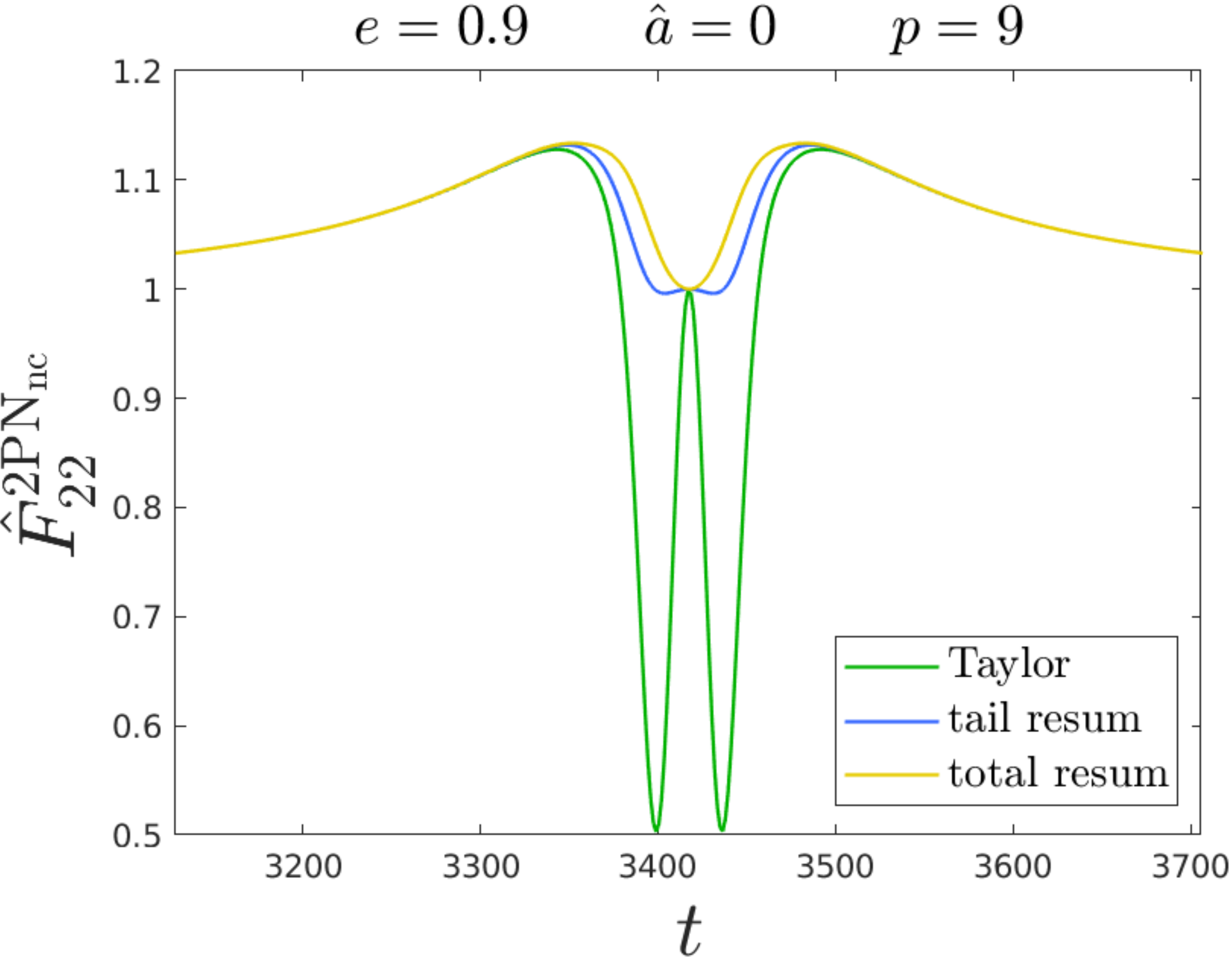}
  \caption{\label{fig:flux_2PN} Trajectories, quadrupolar angular momentum fluxes, 
  and 2PN corrections to the
  flux as discussed in Sec~\ref{sec:2PN_resum}. We show the standard flux from {\TEOBResumS} 
  (red), the result with 2PN resummed corrections (yellow), and the numerical flux (black).
  The corresponding analytical/numerical relative differences reported in the bottom panel
  are shown over the radial period; the corresponding part of the trajectories is 
  highlighted in blue. On the rightmost panels we show the 2PN corrections to the flux in 
  Taylor expanded form, with resummation of the tail factor and with resummation of both the 
  instantaneous and hereditary parts. The latter is used for the 2PN flux of the middle panels.
  For the second configuration with $e=0.9$, the fluxes and the 2PN corrections
  are shown over a time interval that is shorter than the radial period in order to
  highlight the burst of radiation at periastron; the corresponding
  part of the trajectories is highlighted in aqua-green.} 
\end{figure*} 
Again, in the results reported above we expand in $p_{r_*}$ up to order $\mathcal{O}(p_{r_*}^4)$.
The terms $\hat{f}^{\rm PN}_{p_{r_*}^n}$ and $\hat{t}^{\rm 1.5 PN}_{p_{r_*}^n}$ 
are polynomials in the Newtonian-order variable $y=p_\varphi^2 u$. 
For the instantaneous part the coefficients of the polynomials
contain also the symmetric-mass ratio $\nu$ and read
\begin{align}
\hat{f}^{\rm 1PN}_{p_{r_*}^2} = 1+& \frac{2(571-54 \nu) }{281+186 \nu }y+\frac{ 1061-390 \nu }{281+186 \nu }y^2, \\
\hat{f}^{\rm 1PN}_{p_{r_*}^4} = 1+& \frac{ 2(395-366 \nu) }{21 (5-6 \nu )}y+\frac{ 295-234 \nu }{7 (5-6 \nu )}y^2, \\
\hat{f}^{\rm 2PN}_{p_{r_*}^2} = 1+&\frac{2  \left(503861-236326 \nu -42992 \nu ^2\right)}{159697-8324 \nu +81380 \nu ^2}y \cr
+& \frac{6  \left(144635-100862 \nu -59260 \nu ^2\right)}{159697-8324 \nu +81380 \nu ^2}y^2 \cr 
-&\frac{6  \left(119807-50090 \nu +28256 \nu ^2\right)}{159697-8324 \nu +81380 \nu ^2}y^3 \cr
-&\frac{21  \left(26487-19592 \nu +428 \nu ^2\right)}{159697-8324 \nu +81380 \nu ^2}y^4, \\
\hat{f}^{\rm 2PN}_{p_{r_*}^4} = 1 +& \frac{2  \left(1028891+66902 \nu -13616 \nu ^2\right)}{225067+144952 \nu +27572 \nu ^2}y \cr 
+& \frac{6  \left(670405+13630 \nu +8948 \nu ^2\right)}{225067+144952 \nu +27572 \nu ^2}y^2\cr
+& \frac{6  \left(548563+5454 \nu +42912 \nu ^2\right)}{225067+144952 \nu +27572 \nu ^2}y^3\cr
+& \frac{3  \left(478421-54244 \nu +40444 \nu ^2\right)}{225067+144952 \nu +27572 \nu ^2}y^4,
\end{align}
while in the tail factor there are no $\nu$-contributions and the two polynomials are
\begin{align}
\hat{t}^{\rm 1.5PN}_{p_{r_*}^2} = 1&+\frac{19094}{2661}y-\frac{127753}{13305}y^2+\frac{22016}{887}y^3\cr
&-\frac{2569}{2661}y^4-\frac{79250}{2661}y^5+\frac{29231}{887}y^6\cr 
& -\frac{204692}{13305}y^7+\frac{7372}{2661}y^8, \\
\hat{t}^{\rm 1.5PN}_{p_{r_*}^4} =1&-\frac{4222}{443}y+\frac{115273}{11075}y^2-\frac{74904}{2215}y^3
\cr 
&-\frac{15491}{2215}y^4+\frac{91994}{2215}y^5-\frac{77197}{2215}y^6\cr
&+\frac{169412}{11075}y^7 -\frac{7372}{2215}y^8.
\end{align}
Note that the analytical structure of the 2PN correction $\hat{F}^{\rm 2PN_{nc}}_{22}$
is similar to the one of the 2PN corrections to the waveform multipoles
discussed in Ref.~\cite{Placidi:2021rkh}. As already argued for the
waveform, the polynomials in the $y$ variable need to be resummed in order to
provide reliable results in strong field regimes.
We use diagonal Pad\'e approximants for the polynomials in 
the tail and in the 2PN instantaneous part, while we leave 
in Taylor-expanded form the polynomials in the 1PN instantaneous contribution.

\subsection{Testing the 2PN noncircular correction in Newtonian-factorized angular radiation reaction}
\begin{table}[t]
   \caption{\label{tab:mean_differences_2PN} 
   Averaged analytical/numerical relative differences for averaged
   quadrupolar fluxes, 
   $\langle \delta F_{22}  \rangle = \langle |\Delta F_{22}|/F_{22} \rangle$.
   We consider the fluxes computed using NCN, NCN with 2PN corrections in Taylor
   expanded form, and NCN2PN that includes 2PN noncircular corrections in resummed form.
   For each flux we compute the averages over the simulations with same eccentricity, 
   the average for the nonspinning configurations, and the average for all the configurations.
   The values are reported in percentages. The relative difference averaged over
   the circular cases is 
   $\langle \delta J_{22} \rangle_c = \langle \delta E_{22} \rangle_c = 0.07\%$.
   The averages at fixed eccentricity are shown in Fig.~\ref{fig:mean_averaged_2PN}.}
\begin{center}
\begin{ruledtabular}
\begin{tabular}{c|cc|cc|cc}
   \multirow{2}{*}{$[\%]$} 
 & \multicolumn{2}{c|}{\multirow{2}{*}{NCN}}
 & \multicolumn{2}{c|}{\multirow{2}{*}{\shortstack[c]{NCN2PN \\  (Taylor)}}}
 & \multicolumn{2}{c}{\multirow{2}{*}{NCN2PN}}  \\
 & & & & &  \\
 \hline
 & $\langle \delta J_{22} \rangle $ & $\langle \delta E_{22} \rangle $ 
 & $\langle \delta J_{22} \rangle $ & $\langle \delta E_{22} \rangle $
 & $\langle \delta J_{22} \rangle $ & $\langle \delta E_{22} \rangle $ \\
\hline
$e = 0.1$ & 0.31 &  0.39 &  0.26 &  0.34 & 0.24 & 0.32 \\
$e = 0.3$ & 2.03 &  2.52 &  1.70 &  2.23 & 1.47 & 1.98 \\
$e = 0.5$ & 4.70 &  5.41 &  4.24 &  5.17 & 3.40 & 4.26 \\
$e = 0.7$ & 7.41 &  8.10 &  8.75 & 10.05 & 5.45 & 6.49 \\
$e = 0.9$ & 9.66 & 10.36 & 24.36 & 26.91 & 7.37 & 8.44 \\
\hline
$\ha = 0$ & 3.01 &  3.21 &  4.85 &  5.53 & 1.59 & 1.98 \\
\hline
   all    & 4.03 &  4.48 &  6.56 &  7.46 & 3.00 & 3.59 \\
\end{tabular}
\end{ruledtabular}
\end{center}
\end{table}
\begin{figure}[t]
  \center
  \includegraphics[width=0.22\textwidth]{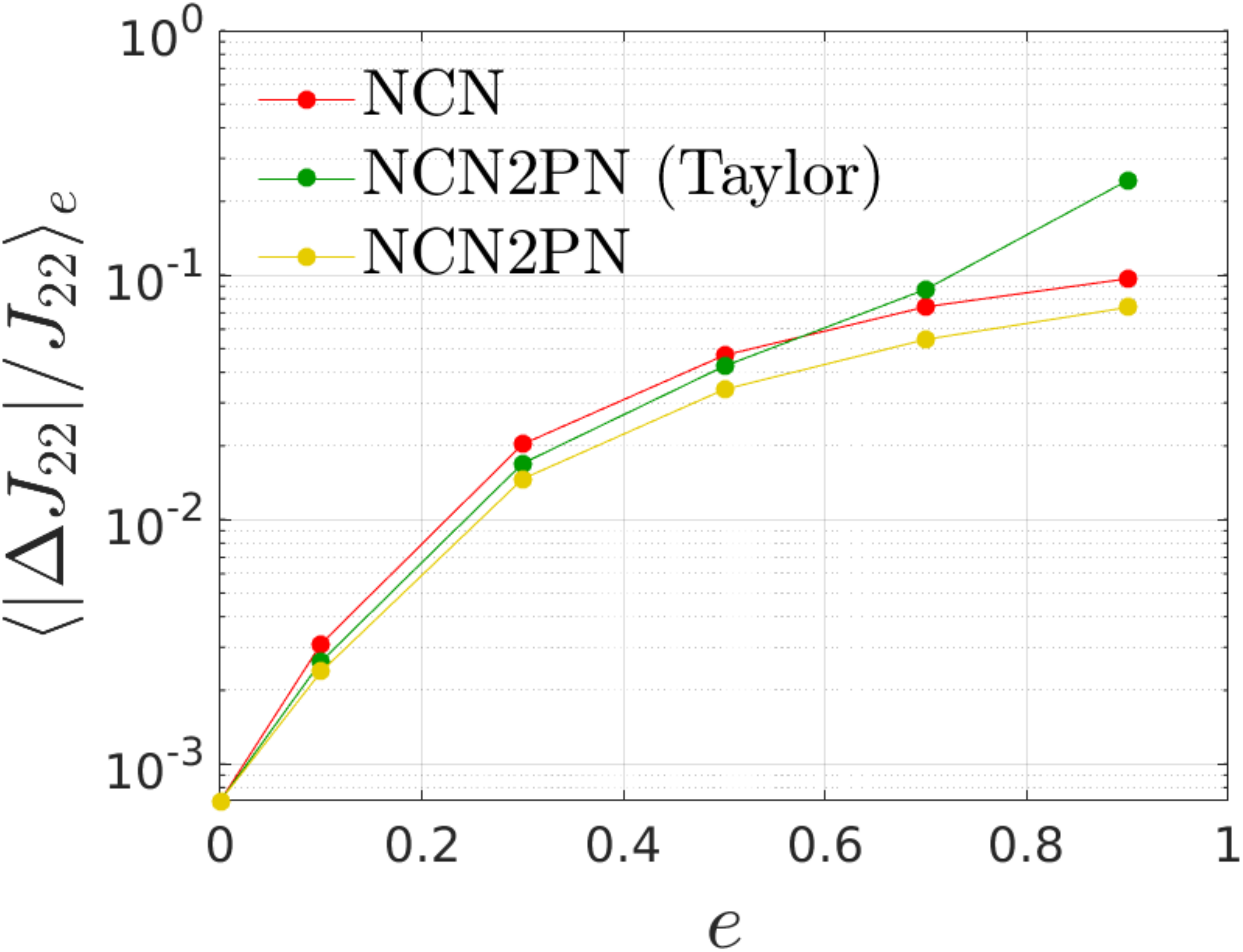}
  \includegraphics[width=0.22\textwidth]{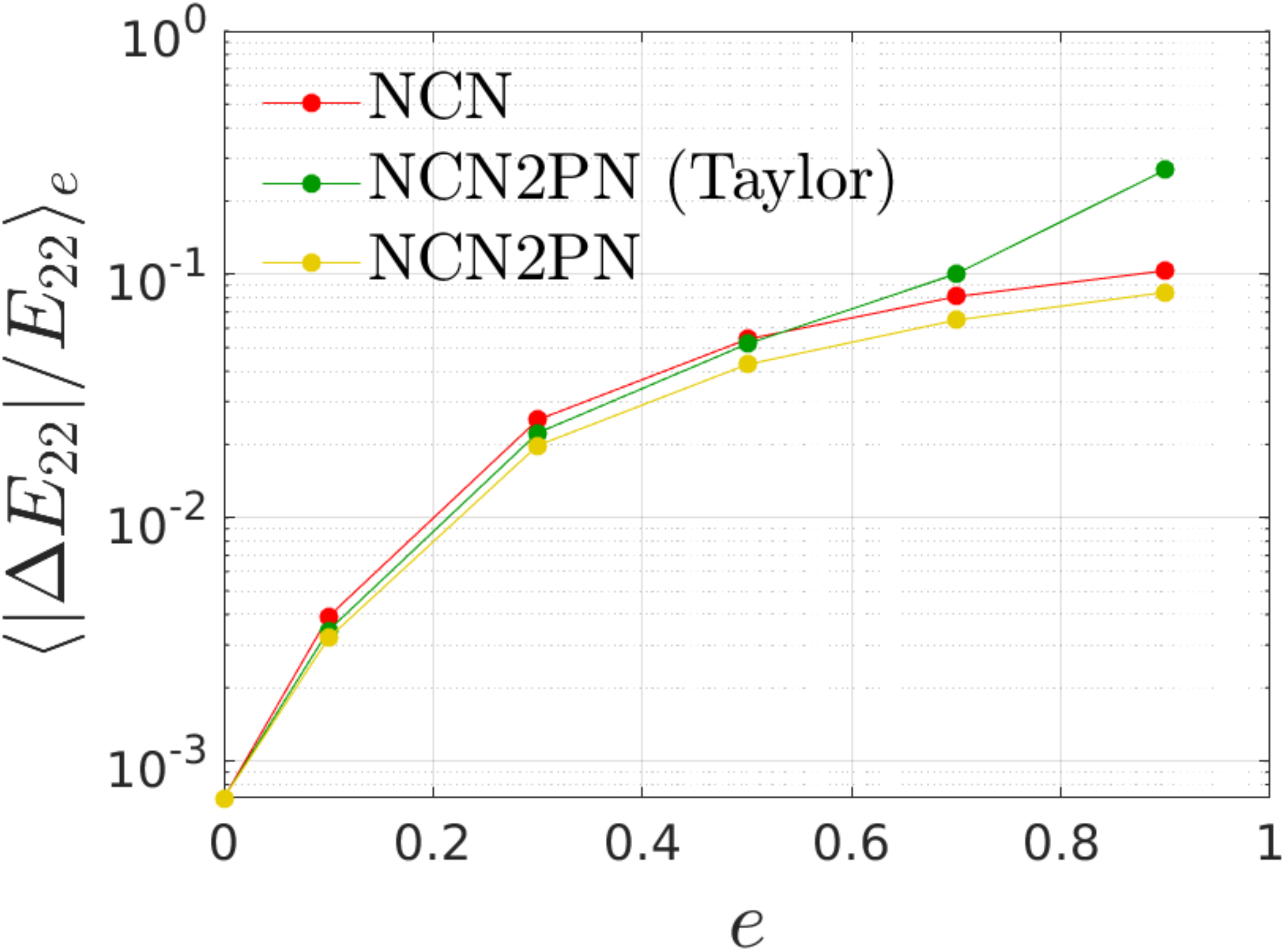}
  \caption{\label{fig:mean_averaged_2PN}
  Means of the analytical/numerical relative differences of the averaged fluxes, 
  graphical representation of Table~\ref{tab:mean_differences_2PN}. 
  The mean $\langle ...\rangle_e$ is performed over all the simulations with
  the same eccentricity. Since we consider two
  semilatera recta for each pair $(e,\ha)$ and $\ha=(0,\pm 0.2, \pm 0.6, \pm 0.9)$, 
  each point is an average over 14 configurations.} 
\end{figure} 
\begin{figure}[t]
  \center
  \includegraphics[width=0.22\textwidth]{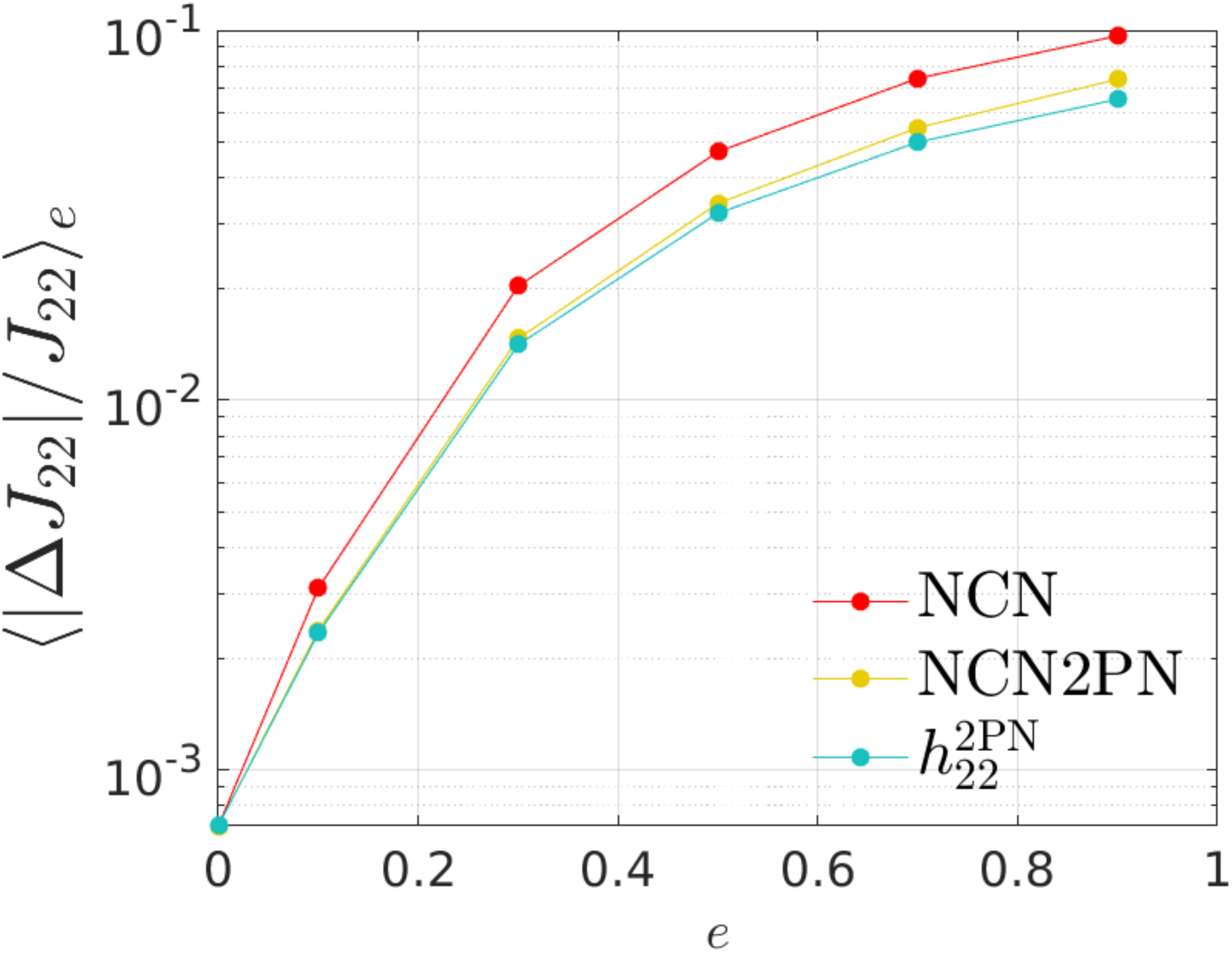}
  \includegraphics[width=0.22\textwidth]{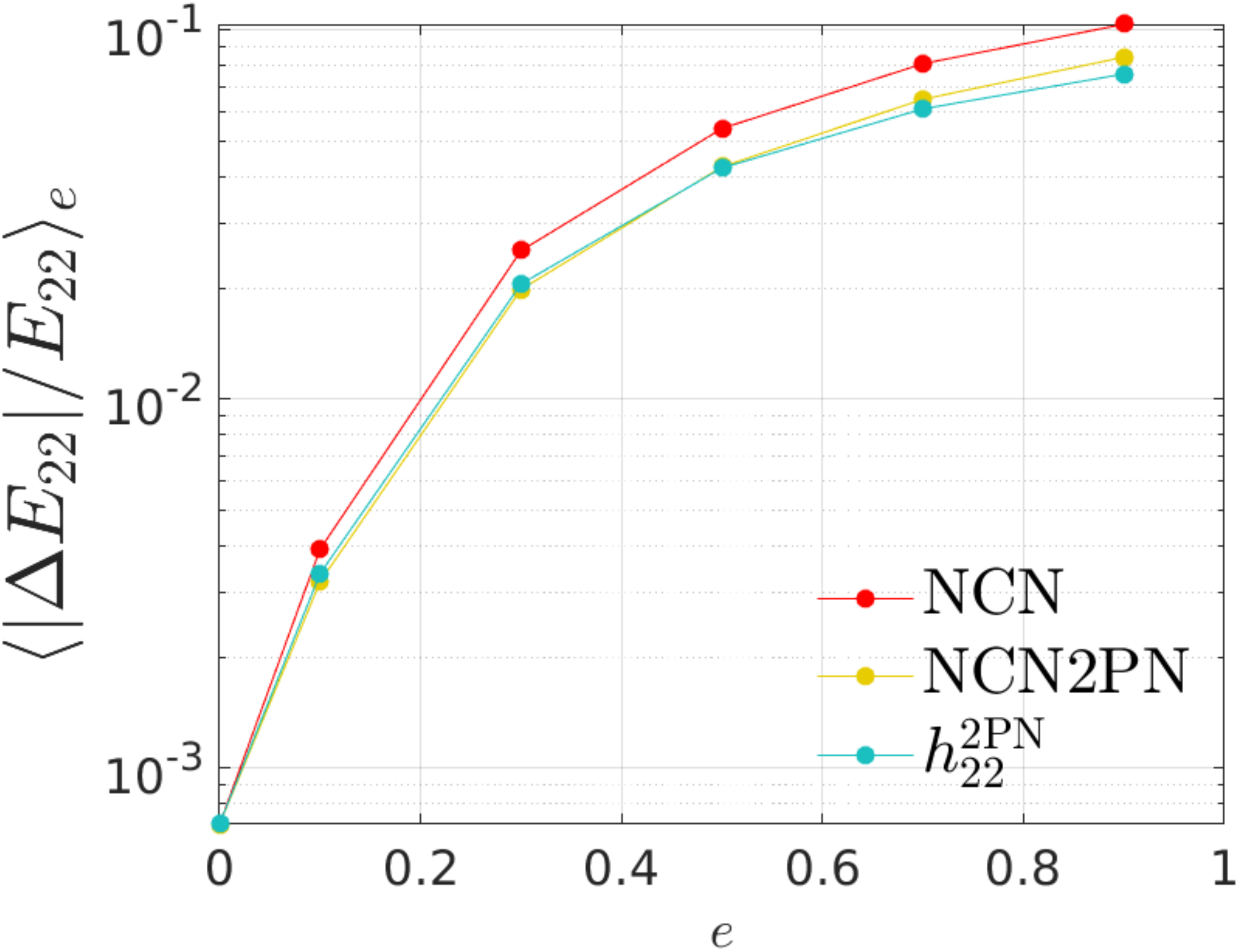}
  \caption{\label{fig:mean_averaged_hlm}
  Means of the analytical/numerical relative differences of the averaged fluxes. 
  The mean $\langle ...\rangle_e$ is performed over all the simulations with
  the same eccentricity. Since we consider two
  semilatera recta for each pair $(e,\ha)$ and $\ha=(0,\pm 0.2, \pm 0.6, \pm 0.9)$, 
  each point is an average over 14 configurations. We report the results
  for the NCN and NCN2PN fluxes, and also for the fluxes
  computed from the 2PN waveform of Ref.~\cite{Placidi:2021rkh} using 
  Eqs.~\eqref{eq:fluxes_infty}.}
\end{figure} 
We test the reliability of the resummed factor $\hat{F}^{\rm 2PN}_{22}$ in the test-mass limit,
focusing on the quadrupolar contributions to the angular momentum and energy fluxes, 
$\dot{J}_{22}$ and $\dot{E}_{22}$.
We start by considering two nonspinning configurations with eccentricities $e=0.1, 0.9$
in Fig.~\ref{fig:flux_2PN}. 
In the rightmost panels we show the 2PN noncircular correction $\hat{F}^{\rm 2PN_{nc}}_{22}$
with different resummation procedures:
in Taylor expanded form (green), with resummation only on the hereditary part 
$\hat{F}^{\rm 2PN_{nc, tail}}_{22}$ (blue),
and with resummation on both $\hat{F}^{\rm 2PN_{nc, inst}}_{22}$
and $\hat{F}^{\rm 2PN_{nc, tail}}_{22}$ (yellow). The latter is used in the 
radiation reaction (and fluxes) that we label as NCN2PN.
In the case with $e=0.1$, the three 2PN circular corrections are similar, 
while in the other configuration
with $e=0.9$ the effects of the resummation become relevant. 
It is also possible to see
that the resummation is more relevant for the hereditary part than for the instantaneous
factor. This is a consequence of the fact that $\hat{t}^{\rm 1.5PN}_{p_{r_*}^n}$ 
are eighth-order $y$-polynomials, while $\hat{f}^{\rm 2PN}_{p_{r_*}^n}$ are fourth-order. 
However, the contribution of the resummed correction to $\dot{J}_{22}$ is small even 
for $e=0.9$, as shown in the middle panels of Fig.~\ref{fig:flux_2PN}.

Deeper insight to the problem is obtained considering the analytical/numerical relative
differences of the averaged quadrupolar fluxes
and averaging over all the simulations with the same 
eccentricity, to obtain mean relative differences for each value of eccentricity, 
$\langle \Delta J_{22}/J_{22} \rangle_e$ and $\langle \Delta E_{22}/E_{22} \rangle_e$.
In Table~\ref{tab:mean_differences_2PN} and Fig.~\ref{fig:mean_averaged_2PN} 
we report these mean differences for three different radiation reactions: the
standard prescription NCN, the prescription that includes in NCN the
2PN noncircular correction $\hat{F}^{\rm 2PN_{nc}}_{22}$ in Taylor-expanded form,
and the prescription NCN2PN that includes the 2PN noncircular correction 
in NCN but with Pad\'e resummation.
As can be seen, the 2PN noncircular correction improves the radiation reaction 
NCN, but the resummation is needed in order to obtain more accurate results 
also for $e\gtrsim 0.6$. Indeed, for high eccentricity the periastron gets closer to the central
black hole, making the $y$-polynomials of $\hat{F}^{\rm 2PN_{nc}}_{22}$ grow too much. 
The resummation prevents this issue and leads better results also for low eccentricity.

However, note that with the factorization scheme here proposed the post-Newtonian 
noncircular corrections cannot improve the analytical/numerical agreement at periastron 
and apastron, since at the two radial turning points we have $p_{r_*}=0$ and
$\hat{F}^{\rm 2PN_{nc}}_{22}$ reduces to unity.
This could be an indication that using the EOB equations of motion at 2PN in the post-Newtonian
noncircular corrections is not the best way to proceed, while leaving the explicit derivatives
of coordinates and momenta could lead to more reliable fluxes. 
A first insight can be obtained computing the analytical fluxes considering
Eqs.~\eqref{eq:fluxes_infty} and the waveform with resummed 2PN corrections\footnote{
More precisely, we consider the waveform with 2PN noncircular corrections where 
the noncircular tail factor and the instantaneous noncircular phase are resummed.} proposed
in Ref.~\cite{Placidi:2021rkh}. As can be seen from Fig.~\ref{fig:mean_averaged_hlm}, 
this procedure is practically consistent with the flux at 2PN order discussed above. 
The exploration of this different procedure and the inclusion of higher multipoles 
is left to future work.

%===========================================================================================
\section{Conclusions}
%===========================================================================================
\label{sec:conclusions}
Future gravitational wave observations will
further explore the astrophysical binary black holes
populations \cite{LIGOScientific:2018mvr,LIGOScientific:2020ibl,LIGOScientific:2021djp}, 
possibly also accessing sources with
intermediate and extreme mass ratios \cite{LISA:2017pwj,Babak:2017tow,Gair:2017ynp}.
One central issue for this science is the availability of faithful
predictions of the complete gravitational waveform from the generic
orbital motion and including the fast motion regime of the merger.
The EOB is currently the only framework capable of 
predicting such waveforms for the entire range of astrophysical
binaries, i.e. for generic mass
ratios~\cite{Yunes:2010zj,Chiaramello:2020ehz,Nagar:2020xsk,Albanesi:2021rby,Nagar:2021gss}.
The radiation reaction force is one building block of any EOB model: 
while robust prescriptions for circularized binaries based on 
the factorized EOB waveform of Ref.~\cite{Damour:2008gu}
exist, less work has been done for generic
orbits, e.g.~\cite{Chiaramello:2020ehz,Nagar:2020xsk,Khalil:2021txt}. 
Improving the current prescriptions for the EOB radiation reaction
requires both new analytical information (PN results and EOB resummations) and
exact numerical data (at least in some regimes). 
The test-mass limit is a critical benchmark: any analytical EOB
structure must remain robust and accurate in this limit since this
permits a continuum deformation of the model in the symmetric mass
ratio and the calculation of the waveforms in different mass-ratio regimes. 
Yet this aspect is often overlooked when EOB models are tested.
In this paper we assessed the performances of different EOB radiation reactions
along generic planar orbits using the exact numerical result in the
test-mass limit, e.g. Refs~\cite{Bernuzzi:2011aj,Harms:2014dqa,Albanesi:2021rby}.
We considered three prescriptions put forward in the recent literature: 
(i) the quasicircular prescription (\QC{}) of {\SEOBNR} proposed for
eccentric mergers in Ref.~\cite{Ramos-Buades:2021adz}, 
(ii) the \QC{} with 2PN noncircular corrections
(\QCtPN{}) proposed in Ref.~\cite{Khalil:2021txt}, 
(iii) the \QC{} corrected by the noncircular Newtonian prefactor
(\NCN{}) proposed in Ref.~\cite{Chiaramello:2020ehz} and used in 
{\TEOBResumS}~\cite{Nagar:2020xsk,Nagar:2021xnh}.
The main finding is that the \NCN{} prescription performs best in
reproducing the numerical results; this is a feature of almost all the
cases analyzed in this work. This result is summarized in Fig.~\ref{fig:mean_averaged}
and Table~\ref{tab:mean_differences}.
The test-mass results appear to carry over to the comparable-mass
regime. Using a mildly eccentric ($\sim 0.2$) numerical relativity simulation of
equal-mass, nonspinning, binary  merger we demonstrated that the \NCN{} better 
captures the waveform phasing and frequency evolution than the \QC{} prescription 
(see Figs.~\ref{fig:eobnr} and~\ref{fig:omg_comp}). In particular,
the key issue is that the \QC{} choice incorrectly circularizes the dynamics
faster than the \NCN{} prescription. By contrast, the latter remains more coherent 
with the NR data up to merger.
The unphysical effect might be small on relatively short waveforms
($\sim 7$ orbits) and practically gauged away in EOB/NR unfaithfulness 
computation thus yielding acceptable values ($\sim 1\%$).
However, it is expected to build up over many GW cycles (as well as
for larger eccentricities) eventually introducing systematics in the
estimate of the parameters.
Our result thus reiterate the importance of a detailed control of
radiation reaction in the construction of EOB waveform models for
eccentric inspirals. 
Given the lack of accurate NR simulations
for eccentric inspirals with tenths of orbits, the use of
test-mass limit data is the only method to validate current analytical
prescriptions.  

We also proposed an improved \NCN{2PN} prescription
that incorporates 2PN corrections in the flux following the prescription of 
Ref.~\cite{Placidi:2021rkh}. The latter has shown that high PN noncircular terms can be introduced
within the waveform as a corrective multiplicative correction to the
generic Newtonian prefactor. Focusing on the $\ell=m=2$ mode we
demonstrated here that applying an analogous procedure to the flux
improves the agreement with the fluxes in the test-mass limit. The extension
to higher modes as well as the inclusion in an improved EOB model with
2PN noncircular flux is postponed to future work.

%===========================================================================================
\section{Acknowledgment}
%===========================================================================================

We thank Astro Vitelli for continuous support during the development of this work.
S.B. acknowledges support by the EU H2020 under ERC Starting Grant, No.~BinGraSp-714626.
M.O. and A.P. acknowledge support from the project ``MOSAICO'' financed by Fondo Ricerca di Base 2020 of the University of Perugia. M.O. and A.P. thank the Niels Bohr Institute for hospitality.
Computations were performed on the ``Tullio'' server in Torino, supported by INFN.

\appendix
%===========================================================================================
\section{Circular flux}
%===========================================================================================
\label{appendix:hatf}
\begin{figure}[t]
  \center
  \includegraphics[width=0.22\textwidth]{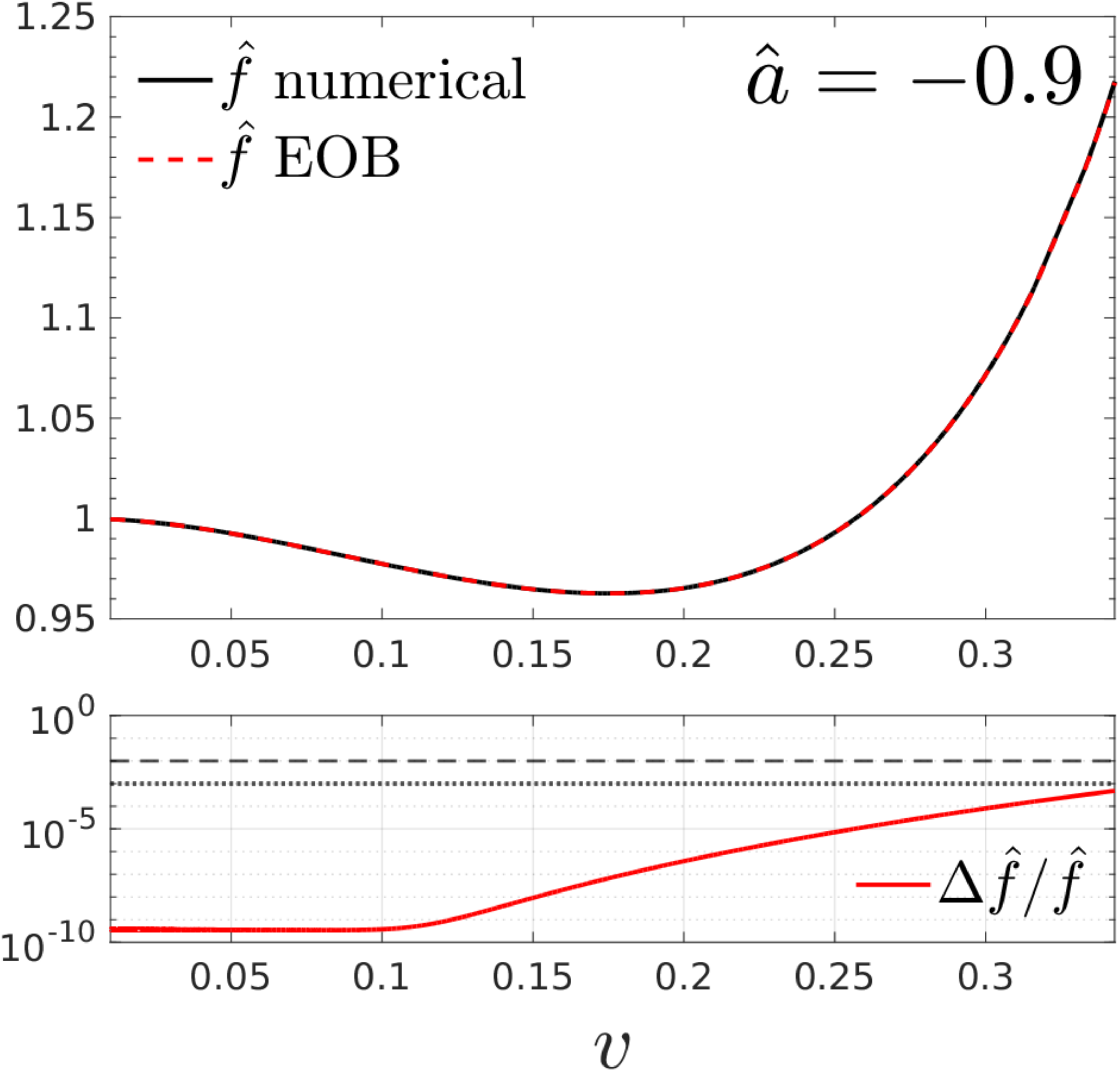}
  \hspace{0.2cm}
  \includegraphics[width=0.22\textwidth]{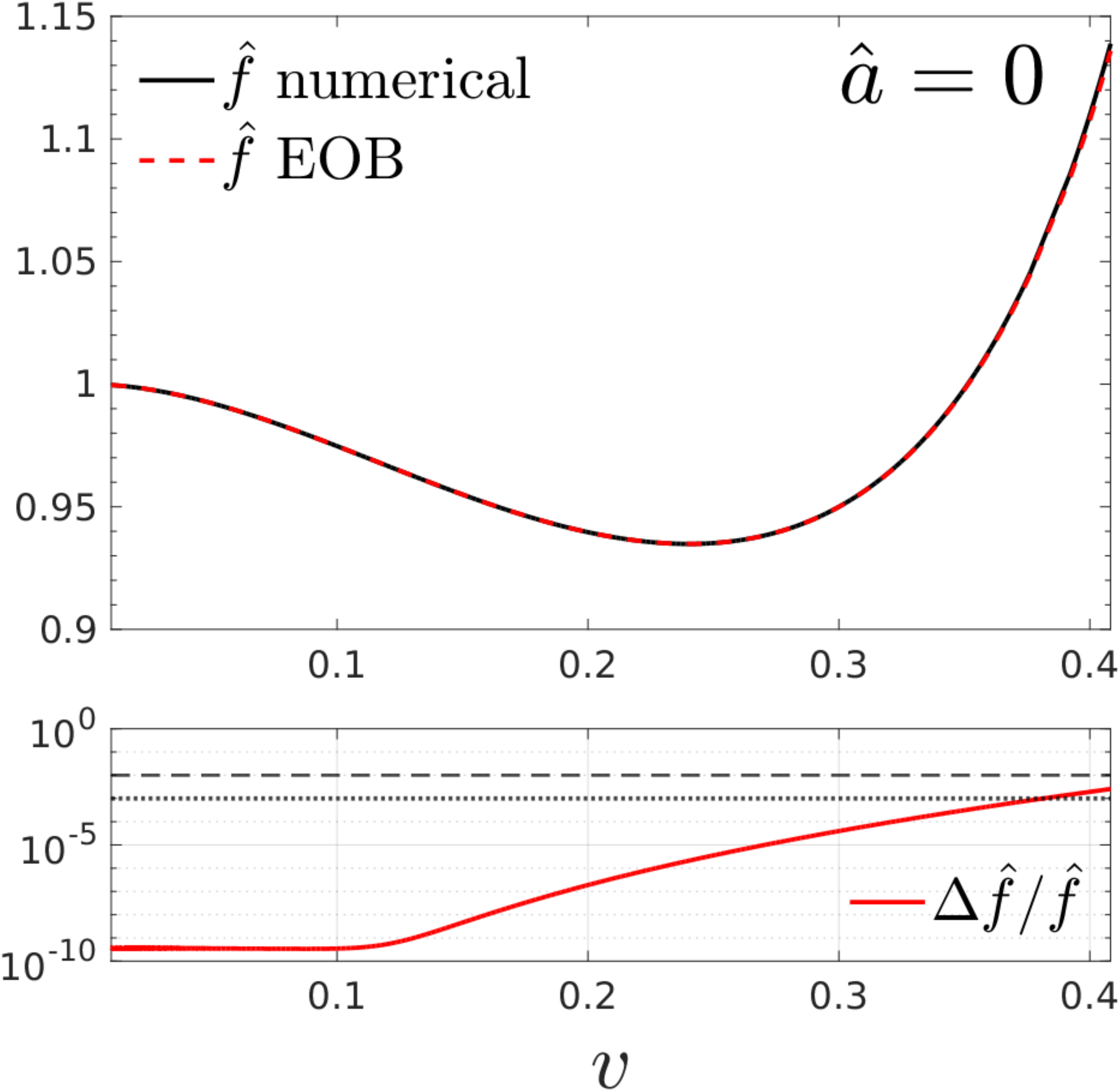}\\
  \includegraphics[width=0.22\textwidth]{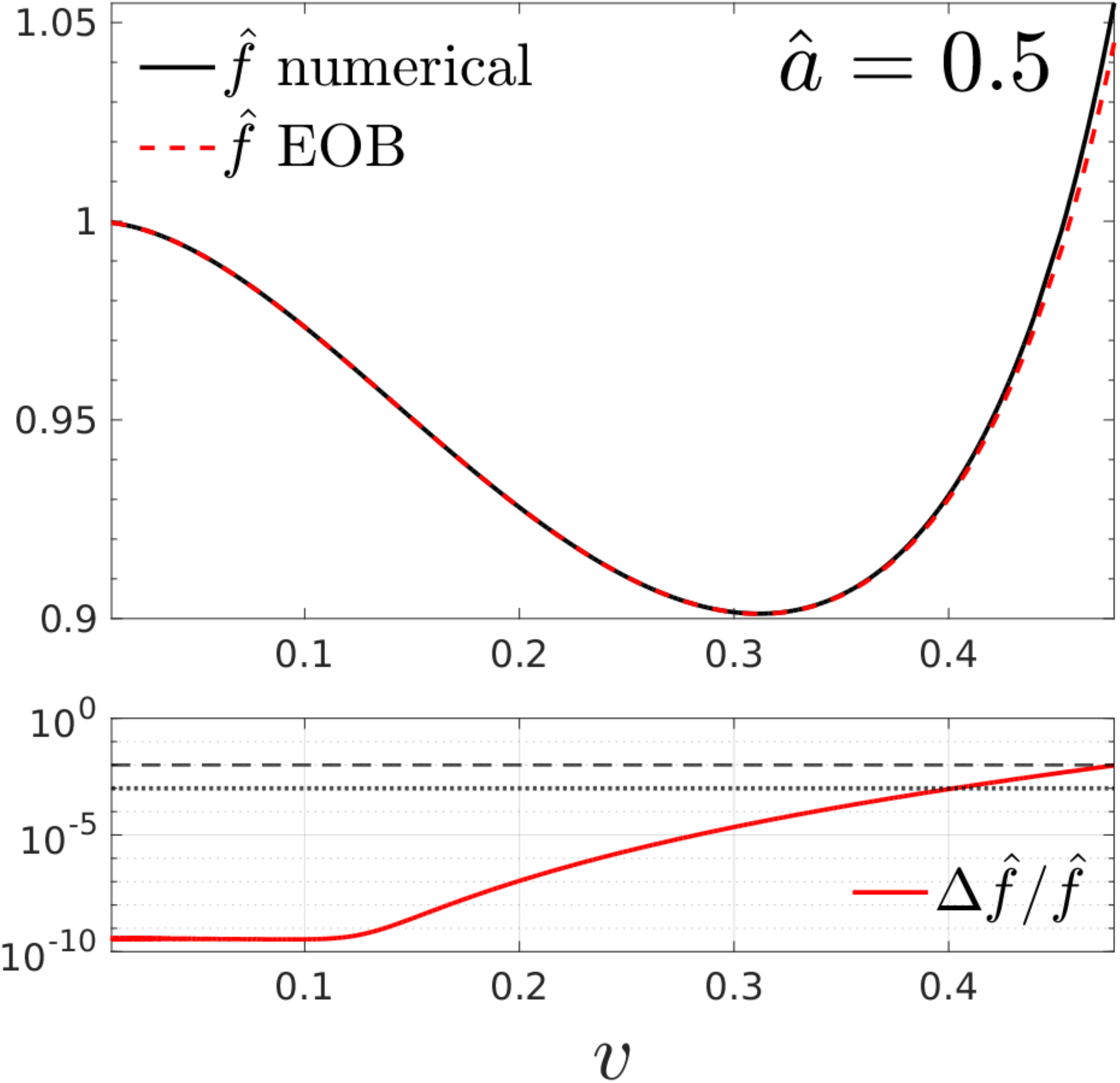}
  \hspace{0.2cm}
  \includegraphics[width=0.22\textwidth]{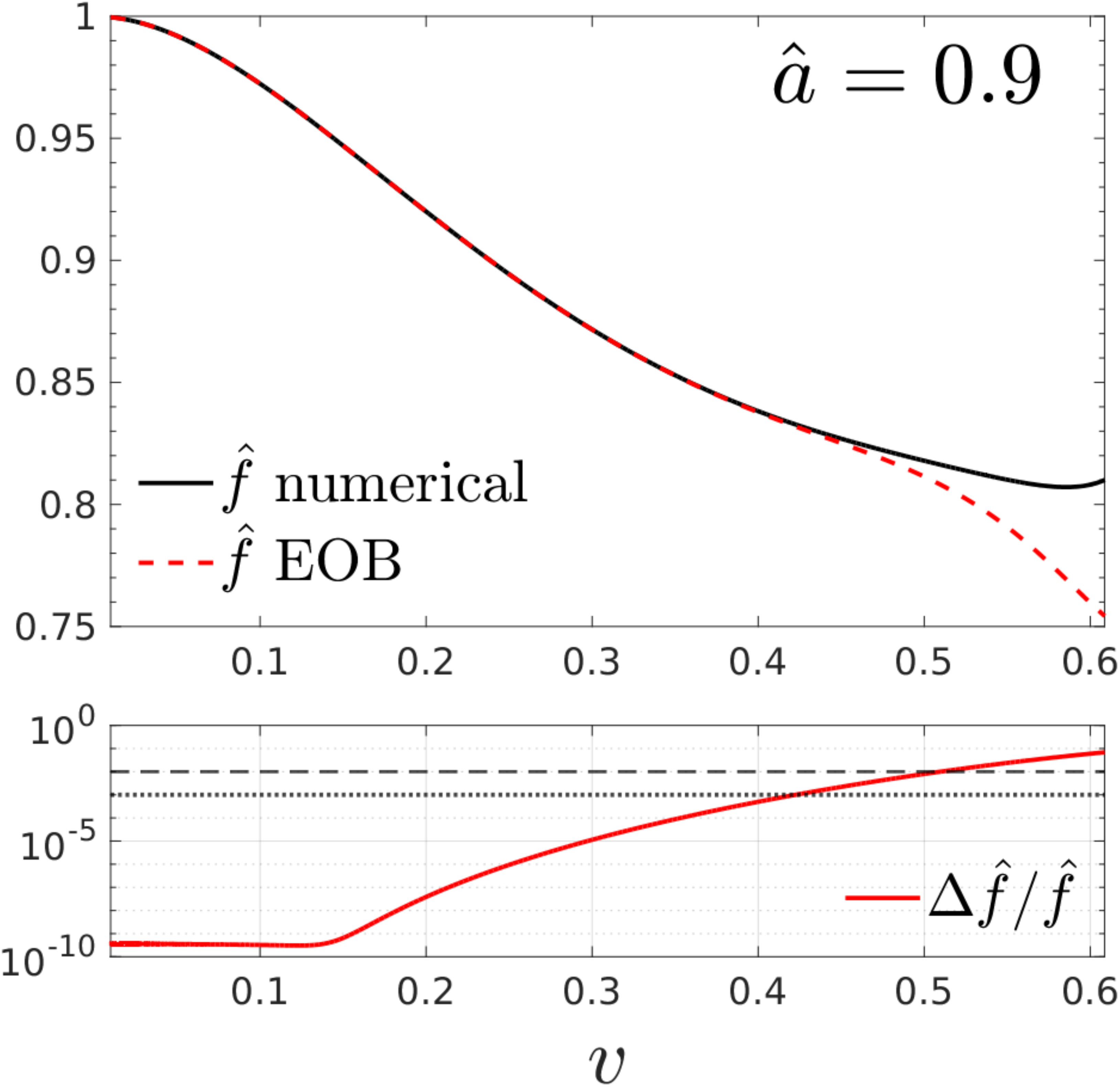}
  \caption{\label{fig:hatf}
  Analytical/numerical comparisons of the
  Newton-normalized circular flux $\hat{f}$ defined in Eq.~\eqref{eq:hatf}
  for four values of the Kerr spin parameter $\ha$. 
  The flux is plotted against $v=\Omega^{1/3}$, where $\Omega$ goes from zero to 
  the orbital frequency of the last stable orbit. 
  The two horizontal lines mark the $0.1\%$ and the $1\%$ thresholds. 
  In the bottom panel we show the relative difference 
  $(\hat{f}_{\rm num}-\hat{f}_{\rm EOB})/\hat{f}_{\rm num}$.}
\end{figure} 

In this Appendix we compare the circular flux with numerical
frequency-domain circular data kindly provided to us by S. Hughes~\cite{Taracchini:2013wfa}.
The comparisons of the analytical $\hat{f}$ as defined in Eq.~\eqref{eq:hatf}
with the numerical results are shown in Fig.~\ref{fig:hatf} for different
values of the spin $\ha$. The frequency interval on which we plot the 
two quantities is $[0,\Omega_{\rm LSO}]$, where $\Omega_{\rm LSO}$ is the 
orbital frequency of the Last Stable Orbit (LSO).
The analytical/numerical relative differences grow monotonically with
the frequency, reaching the $0.25\%$ at the LSO for $\ha=0$. Note
that we are not considering the absolute value of the difference,
so Fig.~\ref{fig:hatf} also shows that the analytical circular flux
is systematically slightly smaller than the numerical one.
We also verified that the resummation of the $\rho_{22}$ used through this paper
for the test-mass fluxes\footnote{
That is the one proposed in Ref.~\cite{Messina:2018ghh} and consists in
resumming the $\rho_{22}$ at 6PN after the factorization of the orbital contribution.} 
leads to better analytical/numerical agreement than the $\rho_{22}$ that is currently used in
the comparable-mass case, which is the 5PN Taylor-expanded 
result with $\nu$-contributions up to 3PN.
Indeed while the former gives a difference at LSO of $0.25\%$, the latter gives
$0.57\%$.
Moreover, we also checked how the differences vary when including the higher multipoles. 
For the cases with $\ha\lesssim 0$, the differences are almost independent of the number
of multipoles used, while for larger positive spins the inclusion of higher modes
significantly changes the agreement.
For example, in the case with $\ha=0.9$, the differences at LSO using only the
(2,2) mode is $2.5\%$, but increases, respectively, to 
$7.4\%$ and $7.0\%$ if all the multipoles up to $\l=5$ and $\l=8$ are considered.

%===========================================================================================
\section{Explicit expressions for $\F^{\rm ecc}_{\varphi,r}$}
%===========================================================================================
\label{appendix:Fecc}
We report here the explicit expressions, in the test-mass limit, of 
the radiation reaction components $\Fphiecc$ and $\Frecc$
that have been introduced in Ref.~\cite{Khalil:2021txt} and that
we have recalled in Eq.~\eqref{eq:FKhalil}.
Separating the PN orders we have
\begin{align}
\Fphiecc &= f^{\rm 0PN}_{\varphi, {\rm ecc}} + f^{\rm 1PN}_{\varphi, {\rm ecc}} +
f^{\rm 1.5PN}_{\varphi, {\rm ecc}} + f^{\rm 2PN}_{\varphi, {\rm ecc}}, \\
\Frecc &= f^{\rm 0PN}_{r, {\rm ecc}} + f^{\rm 1PN}_{r, {\rm ecc}} +
f^{\rm 1.5PN}_{r, {\rm ecc}} + f^{\rm 2PN}_{r, {\rm ecc}}.
\end{align}
where
\begin{widetext}
\begin{align}
f^{\rm 0PN}_{\varphi, {\rm ecc}} = & \frac{29 p_r^2 r-10 r^3 v_0^6+22}{12 r^2 v_0^4}, \\
f^{\rm 1PN}_{\varphi, {\rm ecc}} = & \frac{1}{4032 r^6 v_0^{10}}
\left[ -4928-13888 p_r^2 r-9744 p_r^4 r^2+\left(-28542 r^3-74718 p_r^2 r^4+6824 p_r^4 r^5\right)
   v_0^6+\right.  \nonumber \\
   & \left.\left(27434 r^4+36163 p_r^2 r^5\right) v_0^8+\left(19638 r^6+292 p_r^2 r^7\right)
   v_0^{12}-12470 r^7 v_0^{14}-1132 r^9 v_0^{18} \right], \\
f^{\rm 1.5PN}_{\varphi, {\rm ecc}} = & \frac{\pi}{1440 r^4 v_0^7} \left[ 15390 pr^2 + 10020 pr^4 r - 
 49 pr^6 r^2 + (210 r^2 - 8910 pr^2 r^3) v0^6 - 210 r^5 v_0^{12} \right] + \nonumber \\
  & \frac{1}{288 r^5 v_0^7} \left[ -352 \ha-2336 \ha p_r^2 r-720 \ha p_r^4 r^2+\left(904 \ha r^3+1482 \ha
   p_r^2 r^4\right) v_0^6- 552 \ha r^6 v_0^{12} \right], \\
f^{\rm 2PN}_{\varphi, {\rm ecc}} = & \frac{1}{1152 r^7 v_0^{10}} 
\left[  1408 \ha^2 p_r^2 r+1856 \ha^2 p_r^4 r^2+\left(4908 \ha^2 r^3+12080 \ha^2 p_r^2
   r^4\right) v_0^6+\right. \nonumber \\
   & \left. \left(-4356 \ha^2 r^5-5742 \ha^2 p_r^2 r^6\right) v_0^{10}-2532
   \ha^2 r^6 v_0^{12}+1980 \ha^2 r^8 v_0^{16} \right] + \nonumber \\
  & \frac{1}{12192768 r^{10} v_0^{16}} 
  \left[ 12418560+53625600 p_r^2 r+77051520 p_r^4 r^2+36832320 p_r^6 r^3+\right. \nonumber \\
   & \left. \left(34058304 r^3+226443168 p_r^2
   r^4+248429664 p_r^4 r^5-8358336 p_r^6 r^6\right) v_0^6+\right. \nonumber \\
   & \left. \left(-27653472 r^4-77932512 p_r^2 r^5-54678456
   p_r^4 r^6\right) v_0^8+\left(32614176 r^6+111894048 p_r^2 r^7-\right.\right. \nonumber \\
   & \left.\left. 730761696 p_r^4 r^8-122683680
   p_r^6 r^9\right) v_0^{12}+\left(-346723362 r^7-889467642 p_r^2 r^8+\right.\right. \nonumber \\
   & \left.\left. 58359600 p_r^4 r^9\right)
   v_0^{14}+\left(418059686 r^8+551078677 p_r^2 r^9\right) v_0^{16}+\left(-226596384 r^9-\right.\right. \nonumber \\
   & \left.\left. 428834448 p_r^2
   r^{10}-141429456 p_r^4 r^{11}\right) v_0^{18}+\left(212855418 r^{10}-8664156 p_r^2 r^{11}\right)
   v_0^{20}-\right. \nonumber \\
   & \left. 190027130 r^{11} v_0^{22}+\left(31642128 r^{12}+4003776 p_r^2 r^{13}\right) v_0^{24}-6419556
   r^{13} v_0^{26}+55771632 r^{15} v_0^{30} \right],
\end{align}
\begin{align}
f^{\rm 0PN}_{r, {\rm ecc}} = & \frac{7 p_r^2 r-5 r^3 v_0^6+11}{6 r^2 v_0^4}, \\
f^{\rm 1PN}_{r, {\rm ecc}} = & \frac{1}{2016 r^6 v_0^{10}}
\left[ -2464-5264 p_r^2 r-2352 p_r^4 r^2+\left(-14271 r^3-30621 p_r^2 r^4-1388 p_r^4 r^5\right)
   v_0^6+\right.  \nonumber \\
   & \left. \left(13717 r^4+8729 p_r^2 r^5\right) v_0^8+\left(9819 r^6-4654 p_r^2 r^7\right)
   v_0^{12}-6235 r^7 v_0^{14}-566 r^9 v_0^{18} \right], \\
f^{\rm 1.5PN}_{r, {\rm ecc}} = & \frac{\pi }{576 r^2 v_0}
\left[ 84+9376 p_r^2 r+5 p_r^4 r^2-84 r^3 v_0^6 \right] + \frac{r^5 v_0^7}{144} \left[ -176 \ha-112 \ha p_r^2 r+\right.  \nonumber \\
   & \left. \left(452 \ha r^3+606 \ha p_r^2 r^4\right) v_0^6-276
   \ha r^6 v_0^{12} \right], \\
f^{\rm 2PN}_{r, {\rm ecc}} = &  \frac{1}{576 r^7 v_0^{10}}
\left[ 704 \ha^2 p_r^2 r+448 \ha^2 p_r^4 r^2+\left(2454 \ha^2 r^3+3160 \ha^2 p_r^2
   r^4\right) v_0^6+\right. \nonumber \\
   & \left. \left(-2178 \ha^2 r^5-1386 \ha^2 p_r^2 r^6\right) v_0^{10}-1266
   \ha^2 r^6 v_0^{12}+990 \ha^2 r^8 v_0^{16}  \right] +  \nonumber \\
   & 
   \frac{1}{6096384 r^{10} v_0^{16}} 
   \left[ 6209280+22579200 p_r^2 r+25824960 p_r^4 r^2+8890560 p_r^6 r^3\right.  \nonumber \\
   & \left.+\left(17029152 r^3+107258256 p_r^2
   r^4+105895440 p_r^4 r^5+7160832 p_r^6 r^6\right) v_0^6+\right.  \nonumber \\
   & \left. \left(-13826736 r^4-29538936 p_r^2 r^5-13198248
   p_r^4 r^6\right) v_0^8+\left(16307088 r^6+75893328 p_r^2 r^7\right. \right. \nonumber \\
   & \left.\left. -291847248 p_r^4 r^8-55172880 p_r^6
   r^9\right) v_0^{12}+\left(-173361681 r^7-364399587 p_r^2 r^8\right. \right. \nonumber \\
   & \left.\left. -19976940 p_r^4 r^9\right)
   v_0^{14}+\left(209029843 r^8+133018991 p_r^2 r^9\right) v_0^{16}+\left(-113298192 r^9\right. \right. \nonumber \\
   & \left.\left. -151001928 p_r^2
   r^{10}-82145448 p_r^4 r^{11}\right) v_0^{18}+\left(106427709 r^{10}-53488818 p_r^2 r^{11}\right)
   v_0^{20}-\right. \nonumber \\
   & \left. 95013565 r^{11} v_0^{22}+\left(15821064 r^{12}-15597792 p_r^2 r^{13}\right) v_0^{24}-3209778
   r^{13} v_0^{26}+27885816 r^{15} v_0^{30} \right].
\end{align}
\end{widetext}
The expressions are written using the 0PN orbital velocity $v_0$ that reads 
\begin{equation}
v_0 = \frac{(1 + \dot{p}_r r^2)^{1/6}}{\sqrt{r}}.
\end{equation}

%===========================================================================================
\section{2PN-accurate subleading flux multipoles}
%===========================================================================================
\label{app:2PNfluxes}
In  Eq.~\eqref{eq:F2PN22} of the main text we showed the expression for the Taylor-expanded flux multipole $F^{\rm 2PN}_{22}$ which enters the factorization procedure described in Sec.~\ref{sec:2PN_resum}. Here we list for completeness all the other flux multipoles that are relevant at 2PN accuracy:

\begin{widetext}
	\begin{align}
		F^{\rm 2PN}_{21}&=-\frac{1}{c^2} p_\varphi^3 u^6\frac{1-4 \nu}{9}-\frac{1}{c^4}\Bigg\{p_\varphi^3 u^7 \Bigg[\frac{5}{126}+\frac{13 \nu }{63}-\frac{92 \nu ^2}{63}+p_\varphi^2 u \left(-\frac{85}{126}+\frac{335 \nu }{126}+\frac{10 \nu ^2}{63}\right)\Bigg]\Bigg\},\\
		F^{\rm 2PN}_{31}&= -\frac{1}{c^2}p_\varphi u^3(1-4 \nu)\Bigg[ u^2\Bigg(\frac{1}{24}-\frac{149 p_\varphi^2 u}{2016}+\frac{11 p_\varphi^4 u^2}{336}\Bigg)+p_{r_*}^2  u \Bigg(\frac{1}{336}+\frac{p_\varphi^2 u}{48}\Bigg) - \frac{p_{r_*}^4 }{84} \Bigg]\cr
		&+\frac{1}{c^4}p_\varphi u^4(1-4 \nu) \Bigg\{ u^2 \Bigg[\frac{2479}{6048}-\frac{31 \nu }{189}+p_\varphi^2 u \left(-\frac{863}{1008}+\frac{559 \nu }{2016}\right)+p_\varphi^4 u^2 \left(\frac{445}{1344}-\frac{347 \nu }{4032}\right)\cr
		&+p_\varphi^6 u^3 \left(\frac{47}{504}-\frac{55 \nu }{2016}\right)\Bigg]-p_{r_*}^2 u \Bigg[\frac{73}{336}-\frac{31 \nu }{168}+p_\varphi^2 u \left(-\frac{3779}{4032}+\frac{775 \nu }{1344}\right)+p_\varphi^4 u^2 \left(\frac{51}{224}-\frac{3 \nu }{14}\right)\Bigg]\cr
		&+p_{r_*}^4\Bigg[\frac{59}{672}-\frac{13 \nu }{672}+p_\varphi^2 u \left(-\frac{83}{224}+\frac{51 \nu }{224}\right)\Bigg]\Bigg\},\\
		F^{\rm 2PN}_{32}&=-\frac{1}{c^4} p_\varphi^3 u^6(1-3 \nu )^2\Bigg[\frac{5}{126} u \left(1+7 p_\varphi^2 u\right)-\frac{5}{252} p_{r_*}^2\Bigg],\\
		F^{\rm 2PN}_{33}&=-\frac{1}{c^2}p_\varphi u^3 (1-4 \nu)\Bigg[u^2 \Bigg(\frac{5}{8}+\frac{845 p_\varphi^2 u}{224}+\frac{115 p_\varphi^4 u^2}{112}\Bigg) +p_{r_*}^2 u \Bigg(\frac{5}{112}-\frac{205 p_\varphi^2 u}{112}\Bigg)-\frac{5}{28}p_{r_*}^4\Bigg]\cr
		&+\frac{1}{c^4}p_\varphi u^4 (1-4 \nu) \Bigg\{u^2 \Bigg[\frac{12395}{2016}-\frac{155 \nu }{63}+p_\varphi^2 u \left(\frac{45175}{1008}-\frac{5065 \nu }{288}\right)+p_\varphi^4 u^2 \left(\frac{66385}{4032}-\frac{37445 \nu }{4032}\right)\cr
		&+p_\varphi^6 u^3 \left(\frac{265}{168}-\frac{335 \nu }{672}\right)\Bigg]-p_{r_*}^2 u \Bigg[\frac{365}{112}-\frac{155 \nu }{56}+p_\varphi^2 u \left(\frac{68195}{4032}-\frac{81685 \nu }{4032}\right)+p_\varphi^4 u^2 \left(\frac{985}{224}-\frac{545 \nu }{56}\right)\Bigg]\cr
		&+p_{r_*}^4 \Bigg[\frac{295}{224}-\frac{65 \nu }{224}+p_\varphi^2 u \left(-\frac{545}{96}-\frac{3065 \nu }{672}\right)\Bigg]\Bigg\},\\
		%F^{\rm 2PN}_{41}&=0,\\
		F^{\rm 2PN}_{42}&=-\frac{1}{c^4}p_\varphi u^4 (1-3 \nu )^2\Bigg[u^2 \Bigg(\frac{205}{4536}-\frac{5 p_\varphi^2 u}{294}-\frac{575 p_\varphi^4 u^2}{10584}+\frac{55 p_\varphi^6 u^3}{1764}\Bigg)-p_{r_*}^2 u \Bigg(\frac{25}{588}-\frac{865 p_\varphi^2 u}{10584}\Bigg)\cr
		&+p_{r_*}^4 \Bigg(\frac{10}{441}-\frac{65 p_\varphi^2 u}{1764}\Bigg)\Bigg],\\
		%F^{\rm 2PN}_{43}&=0,\\
		F^{\rm 2PN}_{44}&=-\frac{1}{c^4} p_\varphi u^4 (1-3 \nu )^2 \Bigg[u^2 \Bigg(\frac{205}{648}+\frac{125 p_\varphi^2 u}{42}+\frac{1115 p_\varphi^4 u^2}{216}+\frac{145 p_\varphi^6 u^3}{252}\Bigg)-p_{r_*}^2 u \Bigg(\frac{25}{84}+\frac{1705 p_\varphi^2 u}{1512}\cr
		&+\frac{95 p_\varphi^4 u^2}{42}\Bigg)+p_{r_*}^4\Bigg(\frac{10}{63}-\frac{25 p_\varphi^2 u}{252}\Bigg)\Bigg].
	\end{align}
\end{widetext}
Notice that, at the 2PN accuracy we are considering here, in all of the subleading flux multipoles only the instantaneous part survives. In other words, for these multipoles, the hereditary effects enter at higher orders than 2PN.
\clearpage

\bibliography{refs20220519.bib,refs_loc20220519.bib}

\end{document}